\documentclass{aastex}
\usepackage{rotating}
\usepackage{epsfig}  
\usepackage{epstopdf} 
\usepackage{graphics}
\usepackage{listings}

\begin{document}


\title{The Identification of Hydrogen-Deficient Cataclysmic Variable Donor
Stars$^{\rm 1,2}$}

\author{Thomas E. Harrison}

\affil{Department of Astronomy, New Mexico State University, Box 30001, MSC 
4500, Las Cruces, NM 88003-8001}

\begin{abstract}
We have used ATLAS12 to generate hydrogen-deficient stellar atmospheres to 
allow us to construct synthetic spectra to explore the possibility that the 
donor stars in some cataclysmic variables (CVs) are hydrogen deficient. We 
find that four systems, AE Aqr, DX And, EY Cyg, and QZ Ser have significant 
hydrogen deficits. We confirm that carbon and magnesium deficits, and sodium 
enhancements, are common among CV donor stars. The three Z Cam systems we 
observed are found to have solar metallicities and no abundance anomalies. Two 
of these objects, Z Cam and AH Her, have M-type donor stars; much cooler than 
expected given their long orbital periods. By using the combination of 
equivalent width measurements and light curve modeling, we have developed the 
ability to account for contamination of the donor star spectra by other 
luminosity sources in the binary. This enables more realistic assessments of 
secondary star metallicities. We find that the use of equivalent width 
measurements should allow for robust metallicities and abundance anomalies to 
be determined for CVs with M-type donor stars.
\end{abstract}


\begin{flushleft}
$^{\rm 1}$Partially based on observations obtained with the Apache Point 
Observatory 3.5-meter telescope, which is owned and operated by the 
Astrophysical Research Consortium.\\
$^{\rm 2}$Partially based on observations obtained at the Gemini Observatory, 
which is operated by the Association of Universities for Research in Astronomy, 
Inc., under a cooperative agreement with the NSF on behalf of the Gemini 
partnership: the National Science Foundation (United States), the National 
Research Council (Canada), CONICYT (Chile), Ministerio de Ciencia, 
Tecnolog\'{i}a e Innovaci\'{o}n Productiva (Argentina), and Minist\'{e}rio da 
Ci\^{e}ncia, Tecnologia e Inova\c{c}\~{a}o (Brazil).  \\

\end{flushleft}

\section{Introduction}

Cataclysmic variable stars (CVs) consist of a white dwarf that is accreting 
matter from a late-type secondary star. The standard paradigm for the formation
of CVs postulates that they start out as wide binaries ($a >$ 130 R$_{\sun}$, 
Warner 1995), whose orbital separation depends on the ratio of the initial to 
final mass of the primary, and is also a function of the binary mass ratio 
(see Ritter 2012). For single stars, to produce the typical 0.8 M$_{\sun}$ 
white dwarf found in CVs (Knigge et al. 2011), requires a main sequence dwarf 
with an initial mass of 5 M$_{\sun}$ (Salaris et al.  2009). As the white dwarf
progenitor evolves off of the main sequence, the secondary star suddenly finds 
itself orbiting within the red giant photosphere. During this common envelope 
(CE) phase, most of the angular momentum of the binary is believed to be shed 
due to interactions of the secondary star with the atmosphere of the red giant. During this process, the 
binary period gets shortened, and the red giant envelope gets ejected. 
Depending on the input parameters used in ``standard'' CE models, between 50 
and 90\% of all these close binary stars merge (c.f., Politano \& Weiler 2007).
At the end of this process, 50 to 80\% of the initial mass of the progenitor 
system has been lost (Ritter 1976). After emerging from the CE phase, an epoch 
of angular momentum loss via magnetic braking (e.g., King \& Kolb 1995), 
lasting $\sim$ 10$^{\rm 7}$ yr (Warner 1995), is then required that leads to 
the formation of a ``semi-detached'' binary, and the mass transfer phase that 
signals the birth of a CV.

If this framework were true, the majority of donor stars in CVs would not be 
expected to have undergone any significant nuclear evolution, they would 
perhaps be slightly bloated main-sequence-like stars (see Howell et al.  2001),
but not otherwise unusual.  After a considerable amount of observational data 
had been amassed that showed that many donor stars had cooler spectral types 
than expected for their orbital period (e.g., Friend et al. 1990ab, or
Beuermann et al. 1998; see Fig. 30 in Harrison 2016 for a recent update), had 
masses and radii inconsistent with main sequence stars (e.g., Bitner et al. 
2007, Echevarr\'{i}a et al. 2007, Neustroev \& Zharikov 2008, Rodriguez-Gil et 
al. 2009), and infrared spectra that revealed unusual abundances, especially 
strong deficits of carbon (Harrison et al. 2004a, 2005, 2009), it was clear 
that many CV donor stars bore little resemblance to normal main sequence 
dwarfs. Additionally, it was noted decades ago that the helium emission lines 
in many CVs were much stronger than expected, implying an enhanced abundance of 
helium in the material being transferred from the secondary star
(Williams \& Ferguson 1982). The UV spectra of many CVs also show unusual 
carbon to nitrogen line ratios (e.g., Bonnet-Bidaud \& Mouchet 1987, Szkody 
\& Silber 1996, Mauche et al. 1997, G\"{a}nsicke et al. 2003), suggesting 
deficits of carbon, and enhancements of 
nitrogen. Hamilton et al. (2011) found that CV donor stars that had carbon 
deficits, inferred from weak CO absorption features in $K$-band spectra, also 
had unusually weak C IV emission lines in the UV. Clearly, the donor stars of 
some CVs are chemically peculiar, presumably due to post-main sequence 
evolution prior to contact.

Goliasch \& Nelson (2015) have produced a new population synthesis calculation 
for CVs that includes detailed nuclear evolution, with more massive progenitor 
binaries. They confirm the results of Podsiadlowski et al. (2003), showing 
that there is a large range of masses and radii for donors in CVs with 
P$_{\rm orb}$ $>$ 5 hr when nuclear evolved secondary stars are considered. 
This effect arises due to some donors being zero age main sequence stars at 
the time they become semi-contact binaries, while others have undergone 
significant nuclear processing. The results of our CV temperature/abundance 
survey (Harrison 2016) confirms this conjecture, finding normal main sequence 
star donors in some systems, and secondaries that appear to be highly evolved 
in others. 

Podsiadlowski et al. suggest that the secondaries that have undergone pre-CV 
evolution should show evidence for nuclear processing, in particular, products 
of the CNO cycle. Harrison \& Marra (2017) used moderate resolution $K$-band 
spectroscopy to investigate the isotopic ratio of carbon to test whether there 
was evidence for CNO cycle processes. For the three systems they observed, 
small values of the $^{\rm 12}$C/$^{\rm 13}$C ratio ($\leq$ 15) were found. 
This indicates that all three donor stars had begun to evolve off the main
sequence before becoming semi-contact binaries. Surprisingly, Harrison \& 
Marra also found enhancements in the sodium abundance for all three objects.

In Harrison (2016), unusual abundances of sodium were found in two CVs:
QZ Ser and EI Psc. Both of these objects are short period CVs, but appear
to have K-type secondary stars, instead of the expected M-type (or later)
secondaries. Thorstensen et al. (2002) suggested one possible way to
explain the donor in QZ Ser is that it is hydrogen deficient, with an 
enhancement of helium. Perhaps an atmosphere that is hydrogen-deficient could 
explain many of the abundance anomalies found in Harrison (2016) or Harrison \&
Marra (2017). To test this we have generated a grid of hydrogen deficient
stellar atmospheres using ATLAS12 (Kurucz 2005) for the purpose of creating
synthetic spectra to derive the abundances for CV donor stars. We have found
evidence for hydrogen deficiencies in four systems. In the next section we 
discuss the data used, the construction of the atmosphere models, accounting 
for possible contamination, and the equivalent width measurement process. 
In section 3 we present our analysis of the infrared spectra of CV 
secondary stars, and in section 4 we discuss our results, presenting our 
conclusions in section 5.

\section{Data Reduction and Analysis Techniques}

\subsection{$JHK$ Spectra for the Program Objects}

The observational details of the IRTF/SPEX near-infrared spectroscopic data 
used below are listed in Harrison (2016, ``H16''). In this paper, we only 
analyze the cross-dispersed data for CVs that are expected to have K-type 
donor stars. To the SPEX data set we add observations made with 
GNIRS\footnote{https://www.gemini.edu/sciops/instruments/gnirs/} on Gemini-North, and with TripleSpec\footnote{http://www.apo.nmsu.edu/arc35m/Instruments/TRIPLESPEC/} on the Apache Point 
Observatory 3.5 m. An observation log for the latter two data sets is presented
in Table 1. Both GNIRS and TripleSpec produce crossed-dispersed data across
the $JHK$ bandpasses with a resolution similar to that of SPEX on the IRTF.
The GNIRS and TripleSpec observations were reduced in the usual fashion using
IRAF (see H16). For all of the objects, telluric standards were observed at a 
similar airmass as the program target. The GNIRS spectra were wavelength 
calibrated using arc lamp spectra, while for TripleSpec we used night sky OH 
emission lines.

\subsection{Generation of Hydrogen-Deficient Atmospheres and Synthetic Spectra}

To allow the generation of hydrogen-deficient spectra, we employed ATLAS12 
(Kurucz 2005). Compared to earlier versions of the ATLAS program series,
ATLAS12 uses the ``Opacity Sampling'' method to compute line opacity
in atmosphere models. Models generated
using ATLAS12 can have the abundances of individual elements set while
calculating an atmosphere, versus only being able to change the global 
metallicity in the ``Opacity Distribution Function'' (ODF) method used in 
ATLAS9, and earlier versions of this software. Where to download the codes, 
and how to run the software, are described in Castelli (2005).

To run ATLAS12, one needs to have pre-computed atmospheres for input. As
described in Castelli, these can be ODF models generated using ATLAS9. Thus, 
to generate a grid of ATLAS12 models, requires a set of standard Kurucz
ODF atmospheres covering the desired range of parameters. Given that none
of the objects modeled here appeared to have extremely low metallicities, our 
grid covers the ranges $-$0.5 $\leq$ [Fe/H] $\leq$ $+$0.5 in steps of $\pm$0.1 
(less [Fe/H] = $\pm$0.4), and 4000 K $\leq$ T$_{\rm eff}$ $\leq$ 5500 K in 
steps of 250 K. We computed this range for two different gravities (log$g$ = 
4.0 and 4.5), and at five different H-deficiencies: 0\%, 25\%, 50\%, 75\%, and 
90\%. Obviously, if we change the hydrogen abundance, the actual value of
[Fe/H] also changes. To reduce confusion, and because we are not altering
the abundances of any of the elements when running models of different
metallicities, we keep the labeling consistent: if the ATLAS9 input atmosphere
model had [Fe/H] = $-$0.3, no matter the hydrogen abundance of the ATLAS12
output model, we will label this an [Fe/H] = $-$0.3 atmosphere model.

Before we started, we had to make a choice for the helium abundance.
For a solar abundance, ATLAS12 lists the hydrogen abundance as $log$H = 
$-$0.036, and helium as $log$He = $-$1.106. This is not the standard notation 
by which such abundances are usually expressed, but is the format used by 
Kurucz. Normally the $log$ of the abundance of hydrogen is set equal to 12.0. 
To convert between the two scales one uses the equation $log$(A/N$_{H}$) = 
12.0 + $log$(A/N$_{total}$) $-$ $log$(N$_{H}$/N$_{total}$). Where 
$log$(A/N$_{total}$) is the abundance of the element A in terms of number 
density relative to total number density. The value of 
$log$(N$_{H}$/N$_{total}$) = $-$0.036. 

To test how the He 
abundance actually affects the results we ran three models. The first has 
$log$H = $-$0.3370 (a hydrogen deficit of 50\%), $log$He =  $-$1.106 (the solar 
value), the second has $log$H = $-$0.3370, $log$He = $-$0.8046 (He abundance 
increased by 50\%), and $log$H = $-$0.3370, $log$He =  $-$0.2696 (He abundance 
exactly replaces the lost hydrogen). We ran the programs described below to 
measure the equivalent widths in the near-infrared bandpasses. The last of 
these three models, exact replacement, produces spectra with the largest 
equivalent widths.  For example, the equivalent width measurements for the 
sodium doublet at 22,000 \AA ~in the $K$-band for each of these three models 
(T$_{\rm eff}$ = 5000 K, [Fe/H] = 0.0) are 1.44 \AA, 1.47 \AA, and 1.56 \AA. 
On average, the equivalent width measurements for the exact replacement models 
are $\sim$ 10\% larger for most lines than in either of the other two 
models. This is near the noise level for most of the measurements we present 
below. Thus, for simplicity, we decided to enhance the helium abundance to 
exactly make-up for the hydrogen deficiency, keeping the total abundance of H 
$+$ He constant. If the helium abundance is lower than this, any hydrogen 
deficits found below will actually be slightly larger than measured.

The final grid of atmosphere models, in the form preferred as input to Gray's 
SPECTRUM\footnote{http://www.appstate.edu/~grayro/spectrum/spectrum.html} 
program, can be found at http://astronomy.nmsu.edu/tharriso/hdeficient.  
We then used the program SPECTRUM, as described in H16, to generate synthetic 
spectra using the new hydrogen-deficient atmospheres. As discussed in Harrison 
\& Marra (2017, ``HM17''), two supplemental programs supplied with SPECTRUM, 
{\it avsini} and {\it smooth2}, were used to rotationally broaden and to
match the dispersion of the model spectra to that of the data. Note
that the three instruments whose data we analyze below have slightly different
dispersions, and these were taken into account when generating model grids
for each of the objects.

\subsection{Infrared Light Curve Data for Contamination Estimates}

In H16 and HM17, no attempt was made to examine the issue of contamination of 
the observed spectra due to emission by sources other than the donor star. Such 
a process can strongly affect the derived global metallicities, as it 
simultaneously weakens the apparent strength of all of the absorption features 
in the donor star spectrum. As we will show, it appears to have had little 
affect on the temperatures or relative elemental abundances they derived for the
systems in common which we analyze below. To derive a contamination estimate, 
however, is difficult. It either requires the a priori fixing of the 
metallicity, or the use of some source of information that allows us to check 
whether a particular contamination level is likely. We use light curve modeling 
to supply this additional source of information.

For three targets below, DX And, SS Cyg, and RU Peg, we model $JHK$ light 
curves the data for which were obtained with 
SQIID\footnote{https://www.noao.edu/kpno/sqiid/} on the KPNO 2.1 m telescope. 
The photometry for SS Cyg comes from Harrison et al. (2007). The light curve 
data for RU Peg was collected over three consecutive nights, 3 to 5 August 
2003. The observations for DX And occurred on 7 and 10 August 2003. For each 
of the three CVs, differential photometry was performed relative to stars in 
the field, and calibrated using data from 2MASS. To these data we add the light
curve photometry from $WISE$ (Wright et al. 2010) and $NEOWISE$ (Mainzer et al.
 2011).

For the majority of CVs, we only have $WISE$ and $NEOWISE$ light curves to work 
with. We use the Wilson-Divinney light curve modeling software 
WD2010\footnote{ftp://ftp.astro.ufl.edu/pub/wilson/lcdc2010} to produce
models for the $JHK$ and $WISE$/$NEOWISE$ photometry. See Harrison
et al. (2013a) for details on how WD2010 can be used to model $WISE$
light curves. For light curve modeling, the main influence of a source
of contamination (``third light'') is its affect on the derived orbital
inclination. A source of contamination dilutes the ellipsoidal variations
that arise from the distorted shape of the secondary star, driving the 
apparent orbital inclination to lower values.

\subsection{Equivalent Width Measurements}

For the analysis used below, we change from the $\chi^{\rm 2}$-based 
analysis of H16 and HM17 to the measurement of equivalent widths. The main 
reason for this switch is the fact that different elements have different 
responses to a change in the hydrogen abundance. The increase in the 
complexity introduced by possible hydrogen deficits makes the results
from $\chi^{\rm 2}$ analysis ambiguous, as analysis of broad segments of the 
spectra can produce identical $\chi^{\rm 2}$ values due to either a large 
hydrogen deficit, or a high metallicity. Thus, while individual elemental 
abundances (especially carbon) are derived using the same $\chi^{\rm 2}$ 
analysis techniques used in H16 and HM17, to examine the donor star spectra 
for possible hydrogen deficits requires us to compare the equivalent widths 
of various lines from individual elements in the object spectra to those 
of the model spectra.  A secondary reason for this switch is the fact that
the $\chi^{\rm 2}$ analysis is more sensitive to the quality of the continuum
division than are equivalent width measurements. 

To demonstrate how the spectra change with hydrogen abundance, we 
plot synthetic $K$-band spectra for five different values in Fig. 
\ref{speccomp}. All of the spectra have identical parameters, T$_{\rm eff}$ = 
5000 K, and [Fe/H] = 0.0. As might be expected, diminishing the hydrogen 
abundance results in stronger metal lines in the synthetic spectra. While the 
lines all get stronger as we decrease hydrogen (since we are essentially
increasing the relative metallicity), there are subtle differences 
on how the lines from each of the metals responds due to the changes in
the structure of the stellar atmospheres. Focusing on the spectral 
range from 22500 to 23200 \AA ~we have three large absorption features: the 
Ca I triplet at 22632 \AA, Mg I at 22808 \AA, and the 
$^{\rm 12}$CO$_{\rm (2,0)}$ bandhead at 22937 \AA. It is clear that the CO 
bandhead does not strengthen as quickly as either Ca I or Mg I, and Ca I 
strengthens more quickly than Mg I. In Fig. \ref{eqwbands} we plot the 
equivalent widths for the Na I doublet at 22000 \AA, the Ca I triplet, and the 
Mg I line for the five different hydrogen abundances. The Na I doublet 
strengthens more rapidly with decreasing hydrogen abundance than the lines for 
either of those other two elements.

We will use this type of analysis for all of the objects discussed below.
We calculate the equivalent widths and their errors using the method
discussed in Vollmann \& Eversberg (2006). In Table \ref{eqwwaves} we list the 
spectral limits for each element's equivalent width measurement along with 
those of their continuum bandpasses. Where there are multiple lines from
an element in one of the near-IR bandpasses, we sum the equivalent widths
of those lines, and add their errors in quadrature. Though we have selected
these bandpasses because they are mostly free from strong emission features
from the program CVs, some of them will occasionally be affected by a
stronger-than-normal emission line from the CVs, or from a poor telluric 
absorption correction, such as the red end of the $J$-band. In
cases where there are noise spikes or other features located near a spectral
line of interest, we have resorted to using the {\it splot} routine in IRAF to 
measure equivalent widths. We find that the equivalent width measurements 
produced by our program agree very well with those from {\it splot}. We will 
note these object-specific issues as we encounter them. It is also important 
to note that the equivalent widths for Na I in the $J$-band and $H$-bands are 
often suspect. The Na I doublet in the $J$-band falls on the red edge of a 
strong telluric feature, and the line and continuum regions are always very 
noisy. In the $H$-band, the Na I doublet is quite weak, and falls amongst a 
large number of other absorption lines, so its measurement can be 
unreliable, especially for cooler stars. As shown in HM17, the abundance
of carbon can be derived to a high precision from the CO features in the
$K$-band using the $\chi^{\rm 2}$ method. We will not be using the equivalent 
width measurement techniques to examine its abundance.

To provide a test of the reliability of the equivalent width measurement 
process we return to the spectra of the K dwarfs from the IRTF Spectral Library
(Cushing et al. 2005) used in H16. 
To this data set we add the star 61 Cyg A observed using SPEX (Harrison
et al. 2004a), with the means of its data from the PASTEL catalog (Soubiran et 
al.  2016). We list the mean temperatures and metallicities for these
K dwarfs from the PASTEL catalog in Table \ref{mkdata}. In Fig. \ref{mkeqw}, 
we plot the equivalent width measurements for the eight K dwarfs in the three 
bandpasses against the results from the synthetic spectra for six different 
metallicities: [Fe/H] = $-$0.2, $-$0.1, 0.0, $+$0.1, $+$0.2, and $+$0.5. 
This range in metallicities covers the observed range for the 
K dwarfs listed in Table \ref{mkdata}. 

We used these same objects in H16 to demonstrate the efficacy of the 
$\chi^{\rm 2}$ analysis method in deriving their previously published 
temperatures and metallicities. Thus, it is not surprising that the equivalent 
width measurements for these K dwarfs are consistent with the results for the 
synthetic spectra to within the error bars of the equivalent width 
measurements, assuming the tabulated metallicities (and error bars) listed in 
Table \ref{mkdata}. The equivalent width measurements for the K4V star HD45977, however, {\it generally} appear to be larger than expected given the data in 
Table \ref{mkdata}, and differ from the results in H16. A metallicity closer 
to [Fe/H] = $+$0.2 would be a better match to its measurements. The
PASTEL catalog has log$g$ = 4.3 listed for this star, so perhaps it is  
slightly evolved, and a lower gravity model would provide a better match to 
the observations.

\section{Results}
We present our equivalent width and contamination analysis for the program
CVs below. These are ordered by constellation name, except for the
first object, SS Cyg.  Given the change in our procedure, and the fact that 
the entire process for each object can be quite involved, we use the data set
for SS Cyg to show all of the steps used in assessing whether the 
hydrogen abundance is non-solar.

\subsection{SS Cygni}

HM17 presented a new analysis of SS Cygni using both moderate resolution 
$K$-band spectra from NIRSPEC on Gemini, and cross-dispersed data from SPEX 
on the IRTF. Their best fitting log$g$ = 4.5 model had 
the following parameters: T$_{\rm eff}$ = 4750 $\pm$ 204 K, [Fe/H] = $-$0.3, 
[C/Fe] = $-$0.1, [Mg/Fe] = $-$0.4, [Na/Fe] = $+$0.4, and 
$^{\rm 12}$C/$^{\rm 13}$C = 4. In the following, we will assume an isotopic 
ratio of $^{\rm 12}$C/$^{\rm 13}$C = 4 for all of our models for SS Cyg. The 
weakness of the $^{\rm 13}$CO features, and the lower resolution of the 
cross-dispersed data set, do not allow a robust measurement of the 
$^{\rm 12}$C/$^{\rm 13}$C ratio. As discussed in HM17, the main effect of 
fixing $^{\rm 12}$C/$^{\rm 13}$C = 4 is to raise the value of [C/Fe]
by $+$0.1, when compared to the value of [C/Fe] found using the cosmic value 
($^{\rm 12}$C/$^{\rm 13}$C = 89). We have chosen not to use the red Na I 
doublet in the $K$-band in our analysis procedure as it sits amidst strong 
CO absorption
features, and thus requires a precise value of [C/Fe] to obtain a reliable
measurement. This does not, however, void the need for including the correct
input value for [C/Fe]. Both the level of the blue continuum in the $H$-band, 
and the broad absorption feature in the $J$-band near 1.12 $\mu$m, are shaped 
by CN absorption. Thus, for the model spectra to more closely resemble reality,
we must include the best available value for the carbon abundance when 
generating the synthetic spectra.

Before we describe the equivalent width (hereafter, `EQW') measurements, 
we briefly discuss the possibility of contamination of the spectra by emission 
from sources in SS Cyg other than the donor star. Contamination of the spectrum 
by emission from the white dwarf and its accretion disk can be a serious problem 
as one works towards bluer wavelengths. Bitner et al. (2007) found that at 
5500 \AA, the white dwarf and accretion disk in SS Cyg account for 54\% of the 
system's luminosity. Presumably, this should be smaller in the near-infrared, 
where the spectral energy distributions of K dwarfs peak. As shown in Fig. 1 
of H16, most of the metal lines increase in strength with decreasing 
temperature in the K dwarfs. Thus, if there is significant contamination as 
one progresses from the $K$-band to the $J$-band, the best fitting models 
in the $H$- or $J$-band should, in general, be hotter than those in the 
$K$-band. This allows for the possibility of quantifying {\it differential} 
contamination. We encounter this issue in the following subsection, and address 
the possibility of contamination across the near-infrared in subsection 3.1.2.

\subsubsection{Equivalent Width Measurements}

We first assume that there is no significant contamination of the SS Cyg donor 
star spectrum. We plot results in $JHK$ for both solar (Fig. \ref{sseqwsolar}), 
and subsolar ([Fe/H] = $-$0.3, Fig. \ref{sseqwnonsolar}) models. The strong 
magnesium deficit is apparent for both metallicities in all three bandpasses. 
The subsolar model (with Hd = 0\%) provides the better fit to all of 
the lines across the three wavebands. In the $J$-band, however, the subsolar
fit would be improved if the donor star were slightly hotter: T$_{\rm eff}$ 
$\sim$ 5000 K. Evidence for differential contamination. For the solar metallicity results to reproduce the data for SS 
Cyg, a hotter donor star would be required regardless of bandpass. In the
$H$- and $K$-bands, a donor with T$_{\rm eff}$ = 5000 K produces a reasonable
match between the observations and models. In the $J$-band, however, a
temperature of 5250 K, or higher, would be required for the solar abundance
model to provide a sensible fit. Unless there is a source of contamination, 
or the previously derived donor star temperature is incorrect, it appears that 
a solar abundance for the secondary star in SS Cyg can be excluded.

Using just the results from the $J$-band, we would conclude that T$_{\rm eff}$
$\sim$ 5250 K, and the donor star in SS Cyg has a large enhancement of both 
sodium and carbon. In the $H$-band, the iron line appears to be weaker in SS 
Cyg than found for either abundance, but the continuum bandpasses for this 
line includes both H I Br19 and Br18. In SS Cyg, these high order hydrogen 
lines are sufficiently strong to affect the weak Fe I line. There appears to 
be an aluminum deficit for both metallicities in the $K$-band, but this line 
is corrupted by He I emission in SS Cyg. The measured EQW of the aluminum line 
in the $H$-band agrees with the subsolar metallicity model. The abundance of 
sodium appears to be enhanced in the $K$-band for the lower metallicity model. 
There is an Na I doublet in the $H$-band. This feature is weak compared to 
those in the $J$ or $K$-bands, but will become more relevant, below. For SS 
Cyg, our program was unable to measure an EQW for this doublet due to H I 
emission.

From this analysis we would conclude that the donor in SS Cyg has a 
subsolar metallicity, with a normal abundance {\it pattern}, except 
for the presence of a magnesium deficit, and an enhancement in the 
sodium abundance. These results closely reflect those of HM17, except that 
the sodium enhancement found here is about one half ([Na/Fe] = $+$0.2) of 
what they found. As noted above, one possible cause of this is the fact that 
the $\chi^{\rm 2}$ analysis is more reliant on the relative continuum levels 
between the models and the data, and this requires an accurate continuum 
division for the CV spectrum. This is especially true since they included the 
red Na I doublet in the $K$-band to derive the sodium excess. With CO 
absorption, defining the actual continuum near this feature is challenging. 
The EQW results do suggest that a slightly hotter donor star, T$_{\rm eff}$ 
$\sim$ 4875 K, would produce better fits across the three bandpasses at
[Fe/H] = $-$0.3. 

\subsubsection{A Broader Level of Contamination?}

For the $J$-band, an increase in the level of contamination would
improve the fit to most of the lines for that bandpass. It is impossible to 
quantify the exact level of contamination while allowing the value for the 
metallicity to float without some type of additional constraint. To derive this
constraint we employ the near-infrared light curves of SS Cyg. To the $JHK$ data
from Harrison et al. (2007), we add the 3.4 and 4.5 $\mu$m photometry from 
$WISE$ and $NEOWISE$, and model the combined data set with WD2010. 

For the ellipsoidal variations of Roche-lobe filling CV secondary stars, the 
deeper of the two light curve minima often occurs at phase 0.5. This is
because at this time we are viewing the hemisphere of the donor star that 
extends further from the center of the secondary, and it is cooler. The ratio 
of the two minima is almost completely dependent on the temperature ratio of 
the two stars. A very hot white dwarf can irradiate the secondary, and raise 
the temperature of the distorted hemisphere. As can be seen in Fig. 
\ref{sscyglc}, the observed minima in the $K$-band light curve of SS Cyg have 
similar depths. This implies a {\it small} amount of irradiation of the donor. 
We fix the temperature of the secondary star to T$_{\rm eff_{\rm 2}}$ = 4750 K. 
This then requires T$_{\rm eff_{\rm 1}}$ = 20,000 K to match the depths of the 
two minima. This temperature is lower than that typically reported for SS Cyg 
from modeling UV spectroscopy, T$_{\rm eff}$ $\approx$ 35,000 K (c.f., Urban 
\& Sion 2006).  Such a high temperature for the white dwarf would decrease the 
depths of the secondary minima to smaller than what is observed. With no 
contamination (``no third light'' in the parlance of light curve modeling), we 
derive a best-fitting orbital inclination of $i$ = 35$^{\circ}$ for SS Cyg.

We now add a contaminating source that is present at all phases, and has
no bandpass dependency. For a third light component that supplies 25\% of the 
flux (at $\phi$ = 0), the best fitting model (plotted in red in Fig. 
\ref{sscyglc}), has $i$ = 45$^{\circ}$. For a contamination level of 50\%, the 
orbital inclination has to be $i$ = 55$^{\circ}$ for the model to match the 
observations. The latter level of contamination is what was found by
Bitner et al. (2007) for the visual part of the spectrum. Such a large
contamination level is unreasonable. The spectral energy distribution
of the donor in SS Cyg peaks in the near-IR, thus relative to the optical flux,
it should dominate the infrared luminosity. It is also difficult to reproduce 
the observed near-IR colors of SS Cyg with such a large contamination 
component. We conclude that the contamination level is near 25\%.

Because we do not have $JHK$ light curves for all of our sources, it is
worthwhile to investigate the utility of the $WISE$/$NEOWISE$ light curve
data for estimating contamination.  The same model used above was run for the 
two $WISE$/$NEOWISE$ bandpasses. Assuming
($W1$ $-$ $K$) = 0.07, and ($W2$ $-$ $K$) = 0.0 for the $\sim$ K3.5V secondary
star (see Pecaut \& Mamajek 2013), the resulting model light curves
exactly fit the lower envelope of the $WISE$/$NEOWISE$ data in both bands.
Note that SS Cyg was at minimum for all of the data collected by
$WISE$ and $NEOWISE$ (though some data sets occurred just a few days before 
an outburst, or within a few days of returning to minimum). Thus, the large
amplitude variations above the light curve model are due to intrinsic processes
occurring within the binary system {\it during quiescence}. It is curious that 
the level of variability is much smaller at both inferior {\it and} superior 
conjunctions. The amplitude of the variability in the $W3$ bandpass (11.3 
$\mu$m, not plotted) is about twice that seen in $W2$. Dubus et al. (2004) 
explored the mid-infrared variability of SS Cyg, and had attributed it to 
free-free emission from expanding clouds. Given the radio detection of SS Cyg by 
K\"{o}rding et al. (2008), perhaps SS Cyg is also a synchrotron source
in quiescence.

We now repeat the EQW analysis using a contamination level of 25\% in each
bandpass. As shown in Fig. \ref{sscygcont}, the resulting, best fit 
metallicity for SS Cyg is [Fe/H] = 0.0. The strong magnesium deficit, and the 
enhancement of sodium described above, remain. The remainder of the data set 
is fully consistent with a solar abundance pattern. Increasing the 
contamination level in the $J$-band would aid the fit of the data to the 
models. In the case of SS Cyg, the subsolar metallicity derived by H16 and HM17 
is due to contamination. However, neither their derived temperature for the 
donor in SS Cyg, or the relative elemental abundance anomalies they found, are 
strongly affected by the observed level of contamination. There is no evidence 
to support a non-solar value for the (photospheric) hydrogen abundance.

\subsection{DX Andromedae}

DX And is a U Gem type dwarf nova of long period, P$_{\rm orb}$ = 10.57 hr.
DX And was observed using GNIRS on Gemini North.
Bruch et al. (1997) and Drew et al. (1993) found that the secondary star
has a spectral type near K1. Bruch et al. present an ephemeris for this 
object obtained from a radial velocity study. Their analysis suggested a mass 
ratio of $q$ = 0.66 $\pm$ 0.08, while Drew et al. found $q$ = 0.96. If 
the former is true, the secondary star would have an unusual mass, M$_{\rm 2}$
= 0.53 M$_{\sun}$, for its spectral type (M$_{\rm K1V}$ = 0.8 M$_{\sun}$;
Henry \& McCarthy 1993). The larger value for $q$ would imply a more normal 
mass for the donor star, if the white dwarf mass was similar to that found
in the typical CV ($\langle$M$_{\rm WD} \rangle$ = 0.8 M$_{\sun}$; Knigge et al. 2011). From a 
sparse set of photometry, Drew et al. estimated an orbital inclination of 
45$^{\circ}$, and a contamination level of 15\% in the visual. They also 
derived a rotational velocity of $v$sin$i$ = 79 $\pm$ 5 km s$^{\rm -1}$, and 
we have used this value in generating the model spectra.

In Fig. \ref{dxandlc}, we present a partial $JHK$ light curve, along with
the photometry from $WISE$/$NEOWISE$. Given the poorly known system parameters,
we set T$_{\rm 1}$ = 25,000 K, and T$_{\rm 2}$ = 5000 K, and used the
larger value for $q$. In the model light curve (red) presented in Fig. 
\ref{dxandlc}, we have set the contamination level in all bands to zero.
An inclination angle of $i$ = 45$^{\circ}$ provides a reasonable fit to 
the data. Any flaring in $W1$ and $W2$ bands is much smaller than seen
in most of the CVs of our program, suggesting that the accretion disk is 
relatively unimportant in the near-IR. We will assume no contamination. 

Using equation 2.3b in Warner (1995), the mean density of the donor in
DX And is $\langle \rho \rangle$ $\simeq$ 0.96 gm cm$^{\rm -3}$. Using data
in Boyajian et al. (2012), a K1 dwarf should have $\langle \rho \rangle$ 
$\simeq$ 2.2 gm cm$^{\rm -3}$. The donor in DX And has a density that suggests
it is evolved. The {\it Gaia} Data Release 2 (``DR2'') parallax ($Gaia$ 
Collaboration 2018) gives a distance to DX And of 600 pc. The light curve 
shows that $\langle$ $K$ $\rangle$ = 12.35, and assuming the donor supplies 
100\% of the observed luminosity, implies M$_{\rm K}$ = 3.46. The secondary 
star in DX And is 1.79 times more luminous than a main sequence K1V (Houk et 
al. 1997), suggesting a radius that is 1.34$\times$ that of a K1V. For such an 
object, {\it if} it has the mass of a normal K1V (0.8 M$_{\sun}$), log$g$ = 
4.28, well below that for a main sequence dwarf. 

Using a solar metallicity, we find that all of the lines, except those of 
Mg I, are stronger than expected whether log$g$ = 4.5 or 4.0 (Fig. 
\ref{dxandeqw}). For the dwarf model (log$g$ = 4.5), the best explanation at 
[Fe/H] = 0.0 is that the donor suffers from a hydrogen deficit of at least 
25\%. For the subgiant model (log$g$ = 4.0), the best fit occurs for a 
hydrogen deficit of 50\%. In either case magnesium is deficient, while the
sodium abundance is enhanced.  The only plausible alternative to a hydrogen 
deficit is a large, super-solar metallicity. The highest metallicity we have 
in our grid is [Fe/H] = $+$0.50, and models with this value do a very poor 
job at reproducing the measured equivalent widths, with deficits of some 
elements, and enhancements of others. {\it While the results for both 
gravities are consistent across the three bandpasses for single values of the 
hydrogen deficit}. 

We conclude that the 
donor star in DX And is truly hydrogen deficient. This result explains the 
stronger than expected absorption features found by Drew et al. (1993), and 
enhanced strength of the absorption lines of Ca I and Cr I noted by Bruch et 
al. (1997). The Gemini GNIRS spectrum of DX And is presented in Fig. 
\ref{dxspec}, where we compare it to a model spectrum with T$_{\rm eff}$ = 
5000 K, log$g$ = 4.0, [Fe/H] = 0.0, and Hd = 50\%. Using the $\chi^{\rm 2}$ 
analysis, we derive the following abundances for the log$g$ = 4.0: [C/Fe] = 
$-$0.7, [Mg/Fe] = $-$0.2, and [Na/Fe] = $+$0.5. The model plotted in Fig. 
\ref{dxspec} includes those abundance anomalies. For log$g$ = 4.5, the
abundances are [Fe/H] = 0.0, Hd = 25\%, [C/Fe] = $-$0.5, [Mg/Fe] = $-$0.5, 
and [Na/Fe] = $+$0.3.

\subsection{RX Andromedae}

RX And is a interesting system in that it has dwarf novae eruptions, what appear
to be Z Cam-like standstills, and deep minima like VY Scl stars (Schreiber et
al. 2002). The Ritter-Kolb Catalog (``RKCat'', Ritter \& Kolb 2003) has the 
following parameters listed for the RX And system: M$_{1}$ = 1.14 M$_{\sun}$, 
M$_{\rm 2}$ = 0.48 M$_{\sun}$, $i$ = 51$^{\circ}$, and a spectral 
type of K5 for the donor. Using these parameters we calculate that the 
rotational velocity of the secondary star should be $v$sin$i$ = 101 km 
s$^{\rm -1}$. The SPEX data do not have sufficient resolution to provide a 
robust value for this parameter. Using absorption lines in the $I$-band, 
however, where the dispersion is 2.02 \AA/pix, we find a value of $v$sin$i$ = 
96 $\pm$ 20 km s$^{\rm -1}$ for RX And. To model RX And, we generated a grid 
of synthetic spectra with $v$sin$i$ = 100 km s$^{\rm -1}$.

The temperature of the secondary in RX And appears to be ambiguous.
While the RKCat lists a spectral type of K5, H16 found that the donor appeared
to be an M-type star. The routine they used to derive metallicities and
temperatures for M dwarfs found a value of [Fe/H] = $+$0.07 and a spectral
type of M2. RX And has strong Na I and Ca I absorption, with equivalent
widths of 5.35 \AA ~and 4.18 \AA, respectively. For main sequence dwarfs,
the equivalent width of the Na I doublet does not become larger than that for
the Ca I triplet until the spectral type is later than $\sim$ M1V. H16 found
that if the spectral type is earlier than M2, the donor star has a subsolar
metallicity. An alternative explanation is that the donor star has a large 
excess of sodium, and its flux is diluted by emission from the non-stellar
components in the system.

We use the $WISE/NEOWISE$ light curves to explore the issue of contamination.
We only use those data for when RX And was at a minimum (two of the six 
NEOWISE epochs occurred during maxima). The $WISE/NEOWISE$ light curves in 
the W1 and W2 bands both show random variations of $\Delta$m = 0.5 mag. It is 
likely that there is a significant level of contamination. The orbital 
ephemeris for RX And has been published by Kaitchuck (1989), and we phased
the $WISE/NEOWISE$ photometry using their result.  We used WD2010 to generate 
a light curve setting T$_{\rm eff_{\rm 2}}$ = 4500 K (K5V). Sepinsky et al. 
(2002) quote a temperature for the white dwarf of 34,000 K. With $i$ = 
51$^{\circ}$, a light curve assuming no contamination has too large of an 
amplitude to explain the observed photometry. If we use a contamination level 
of 50\%, we can better match the lower envelope of the W1 and W2 data sets. 
If we use T$_{\rm eff_{\rm 2}}$ = 3500 K (M2V) we again find that the amplitude
of the variations is too large for $i$ = 51$^{\circ}$ without a third light 
component. In this case, a contamination level of 30\% is consistent with 
the lowest values in the $WISE/NEOWISE$ photometry throughout the orbit.

The EQW analysis for RX And, assuming that T$_{\rm eff}$ $\geq$ 4000 K,
is wildly inconsistent. It requires a contamination of 30\% to get calcium,
aluminum and iron within the error bars of the measurements for RX And in
the $K$-band. In this scenario sodium would have an enormous excess, while 
magnesium has a very large deficit. In the $J$-band, a 30\% contamination 
level is only consistent with the Al I datum. Sodium, potassium and 
manganese would have excesses, while silicon, magnesium and iron would have 
deficits. {\it The donor in RX And is not a K-type dwarf.}

To attempt to constrain the secondary in RX And, we decided to apply our
EQW measurement programs to the M dwarfs in the IRTF Spectral Library.
While blindly applying our programs to these M dwarfs could be quite dangerous
due to the number of molecular features found in the spectra of M dwarfs, 
we think the results are quite illuminating. In Fig. \ref{rxmdwf} we present 
the results for the three bandpasses, plotting EQW versus spectral type. In 
addition, we separate the M dwarfs into those with [Fe/H] $>$ 0 (red), and 
those with [Fe/H] $<$ 0 (blue). In the $J$-band, assuming the donor is an M2,
the only anomalies are deficits of magnesium and aluminum, and an enhancement
of manganese. Given the possibility of intrinsic water vapor absorption at the
red end of the $J$-band, peculiar values for the EQWs of manganese might be
expected. In the $H$-band, the only peculiarity is a deficit of magnesium. 
This is also true in the $K$-band {\it if} the donor in RX And has a solar 
metallicity.

Given that there is probably significant contamination of the spectrum of
the donor in RX And, the true EQW measurements would be somewhat larger than 
plotted in Fig. \ref{rxmdwf}. A 25\% contamination level results in EQW 
measurements that are 75\% the depth of those without contamination. We 
conclude that the best fit spectral type for the secondary in RX And is an M2, 
and it has a near-solar abundance. It appears to have normal levels of sodium, 
with a deficit of magnesium. An M2 spectral type with solar abundance requires 
a sizeable carbon deficit, [C/Fe] $\leq$ $-$0.3.

\subsection{AE Aquarii}

HM17 found that the donor in AE Aqr has a very low carbon abundance, and 
likely has a small value for the isotopic ratio of carbon: 
$^{\rm 12}$C/$^{\rm 13}$C = 4. They found for their log$g$ = 4.5 (``dwarf'') 
model values of T$_{\rm eff}$ = 5000 K, [Fe/H] = 0.0, [C/Fe] = $-$1.4, [Mg/Fe] 
= 0.0, and [Na/Fe] = $+$0.5. For the log$g$ = 4.0 (``subgiant'') model they 
found T$_{\rm eff}$ = 4750 K, [Fe/H] = 0.0, [C/Fe] = $-$1.7, and [Na/Fe] = 
$+$0.52. We adopt the HM17 carbon abundances
for the two gravities in the following. The equivalent width analysis for
the dwarf model with [Fe/H] = 0.0 leads to the conclusion that the donor 
in AE Aqr has a hydrogen deficit of 25\%, a small deficit of magnesium,
and an excess of sodium (Fig. \ref{aeaqrEQWdw}). In all three bandpasses, 
slightly better fits of the data to the models (less magnesium) occurs if we 
lower the temperature of the donor in AE Aqr to T$_{\rm eff}$ $\simeq$ 4750 K.
A model with [Fe/H] = $+$0.1 also reduces the need for a hydrogen deficit at 
T$_{\rm eff}$ = 5000 K in the $H$- and $K$-bands, but worsens the fit in the 
$J$-band. Obviously, a higher metallicity lowers any sodium excess, while 
making the magnesium deficit more severe. 

Given the long orbital period of AE Aqr, 9.88 hr, it is likely that the
secondary star has a lower gravity than that of a dwarf (see HM17). Using a 
solar abundance, subgiant model, we find that the most likely solution 
(see Fig. \ref{aeaqrEQWsg}) at T$_{\rm eff}$ = 4750 K is one where the donor 
star has a 25\% deficit of hydrogen. In this scenario the abundance of sodium 
is enhanced, while the magnesium abundance is close to normal. The one 
deviation from this result is that potassium seems to be underabundant. 
Increasing the metallicity to [Fe/H] = $+$0.2 creates a better match between 
models and the data for Mg I, Ca I, and Al I in the $H$- and $K$-bands. It 
also reduces the excess of sodium. The fit of this solution to the observed 
$J$-band equivalent widths is much worse. A significant level of 
contamination in the $J$-band would be needed to make such a model work. 

It appears that either the donor in AE Aqr is hydrogen deficient, or that
the metallicity is super-solar. It is impossible to decide between these
alternatives without knowing the contamination level. Welsh et al. (1995) 
found that the contamination of the donor star by the white dwarf and 
accretion disk was very low for AE Aqr in the red ($\sim$ $\lambda$6600 \AA), 
with the secondary supplying 94\% of the luminosity for the donor star 
temperatures we have derived. This would suggest that it is unlikely that the 
$J$-band suffers from a large contamination component. Unfortunately, there 
has been very little near-infrared photometry published for AE Aqr. The only 
near-IR light curve that we have found is from Tanzi et al. (1981); 
they observed it in the $K$-band, covering 30\% of an orbit. That light curve 
is consistent with pure ellipsoidal variations. 

It is well known that AE Aqr is variable in the mid-infrared (see Dubus et al. 
2004, Abada-Simon et al. 2005). This variability has been ascribed to 
free-free emission from expanding clouds, a synchrotron source, or a 
combination of both. Fortunately, sufficient $WISE$ and $NEOWISE$ observations 
exist to better probe the infrared variability of this source. In Fig. 
\ref{aeaqrlc} we plot the
$WISE$ four band photometry, and the $NEOWISE$ two band data. There does
not appear to be any offset between the overlapping data sets. Using the 
system parameters derived by Hill et al. (2014), we have used WD2010 (with
$i$ = 50$^{\circ}$) to generate a set of light curve models for AE Aqr. The 
$W1$ bandpass shows clear ellipsoidal variations, but with excursions of 
$\sim$ 20\% from the model baseline. The amplitude of the random variations is 
larger in the $W2$ band, and a non-stellar source dominates the light curves 
in the $W3$ and $W4$ (22 $\mu$m) bands.

The strong wavelength dependence suggests that the source of the variation
is relatively unimportant in the $J$-band, while contributing $\sim$ 10\%
to the $K$-band flux during flaring events. Ignoring the possibility of
contamination, for the dwarf model, a value of [Fe/H] = 0.0 with Hd = 25\%
reproduces most of the observations {\it if} the temperature is T$_{\rm eff}$ 
= 4750 K. In this solution, the sodium abundance is super-solar, and magnesium 
is deficient. For the subgiant models, the best fit occurs for a hydrogen
deficit of 25\%, with an enhanced abundance of sodium, but a normal abundance 
for magnesium.  HM17 calculated that the expected gravity for 
the donor in AE Aqr was intermediate between the dwarf and subgiant models 
that we have used here, suggesting the true solution lies somewhere between 
our two results. It does appear, however, that the donor in AE Aqr has
a modest hydrogen deficit.

\subsection{Z Camelopardalis}

Z Cam is the proto-type of a small family of CVs that show ``standstills''
in their outbursts (see Simonsen et al. 2014). The RKCat lists the following
parameters for the system: P$_{\rm orb}$ = 0.2898406 d, M$_{\rm 1}$ = 1.00 
$\pm$ 0.06 M$_{\sun}$, M$_{\rm 2}$ = 0.77 $\pm$ 0.03 M$_{\sun}$, and $i$ = 
57$^{\circ}$. The last three parameters come from Shafter (1983). While Kraft 
et al. (1969) 
believed the secondary star had a G spectral type, Wade (1980) lists the 
spectral type as K7. Thorstensen \& Ringwald (1995) have published the most 
recent ephemeris for Z Cam. We downloaded the $WISE/NEOWISE$ data to evaluate 
the orbital inclination. There were eight epochs of observations, but the AAVSO 
data base shows that only two, those for 2014 October 10 and 2015 October 10, 
were obtained during minimum light ($V$ $\geq$ 13). We plot the data for these 
two epochs in Fig. \ref{zcamlc}, with the data for 2014 plotted in red, and that 
for 2015 plotted in green. The photometry for the 2014 observation appeared to 
be 0.17 mag brighter than that for 2015. This mirrored the visual light curve 
data that showed the system to be $\sim$0.2 mag brighter in the 2014 epoch
versus the 2015 epoch.

We used the system parameters to generate model light curves using WD2010.
Hartley et al. (2005) found that the white dwarf in Z Cam appeared to be
very hot, T$_{\rm eff}$ = 57,000 K, though a model with T$_{\rm eff}$ = 26,000
K also appeared to fit the data. We set the white dwarf temperature to 26,000 K 
as there is little evidence for intense irradiation in the mid-infrared light
curves. As we will discuss below, the secondary star in Z Cam appears to
be an early M dwarf. We set the temperature of the secondary star to 3575 K. An 
orbital inclination of 57$^{\circ}$ is not consistent with the $NEOWISE$ data 
unless there is a contamination level of $\sim$ 30\% in each band (see Fig. 
\ref{zcamlc}). If we assume that there is no contamination, we
find that the best fit for the orbital inclination is $i$ = 40$^{\circ}$. The 
2MASS photometry of Z Cam was obtained during minimum light and, at an orbital 
phase of $\phi$ = 0.55, had the following colors: ($J - H)$ = 0.53, and 
($H - K$) = 0.18. These suggest a K4 spectral type, and such an object should 
have ($K - W1$) = 0.08. For Z Cam, at $\phi$ = 0.55, ($K -  W1$) = 0.12. A 
secondary with T$_{\rm eff}$ = 3575 K would have ($J - H)$ = 0.66, ($H - K$) = 
0.19, ($K - W1$) = 0.14. For the M spectral type to agree with the photometry 
requires some level of contamination in the near-infrared.

To construct a synthetic spectral grid for Z Cam requires a value of $v$sin$i$.
Using the parameters listed above, for $i$ = 57$^{\circ}$, we calculate
$v$sin$i$ = 117 km s$^{\rm -1}$. For $i$ = 40$^{\circ}$, we get $v$sin$i$ = 89 
km s$^{\rm -1}$. Unfortunately, the resolution of our spectra is insufficient
to decide between these two. Using four isolated lines in the $J$-band,
we get $v$sin$i$ = 91 $\pm$ 26 km s$^{\rm -1}$. Hartley et al. argue that a
higher inclination than $i$ = 57$^{\circ}$ would produce a better match
to their UV data. We will assume an inclination of 57$^{\circ}$, and use
$v$sin$i$ = 117 km s$^{\rm -1}$ for generating the grid.

The results of the EQW analysis are inconsistent. The main 
deviations appear to be a significant enhancement of sodium and a deficit of 
magnesium. The data appear to be most consistent with an object that has 
T$_{\rm eff}$ = 4500 K, and a small hydrogen deficit. If the temperature of the 
donor is slightly cooler, T$_{\rm eff}$ = 4250 K versus T$_{\rm eff}$ = 4500 K, 
the need for a hydrogen deficit is diminished. The strong sodium excess and 
magnesium deficit, however, remain. At temperatures much higher or lower than 
4500 K, the match to the CO features becomes very poor. We also attempted more 
metal-rich models, and found that synthetic spectra with T$_{\rm eff}$ = 4500 K,
Hd = 0\%, and [Fe/H] = $+$0.3 do a reasonable job at reproducing the data. In 
this case, however, the CO features are much stronger than seen in Z Cam, and 
would require a large carbon deficit. There does not seem to be a robust 
solution assuming the donor is a K dwarf. 

If we apply the metallicity program used for the M-dwarfs in H16 to the $K$-band
spectrum of Z Cam, we derive a spectral type of M1.6, and [Fe/H] = $+$0.03. The 
spectral type determined by this routine uses a ``water vapor index'' (see 
Covey et al. 2010) that is calculated using the ratios of three different 
continuum bandpass fluxes (21800 $-$ 22000 \AA, 22700 $-$ 22900 \AA, and 23600 
$-$ 23800 \AA). The only strong line in these bandpasses is Mg I at 22820 \AA. 
As we did for RX And, we compare the EQW measurements for Z Cam to those of M 
dwarfs in Fig. \ref{zcammdwarfs}. The results are fully consistent across the
three bandpasses. The only deviation appears to be the enhanced strength
of Mn I, but this is almost certainly due to the continuum division process,
with the added complication of water vapor absorption in the donor.

In Fig. \ref{zcamjhkspec}, we compare the $JHK$ spectra of an M1V (HD 42581)
from the IRTF Spectral Library to that of Z Cam. Using the metallicity
program, we derive a value of [Fe/H] = $+$0.12, and a spectral type of M1.3
for HD 42581. The match between the two sources is excellent, including
the CO features in the $K$-band. The donor star of Z Cam does not appear to 
show {\it any} abundance anomalies. It is, however, surprising that an
object with a longer orbital period than SS Cyg, has a donor that is 
1000 K cooler. The parallax of Z Cam from the $Gaia$ DR2 is $\pi$ = 4.467 $\pm$
0.05 mas. With $K$ = 10.85 from the 2MASS catalog, M$_{\rm K}$ = 4.11. An M1.3V
should have M$_{\rm K} \sim$ 5.55. Note that due to its brightness, ``G'' $<$ 
12, systematic effects in the $Gaia$ data analysis could lead to an actual 
error bar of $\pm$ 0.3 mas for Z Cam (Lindegren et al.  2018). Even assuming 
this large of an error bar, and moving Z Cam to 186 pc ($\pi$ $+$ 3 $\sigma$), 
the system has M$_{\rm K}$ = 4.5. The donor star in Z Cam is much more 
luminous than a main sequence M dwarf.

\subsection{SY Cancri}

The long period Z Cam system SY Cnc is unusual in that it has a mass ratio
greater than one; Casares et al. (2009) found $q$ = 1.18 $\pm$ 0.14. Casares 
et al. refined the orbital period and ephemeris for SY Cnc, measured 
$v$sin$i$ = 75.5 km s$^{\rm -1}$, and derived a spectral type of G8 +/- 2 
(T$_{\rm eff}$ = 5500 K) for the donor. H16 found that the donor in SY Cnc was 
completely normal, with T$_{\rm eff_{\rm 2}}$ = 5500 K, [Fe/H] = 0.0, and [C/Fe]
 = 0.0.  At the epoch of the observations used by H16, the system had $V$ = 12, 
and was declining from an outburst that had peaked four days earlier. That 
spectrum must have a considerable level of contamination, weakening the depth 
of any absorption features. It is likely that the spectral type of the donor is 
cooler than derived in H16. The new cross-dispersed spectrum obtained with 
TripleSpec, analyzed below, occurred when the system had $V$ = 11.8.

There is only a rough estimate for the orbital inclination. Casares et al.
use the mass function, M$_{\rm 1}$sin$^{\rm 3}i$ = 0.13 $\pm$ 0.02 M$_{\sun}$,
to derive the following limits: 26$^{\circ}$ $\leq$ $i$ $\leq$ 38$^{\circ}$.
We are not aware of any infrared light curves of this source, so we downloaded
the $WISE/NEOWISE$ data to investigate the system. There are three epochs
of data when SY Cnc was in quiescence ($V$ $\gtrsim$ 13): 2010 April 28, 2016 
April 18, and 2016 November 4. The latter of these was just as SY Cnc began an
outburst, having $V$ = 13.2. The light curves, Fig. \ref{sycnclc}, in both
bandpasses are quite noisy, though there are clearly ellipsoidal variations 
present. 

We use WD2010 with T$_{\rm eff_{\rm 1}}$ = 20,000 K, and T$_{\rm eff_{\rm 2}}$ 
= 5500 K, $q$ = 1.18, and M$_{\rm 2}$ = 0.9 M$_{\sun}$ to generate model
light curves. We find that the best 
fitting inclination is $i$ = 55$^{\circ}$. Given the observed variability, it 
is likely that there is a small amount of contamination, and the true 
inclination is likely higher. This might pose a conundrum. Given the radial 
velocity solution by Casares et al., if our derived inclination was correct, the
white dwarf would have an extremely low mass: 0.24 M$_{\sun}$! As shown in Fig.
\ref{sycnclc}, the 38$^{\circ}$ upper limit established by Casares et al. {\it
 is} consistent with the envelope of the faintest data points, but {\it only if}
there is no contamination. This latter inclination implies a white dwarf mass 
of 0.58 M$_{\sun}$, somewhat lower than typically found in CVs. This then 
leads to M$_{\rm 2}$ = 0.66 M$_{\sun}$, a mass more typical of late K-type 
main sequence stars, than late G (note,
changing the input masses for WD2010 to these lower values increases the 
amplitude of the variations by 0.01 mag).

Given that our spectroscopy was not obtained during quiescence, we need to
estimate a contamination level. The $NEOWISE$ observation that began on
2015 November 6 occurred at a time when $V$ = 11.7. At $\phi$ = 0, the
$W1$ and $W2$ magnitudes in outburst were 0.48 mag brighter than their
quiescent values. This suggests that at this time, the donor star only
contributed 40\% of the infrared flux. We assume this contamination level,
and find {\it globally} that the EQW measures of SY Cnc, Fig. \ref{sycnceqw}, 
are consistent with T$_{\rm eff}$ = 5500 K, and [Fe/H] = 0. Magnesium appears 
to be slightly underabundant. The presence of numerous emission/noise features 
corrupts several of the measurements, including Al I and Fe I in the $K$-band, 
and Al I, K I, Si I, and Fe I in the $J$-band. 

Plotting the spectrum of SY Cnc versus the best fitting model from the EQW
analysis, finds that the CO features in the synthetic spectrum are too weak. 
Assuming the same contamination level, we find that synthetic spectra with a 
temperature of T$_{\rm eff}$ = 5250 K are a better match to the data in 
the $K$-band. However, such a spectrum is not as good of a match to the
$J$ and $H$-band spectrum without additional contamination. This can be
seen in the EQW plots, where the observed Na I and Ca I measures are too large 
in the $K$-band, but not overtly so in the $J$ and $H$-bands. 
For the synthetic spectrum in Fig. \ref{sycncjhk}, we have used a contamination 
level of 60\% in the $K$-band, 70\% in the $H$-band, and 75\% in the $J$-band, 
to achieve a reasonable match to the observations. The prominence of the donor 
and brightness of the system in the near-IR would enable high resolution 
observations, allowing a more robust measurement of $v$sin$i$ to confirm the 
nature of this unusual system. Phase-resolved $JHK$ photometry during
quiescence is needed to better constrain the orbital inclination.

\subsection{EM Cygni}

H16 only had $K$-band spectra of EM Cyg, finding a subsolar metallicity
and a small carbon deficit. Here we have cross-dispersed $JHK$ spectra, 
plotted in Fig. \ref{emspec}, obtained with GNIRS on Gemini. The AAVSO light
curve for EM Cyg shows that it was in quiescence with $V$ $\simeq$ 13.4 at the
epoch of the Gemini observations. North et al. (2000) showed that there 
is a third star, of similar spectral type as the donor star,
that contaminates the light of the system. By subtracting this star out,
they were able to solve for the system parameters. We have used the values
they found (M$_{\rm 1}$ = 1.12 $\pm$ 0.08 M$_{\sun}$, M$_{\rm 2}$ = 0.99 $\pm$ 
0.08 M$_{\sun}$, $q$ = 0.88, and $i$ = 67$^{\circ}$) to set up WD2010
for modeling the $WISE$ and $NEOWISE$ light curves of EM Cyg. North et al.
found that the donor and contaminating star had very similar contributions
to the luminosity at 6500 \AA, with the donor supplying 7\% more flux. For 
light curve modeling, we add a third component that has exactly 50\% of the 
total light. We set the white dwarf temperature to 20,000 K, and the
secondary to 4500 K.

The resulting light curve is shown in Fig. \ref{emcyglc}. During the $WISE$
observations, the AAVSO data base show that EM Cyg had $V$ = 13.2, a 
few tenths of a magnitude brighter than its typical minimum light value,
though well above its rare, deep minima at $V$ = 14.5. There were
six epochs of $NEOWISE$ data, but only two of those were during times
when $V$ $\sim$ 13.0. In Fig. \ref{emcyglc}, the $WISE$ photometry is plotted
in black, and the two separate $NEOWISE$ epochs are in green (2014 Oct 24)
and red (2015 Oct 18). To achieve overlap between the light curves, we added 
0.06 mag to the $W1$ photometry for the 2014 observations, and
subtracted $-$0.095 from the 2015 data. For $W2$ these offsets were 0.0 mag, 
and $-$0.13 mag, respectively. The solid line in this figure is the model
light curve for a 50\% contamination by the interloping star. The
dashed line is with no third light component. To explain the light curves 
requires a contaminating source that supplies $\sim$ 50\% of the total 
mid-infrared flux. 

The excursions from the model light curves, $\Delta$m $\sim \pm$ 0.1 mag, during
the epochs when $V$ $\sim$ 13, suggest a low (non-stellar) contamination level. 
When adding-in the $NEOWISE$ data from the epochs when EM Cyg was brighter 
than $V$ = 13, the $W1$ photometry has maxima that reach to $W1$ = 10.8, while
maintaining the same floor of $W1$ = 11.2. While there are few data points,
the minimum at $\phi$ = 0 in the $W1$ light curve is suggestive of an 
eclipse. Perhaps the inclination angle is larger than derived by North et al.

North et al. found a rotation velocity of $v$sin$i$ = 140 $\pm$ 6 km 
s$^{\rm -1}$.  We have broadened our model grid using this value. 
The EQW measurements for EM Cyg, Fig. \ref{emeqw}, are completely consistent 
with a solar abundance for T$_{\rm eff}$ = 4500 K. There are several
deviant values that can be resolved by looking at the spectra in Fig. 
\ref{emspec}. The EQW for Al I in the $K$-band suggests a strong
deficit, but this feature is compromised by He I emission.
The EQW for the Ca I triplet suggests an enhanced abundance. The spectra
show a slightly deeper Ca I triplet than the model, perhaps due to the 
presence of narrow absorption features from the contaminating star. The 
profile of the Mg I line in the $K$-band  was corrupted during the telluric 
correction, and it appears broader and deeper than it actually should be. The 
Mg I line in the $H$-band suggests a deficit, while that in the $J$-band 
suggests a normal abundance. The Al I doublet in the $J$-band has a much 
smaller equivalent width than expected for a solar abundance. The Al I line 
in the $H$-band, however, appears to be normal. Presumably this contradiction
is simply due to low S/N and a poor telluric correction at the red end of 
the $J$-band.

With the cross-dispersed data, we use the techniques in H16 to construct a 
model for EM Cyg. The best fitting temperature is T$_{\rm eff_{\rm 2}}$ = 
4500 $\pm$ 150 K. We find that carbon and magnesium suffer identically sized 
deficits: [C/Fe] = $-$0.5 $\pm$ 0.02, and [Mg/Fe] = $-$0.5 $\pm$ 0.1
(using just the $H$-band spectrum). If the contaminating star is a normal 
field dwarf of solar abundance, than the carbon and magnesium deficits in
the secondary of EM Cyg would be about twice those just found. It will take 
higher resolution spectroscopy for a more robust measurement of the
abundances of the donor in EM Cyg.

\subsection{EY Cygni}

EY Cyg is a long period (P$_{\rm orb}$ = 11.0 h) U Gem-type dwarf nova.
Echevarr\'{i}a et al. (2007) conducted a radial velocity study of the system
and found M$_{\rm 1}$ = 1.1 $\pm$ 0.09 M$_{\sun}$, and M$_{\rm 2}$ = 0.49 
$\pm$ 0.09 M$_{\sun}$. They estimated a spectral type of K0 for the secondary, 
and limited
the orbital inclination to $i$ = 14$^{\circ}$ $\pm$ 1. Echevarr\'{i}a et al.
note that the secondary star has a much larger radius than a main sequence
dwarf. Using their parameters, we estimate log$g$ $\simeq$ 4.2. As might be
expected, the $WISE/NEOWISE$ light curves show no evidence for ellipsoidal 
variations, with only random, small scale variability of $\leq$ $\pm$ 0.2 mag 
in both $W1$ and $W2$. We will assume that the contamination level is
insignificant.

The results from H16 were T$_{\rm eff_{\rm 2}}$ = 5250 K (= K0), [Fe/H] = 0.0, 
and [C/Fe] = $-$0.5. Echevarr\'{i}a et al. measured $v$sin$i$ = 34 $\pm$ 4 km 
s$^{\rm -1}$, and we have used this value to generate the synthetic spectral 
grid.  From that grid we find that the donor star in EY Cyg has a significant
hydrogen deficit. For the dwarf gravity, we find that Hd $\geq$ 25\%,
while for the subgiant gravity (see Fig. \ref{eycyg}), Hd $\simeq$ 50\%.
At both gravities, magnesium is slightly underabundant, and sodium has
a significant excess. Sion et al. (2004) found a large N V/C IV ratio.
The carbon deficit we find is consistent with this result, and suggests
a nuclear evolved donor. Sion et al. also found a very low abundance of 
silicon (10\% solar) from $FUSE$ and $HST$ spectroscopy, but from our
measurements, the silicon abundance does not appear to be unusual. Due to the 
lower S/N ratio of the SPEX data for EY Cyg, the error bars on the EQW 
measurements are significant. However, the results were consistent across all 
three bandpasses. We attempted to fit higher metallicity models but, like DX 
And, the measured EQWs were discordant. The best fit of any of the 
higher metallicity models for log$g$ = 4.0 occurred with [Fe/H] = $+$0.3. 

Smaller hydrogen deficits are possible if the donor star
has a much lower temperature; at T$_{\rm eff_{\rm 2}}$ = 4750 K, there is no
hydrogen deficit. Such a temperature is substantially different from that 
derived by H16, or the K0 spectral type assigned to the donor by both 
Echevarr\'{i}a et al. and Kraft (1962). Given the low inclination, it is
worth noting that we are looking at the pole of the secondary in EY Cyg,
and at this viewing angle, the donor probably looks hotter than it would
if it were viewed at a more equatorial angle (see H16). With $K_{\rm 2MASS}$
= 12.62, and a $Gaia$ DR2 distance of 637 pc, the system has M$_{\rm K}$
= 3.58. This is 0.4 mag more luminous than a K0V.

\subsection{V508 Draconis (SDSS J171456.78+585128.3)}

V508 Dra is a poorly studied long period CV. Ag\"{u}eros et al.
(2009) confirm the CV nature of the system, found a K4 spectral type for
the donor, and believed the orbital period to be about 10 hours. Wils (2011) 
shows that the period is in fact 20.113 hr, and have published an ephemeris
for the system. We downloaded the $WISE/NEOWISE$ light curves and present
them as Fig. \ref{v605dralc}. We have phased them using the ephemeris of Wils,
but have shifted the time of phase 0 by $\Delta \phi$ = $-$0.52 to place the
deeper of the two minima at $\phi$ = 0.5, (i.e., inferior conjunction of the
donor at $\phi$ = 0.0). Regardless of the nature of the donor, the large 
amplitude of 
these variations suggest a highly inclined system. If we assume a K4 spectral 
type, $q$ = 0.8, and M$_{\rm 1}$ = 1.0 M$_{\sun}$, we find a best fitting 
inclination of $i$ = 75$^{\circ}$. Changing the masses of the components in 
the system, or $q$, would change this inclination, but only by a few degrees.
The scatter around the light curve model is minimal, so we expect very
little contamination.

Using the $q$ = 0.8 model, the mass of the secondary is M$_{\rm 2}$ = 0.8 
M$_{\sun}$, and WD2010 calculates its radius as 1.64 R$_{\sun}$, and log$g$ =
3.9. Using these parameters we estimate a rotational velocity of 95 km 
s$^{\rm -1}$. We will assume that the donor star is a subgiant. Our EQW analysis
finds that for T$_{\rm eff}$ = 4500 K, there are no abundance anomalies. There 
is, however, a significant carbon deficit, and the $\chi^{\rm 2}$ analysis gives
[C/Fe] = $-$0.3 $\pm$ 0.02. The TripleSpec data are shown in Fig. 
\ref{v605drajhk}, where a model with T$_{\rm eff}$ = 4500 K, log$g$ = 4.0, 
[Fe/H] = 0.0, and [C/Fe] = $-$0.3 is overplotted. The carbon deficit is 
necessary to match both the CO features in the $K$-band, as well as the depth
of the broad CN feature in the $J$-band.

\subsection{AH Herculis}

AH Her is another long period (P$_{\rm orb}$ = 6.19 hr) Z Cam system. H16
found T$_{\rm eff_{\rm 2}}$ = 4500 K, [Fe/H] = $-$0.7, and [C/Fe] = $-$0.4.
Our observations with TripleSpec on 2018 Feb 2 appear to have occurred during
an outburst. Unfortunately, the AAVSO data base has sparse coverage, with
the average of the two nearest points giving $V$ = 11.8. At visual maximum,
AH Her reaches $V$ = 11. The conditions were non-photometric for our 
observing run, but by averaging the fluxes for AH Her and the telluric
standard (HD 153808), we estimate $K$ $\sim$ 11.7. The 2MASS observations
were obtained at the peak of an outburst of AH Her, and had $K$ = 11.4.
There were eight epochs of $WISE$ and $NEOWISE$ observations, none of them
was centered on an outburst. Except for the $NEOWISE$ observations on
2016 February 22, when AH Her was at minimum ($V$ = 14.5), all of these 
occurred at an intermediate brightness level, with 12 $\leq$ $V$ $\leq$ 13. 
This makes it difficult to estimate the contamination level for our spectra.

To construct the $WISE/NEOWISE$ light curves, we took the mean values for
each epoch, and offset them to match that found for the first $WISE$ epoch.
We plot the result in Fig. \ref{ahherlc}, where each epoch is color coded.
We phase these data using the updated ephemeris in North et al. (2002).
Three of the epochs of $NEOWISE$ data have been excluded from this plot, those 
for 2014 August 24, 2015 February 23, and 2016 August 10. The intrinsic scatter 
in those three data sets was twice that observed for the other epochs. 
Construction of the mean for when AH Her was in quiescence finds $\langle W1 
\rangle$ = 11.54, and $\langle W2 \rangle$ = 11.48. For the first $WISE$ epoch, 
we found $\langle W1 \rangle$ = 11.17, and $\langle W2 \rangle$ = 11.11. AH Her 
is about 0.4 mag brighter in the $WISE$ bandpasses during standstill.

North et al. list the parameters for AH Her as M$_{\rm 1}$ = 0.95 M$_{\sun}$,
M$_{\rm 2}$ = 0.76 M$_{\sun}$, $i$ = 46$^{\circ}$, and a donor spectral type
of K7V (T$_{\rm eff}$ = 4250 K). Inputting these values into WD2010, we get the 
light curve model plotted as the solid line in Fig. \ref{ahherlc}. This does a 
reasonable job of fitting the lower bounds of the data, {\it assuming there is 
no contamination.} If we add a source that supplies 30\% of the total flux, we 
need to increase the inclination to $i$ = 70$^{\circ}$ (dashed line) to achieve 
a similar result. Given that the change in the visual magnitude between 
quiescence and standstill is $\Delta V$ = 2, while $\Delta W1$ = 0.4, we expect 
that the contamination level in our $JHK$ spectra will be even higher, of order 
40 $-$ 50\%.

The results of the EQW analysis for AH Her, like those for RX And and Z Cam, 
are not consistent with those of the synthetic spectra even with a 40 or 50\% 
level of contamination. Again, some elements have strong excesses (sodium), 
while others have huge deficits (magnesium). We turn to the spectra, shown in
Fig. \ref{ahherjhk}, to attempt to better characterize the donor. We find 
that the spectrum of HD 42581 (M1.3V with [Fe/H] = 0.12) with a 40\% 
contamination level is a near-perfect match for the spectrum of AH Her. The 
only deviation is that the Al I doublet in the $J$-band spectrum of AH Her is 
much weaker than seen in the M1V. Perhaps, like SY Cancri, there is an emission 
line at this position that corrupts the depth of this feature. Though, there is 
no sign of an emission feature here. The Al I feature in the $K$-band appears 
to be relatively normal, as the He I emission line that normally corrupts this 
feature is weak in AH Her.

The other interesting aspect of the spectra can be seen in the $H$-band:
there are weak absorption lines from the H I Brackett series superposed on
the late-type star spectrum. This is the first time we have seen ``disk'' H I
absorption features in any of our near-IR spectra of CVs, showing that the 
system must have been in outburst at the time of our observations. We
repeat the analysis that we used for Z Cam to check whether a donor with a
spectral type of M1.3 can explain the observations of AH Her. In Fig. 
\ref{ahhermdwarf}, we plot
the EQW measures for the M dwarfs. Given the uncertainty in the donor spectral
type, instead of plotting the measurements of AH Her as data points with error 
bars, we plot the EQW measures as a horizontal solid lines. We also plot what 
the EQWs should be if the contamination level was 40\% as dashed lines. 

It is clear that throughout the three bandpasses a 40\% 
contamination level works well if the donor is an early M-type star. The main 
deviations are the Al I measure in the $J$-band and the Na I feature in the 
$H$-band. We noted the issue with the Al I line above. The reason the Na I 
doublet in the $H$-band is stronger than expected is that this feature is 
convolved with the absorption feature due to H I Br12. We conclude that the 
donor star in AH Her is an early M-type star with a solar abundance pattern.
The hotter temperatures estimated for this source are due to contamination
issues, as is the low value of [Fe/H] found in H16. AH Her does not have
a carbon deficit. Modeling the light curve using a donor star temperature of 
3600 K reduces the amplitude of the ellipsoidal variations by about 0.02 mag. 
It is a certainty that the orbital inclination angle is larger than 
46$^{\circ}$.

\subsection{RU Pegasi}

H16 and HM17 found for log$g$ = 4.5 that the donor star in RU Peg had
T$_{\rm eff}$ = 5125 K, [Fe/H] = $-$0.28, [Mg/Fe] = $-$0.20, [C/Fe] = $-$0.2,
[Na/Fe] = 0.5, and $^{\rm 12}$C/$^{\rm 13}$C = 15. Given the long orbital
period, it is likely that the donor star in RU Peg has a lower gravity
than a main sequence star. For log$g$ = 4.0, HM17 found T$_{\rm eff}$ = 4750 K,
[Fe/H] = $-$0.2, [C/Fe] = $-$0.6, [Mg/Fe] = $-$0.18, [Na/Fe] = $+$0.3, and
$^{\rm 12}$C/$^{\rm 13}$C = 6. We will consider both gravities, and assume
the derived values for the carbon abundance in our analysis.

For the log$g$ = 4.5 (``dwarf'') models, a metallicity of [Fe/H] = $-$0.3 fits
 best in each of the wavebands. In the $K$-band, there is a small deficit of 
magnesium, and apparent enhancements of aluminum and sodium. The aluminum 
abundance is normal in the $H$-band for the subsolar metallicity. The same 
result attains in the $J$-band, where silicon appears to have a normal 
abundance at [Fe/H] = $-$0.3, compared to the fit seen in the 
$K$-band. There is no evidence for a sodium enhancement in the $J$-band. 
A slightly better fit results when using hotter donor, T$_{\rm eff}$ $\geq$ 
5000 K, in agreement with the results of HM17.

For the log$g$ = 4.0 models, a temperature of T$_{\rm eff}$ = 4750 K works 
poorly at any metallicity. For the $K$-band, a model with T$_{\rm eff}$ = 
5000 K and [Fe/H] = $-$0.3 provides a reasonable fit to Mg I, Ca I, and Fe I. 
It appears that both sodium and aluminum are enhanced, but the magnesium 
abundance is normal within the error bars of the measurement. Silicon appears 
to have a deficit. In the $H$-band, magnesium appears to be enhanced, and 
aluminum normal.  In the $J$-band, there are no statistically significant 
deviations from a solar abundance pattern at a subsolar metallicity. The 
largest offsets imply a small enhancement of sodium, and a small deficit of 
silicon.

These results for RU Peg are similar to those derived in HM17. The difference, 
however, is that here we find that a subgiant model fits the measured 
equivalent widths better than a dwarf model. In the subgiant scenario, the 
deficit of magnesium is much smaller than found in the dwarf model. Other than
carbon, the only deviations from a solar abundance pattern for the subgiant
model are the enhanced levels of sodium, a possible underabundance of silicon 
and, of course, the low value of the $^{\rm 12}$C/$^{\rm 13}$C ratio derived in
HM17. It is interesting to note that Godon et al. (2008) found that
silicon and carbon were underabundant from their modeling of the UV spectra of 
RU Peg.

The subsolar abundance of RU Peg, like that for SS Cyg, argues for 
contamination. In Fig. \ref{rupeglc} we present the $JHK$ plus $WISE$
light curves of RU Peg. Dunford et al. (2012) have provided an updated
ephemeris for RU Peg, and solved for the system parameters: M$_{\rm 1}$ = 1.06
M$_{\sun}$, M$_{\rm 2}$ = 0.96 M$_{\sun}$, and $i$ = 43$^{\circ}$. The
$JHK$ light curves are not clean ellipsoidal variations, and confirm that
there is a contaminating, non-stellar, highly variable source in the system. 
The smaller depth of the $\phi$ = 0.5 minimum argues for a heavily 
irradiated hemisphere. Assuming T$_{\rm eff_{\rm 2}}$ = 5000 K, we find that
the white dwarf in the model has to have T$_{\rm eff_{\rm 1}}$ = 45,000 K to 
get the ratio of the minima to match observations. The $JHK$ observations
were obtained about a week after RU Peg had returned to minimum light following
an eruption. This might partially account for the high temperature of the white
dwarf we find here, however, Sion \& Urban (2002) used UV observations to find
a quiescent white dwarf temperature of T$_{\rm eff}$ = 50,000 $-$ 53,000 K. 

Without a contaminating source, we find that the best fit orbital inclination
is 40$^{\circ}$, consistent with the value from Dunford et al. A third light 
component that supplies 25\% of the flux at inferior conjunction requires $i$ 
= 50$^{\circ}$, while reproducing the lower envelope of the light curves in all
five bands. The 5 to 10\% random variations seen in $JHK$ give way to 20\% 
variations in $W1$ and $W2$. The total amplitude of the variations in the $W3$
band (not plotted) is even larger: $\Delta W3$ = 0.9 mag.

The $JHK$ spectra were obtained near inferior conjunction, $\phi$ = 0.85, about
two weeks after RU Peg had returned to minimum light. While a contamination 
level of 25\% is probably reasonable in the $K$-band, it appears that the 
contamination in $J$ and $H$ might be even higher. Assuming that the donor 
supplies 75\% of the flux in each of the bandpasses, we find that the best 
fitting model at both gravities has [Fe/H] = 0.0. At this metallicity, the 
subgiant model fits the data slightly better than the dwarf model. If, 
however, the contamination level is increased to 35\%, the dwarf model would 
fit equally well. A higher temperature donor, T$_{\rm eff_{\rm 2}}$ = 5250 K 
(at 25\% contamination), also results in a better fit for the dwarf model. In 
all scenarios, the only significant abundance anomaly is the large excess of 
sodium. Magnesium has a small deficit in the dwarf model. 

Given the large mass for the donor found by Dunford et al., its surface 
gravity is log$g$ = 4.4, a value consistent with that of a main sequence dwarf.
That mass, however, is not consistent with the temperature we have derived, 
being more appropriate for an early G-type dwarf, than an early K-type dwarf. 
If one uses an inclination of 50$^{\circ}$ instead of 43$^{\circ}$, with
$q$ = 0.88, the masses of both components in the system are lower: M$_{\rm 1}$ 
= 0.76 M$_{\sun}$, and M$_{\rm 2}$ = 0.67 M$_{\sun}$. The resulting secondary 
star mass would be much more consistent with the effective temperature found 
above. The $Gaia$ DR2 parallax is in agreement with the $HST$ parallax 
(Harrison et al. 2004b) and, as noted in H16, RU Peg is 1.2 mag more luminous 
than a main sequence star of its observed spectral type.

\subsection{GK Persei}

For the long-period, old classical nova GK Per, H16 found (for log$g$ = 4.0):
T$_{\rm eff}$ = 5000 K, [Fe/H] = $-$0.3, [Mg/Fe] = $-$0.3, and [C/Fe] = $-$0.5.
We generated our synthetic spectral grid using a value of $v$sin$i$ = 55 km 
s$^{\rm -1}$ from Harrison \& Hamilton (2015), and with [C/Fe] = $-$0.5. We 
assume that the subgiant donor star dominates the near-IR spectral energy 
distribution (see Fig. 16 in Harrison et al. 2013b). Our 
equivalent width analysis finds an identical result to that of H16. There are 
no significant abundance anomalies except that of magnesium, and no evidence 
for a hydrogen deficit. Silicon and potassium appear to be slightly 
underabundant, but the spectra for GK Per have lower S/N than most of the 
other targets modeled here.

\subsection{QZ Serpentis}

The oddest object in the survey of H16 was QZ Ser. It has an orbital period
near two hours, yet the donor has a spectral type near K5. H16 generated
spectra where they altered the hydrogen abundance, and concluded that
a hydrogen deficiency was the best explanation for its unusual spectrum. They
found that sodium was highly overabundant ([Na/Fe] = $+$1.5), and aluminum
also had a greatly enhanced abundance ([Al/Fe] = $+$0.5). Carbon was nearly 
undetectable,  [C/Fe] = $-$1.7. No other abundance anomalies 
were obvious in their analysis. In the following we will assume that
the donor star has a temperature of 4500 K, and that there is no 
contamination from other sources in the system (see Thorstensen et al.
2002).

To derive an estimate of $v$sin$i$ for QZ Ser, we velocity broadened a
synthetic spectrum (T$_{\rm eff}$ = 4500 K, [Fe/H] = 0.0, Hd = 90\%) until
we achieved a good match between the lines across the three bandpasses. 
Rotation velocities near $v$sin$i$ = 150 km s$^{\rm -1}$ worked reasonably
well, given the unusual abundance pattern. A grid was generated with
this rotation velocity, and [C/Fe] = $-$1.7, with our EQW programs applied
to the resulting data set. The values for the EQWs for QZ Ser across
the three bandpasses plotted against the synthetic grid is presented as Fig. 
\ref{qzser}. The enormous
sodium overabundance is obvious in each of the three bandpasses. In the
$J$ and $K$ bandpasses, the Na I doublet is about twice as deep as the
Hd = 90\% model. In the $H$-band, the sodium doublet is about three times
stronger than the 90\% model. This difference was noted in H16.

The measurement of EQWs in QZ Ser is hampered by the presence of a large
number of absorption features caused by the peculiar nature of the donor.
If we use the higher atomic mass elements, K, Ca, Fe, and Mn, the best
fit occurs for Hd = 90\%. Assuming this is correct, it appears that magnesium
is deficient. Though the EQW of the Mg I line in the $K$-band is 
consistent with Hd = 90\%, the continuum region from 22500 to 22850 \AA,
where the Mg I line is located, appears to be depressed compared to nearby 
segments of the spectrum. It is likely that other absorption features are 
complicating the extraction of the true value of the EQW for the Mg I line
in the $K$-band.

H16 found an enhanced aluminum abundance, and the results in the $J$ and
$H$-bands are consistent with this conclusion. In the $K$-band, the 
combination of He I emission, and another segment of depressed continuum,
makes the EQW measurement of the Al I feature unreliable. We cannot determine 
the silicon abundance using our EQW programs, as two of the four lines in the 
$J$-band are corrupted by what appear to be emission features, and one of the 
other lines is convolved with a nearby absorption feature. This is also true
for the Si I feature in the $K$-band. As can be seen in Fig. 28 of H16,
there are several strong absorption features in the $K$-band between 21200 
and 22000 \AA. The reddest of these, centered near $\lambda$21900 \AA ~appears
to be a convolution of absorption from Ti I at 21789 \AA, and Si I at 21903 \AA. 
Using IRAF, measurement of this line gives a value of 8.3 \AA ~for the 
equivalent width. For the synthetic spectrum (with Hd = 90\%), we find EQW = 
1.1 \AA. If the silicon abundance was this strongly enhanced, we would be 
able to see it in the Si I line at $\lambda$12264 \AA, the only clearly 
visible Si I line in the $J$-band. There is a Si I absorption line in the
$H$-band at $\lambda$15888 \AA, and for QZ Ser we measure EQW = 6.8 \AA. For 
the synthetic spectrum, we find EQW = 6.5 \AA. Silicon appears to have a normal 
abundance.

Another one of the strong absorption lines in the region near H I Br$\gamma$
in the $K$-band is the feature at $\lambda$21789 \AA. There is a Ti I line at 
this position. We measure EQW = 4.5 \AA ~in QZ Ser, and EQW = 1.2 in the 
synthetic spectrum. There are two other strong lines to the blue of previous 
feature, one at $\lambda$21450 \AA, and the other at $\lambda$21618 \AA. Using 
the NIST Spectral 
Database\footnote{https://physics.nist.gov/PhysRefData/ASD/lines\_form.html}, 
there are absorption lines from both Na I and Ti I at $\lambda$21450 \AA, 
and a Ti I line at $\lambda$21604 \AA. The bluer of these two lines has
EQW = 5.4 \AA, compared to EQW = 1.17 in the synthetic data. There does
not seem to be a counterpart in the model spectrum for the redder line.
Unless there are absorption lines of other elements contaminating these
various features, it appears that titanium has an enhanced abundance.

Our analysis for the spectrum of QZ Ser confirms the results from H16:
it appears that the donor star in this system suffers an extreme hydrogen
deficit ($\geq$ 90\%), larger than any of the other program objects. The
enormous overabundance of sodium, enhanced aluminum, and the deficit of 
magnesium found by H16 are confirmed. In addition, there appears to be a large 
enhancement of titanium, [Ti/Fe] $\sim$ $+$0.5. 

\section{Discussion}

We have used ATLAS12 to generate hydrogen-deficient stellar atmospheres
for K dwarfs to enable us to generate synthetic spectra to determine if the
donor stars of any long period cataclysmic variables have evidence for
photospheric hydrogen deficiencies.  It is clear that several systems show 
strong evidence for such anomalies: AE 
Aqr, DX And, EY Cyg, and QZ Ser. For the first three of these, the hydrogen
deficits are in the 25 to 50\% range, while QZ Ser appears to have a hydrogen
deficit of 90\%. Unfortunately, we cannot rigorously quantify the hydrogen 
deficits due to spectral type uncertainty, the limited number of uncontaminated
spectral features, intrinsic abundance variations, contamination issues, the 
presence of emission lines of H and He, insufficient S/N in the data, and the 
vagaries of the near-infrared spectral reduction process. Fortunately, all four
systems with identified hydrogen deficits appear to have low levels of 
contamination, and have donors with well-determined spectral types. We 
summarize our results for the program objects in Table \ref{results}.

We had previously shown that AE Aqr was unusual in that its carbon deficit was 
very large, [C/Fe] $\leq$ $-$1.4. It has also been proposed that AE Aqr is a 
relatively young CV (Schenker et al. 2002), having only recently exited the 
``post-thermal-time-scale mass transfer state.'' They predicted that it might 
have a small $^{\rm 12}$C/$^{\rm 13}$C ratio, which appears to have been 
confirmed by HM17. Echevarr\'{i}a et al. (2007) concluded that the donor star 
in EY Cyg was 30\% larger than a main sequence star of the same spectral type, 
suggesting it is evolved. A similar conclusion was reached by Drew et al. (1993)
for DX And. Thus, it is not too surprising that these three systems might
show evidence for hydrogen deficits. As described in H16, Thorstensen et al. 
(2002) had already suggested that the donor in QZ Ser might be hydrogen 
deficient. It certainly remains the poster child for such systems. There 
appeared to be several short period CVs in the survey of H16 that had similar 
properties to QZ Ser, but which had M-type secondaries. 

It is worthwhile to examine the outburst intervals for the hydrogen deficient
dwarf novae (AE Aqr is not in this category). Using the AAVSO data 
base, we find that in the past five years DX And, EY Cyg and QZ Ser have had 
very few dwarf novae outbursts. For DX And there have been three observed 
eruptions, two for QZ Ser (one of which appears to have been a 
``superoutburst''), and only one for EY Cyg. While RU Peg, a solar abundance 
pattern system with a similar orbital period to DX And and EY Cyg, has had 21 
eruptions! Many CVs with orbital periods like that of QZ Ser are SU UMa systems 
that exhibit frequent dwarf novae outbursts, and more infrequent 
superoutbursts. Other CV systems with similar orbital periods, however, can 
also have infrequent outbursts (see Kato et al. 2003). Establishment of any 
tendencies will require a larger survey, but starting by selecting infrequently
outbursting dwarf novae for study might prove fruitful for finding additional 
hydrogen deficient objects.

As in H16, we find that carbon and magnesium deficits are the most common 
abundance anamolies, whether or not the system has a hydrogen deficit. Aside
from sodium, there are few other elements in the program objects that have odd 
abundances. RU Peg may have a small deficit of silicon. In QZ Ser, both 
aluminum and titanium appear to have enhanced abundances. One of the most
surprising results from our survey is that the three long period Z Cam systems 
(Z Cam, SY Cnc, and AH Her) have solar metallicities and abundance patterns. 
The origin of the Z Cam phenomenon does not appear to be related to 
{\it unusual} abundance issues. It is interesting that RX And appears to exhibit
some Z Cam-like behavior, and its donor has a solar metallicity. However, 
the donor in RX And also has deficits of both magnesium and carbon, unlike
the three Z Cam systems studied here.

The consideration of the contamination of the infrared spectra by non-stellar
sources has led to the dramatic revision in the metallicities for both SS Cyg 
and RU Peg. Both systems appear to have [Fe/H] = 0.0, though the abundance
anomalies identified in H16 and HM17 for these two objects remain. It is likely 
that
many of the objects identified as having low metallicity in that survey actually
have spectra that suffer from significant contamination issues. The combination 
of EQW measures, light curve modeling of the $WISE/NEOWISE$ data, and matching 
of (contaminated) synthetic/template spectra to the observed $JHK$ spectra, 
appears to allow for excellent estimates of the contamination level. Obviously, 
it would be better to have quiescent $JHK$ light curves of every source for
light curve modeling, but given that such data do not exist, the $WISE/NEOWISE$ 
photometry provides a useful alternative. As demonstrated above, those data
were sufficient to determine/revise the orbital inclinations for SY Cnc, V508
Dra, AH Her, and RU Peg.

Our new methods allow us to examine the reliability of the spectral type
determinations listed in H16, and elsewhere. For the majority of the
systems, the $\chi^{\rm 2}$ techniques used in H16 produced reliable spectral
types. We confirm their assignment of an M-type donor in RX And. The only
significantly different spectral type reassignment is that for AH Her. Not
surprising given that H16 only had a $K$-band spectrum to work with. The same 
cannot be said for most of the spectral types assigned using optical data. Only 
in those systems where the contamination in the visual is insufficient to mask 
the donor star (e.g., AE Aqr, DX And or SS Cyg), and/or where line ratio
analysis was performed, are optically-derived spectral types accurate.
Perhaps the most surprising result in this vein is the late spectral
type we find for the donor in Z Cam. This system has an orbital period close 
to seven hours. The mass of the donor, M$_{\rm 2}$ = 0.77 M$_{\sun}$, is
similar to that of a main sequence K0V. Perhaps, like the results in H16
for U Gem, it is the large inclination angle that makes the system appear to
have a such a cool temperature.

The success of using the equivalent widths of field M dwarfs in ascertaining
the nature of the donors in RX And, Z Cam and AH Her suggests that we can use 
this technique to validate our analysis regimen. In Fig. \ref{sscygmdwarfs}, we 
plot the EQW measurements for all of the K and M dwarfs discussed above. The 
x-axis in this plot now extends to $-$8, which is the coding for a K0V. We plot 
the values of the EQW measures for SS Cyg as a horizontal solid line in each 
plot. We also plot what the equivalent widths would be if the contamination
was 25\% (dashed lines), and 50\% (dotted lines). Our results for SS Cyg put
its spectral type at $-$4.5. It is obvious that a 25\% contamination is the
best fit to the majority of the EQW measures near this spectral type. The only
significant deviation is the large magnesium deficit. This confirms the results 
derived above, and lends confidence to the entire process. If the metallicity
of the CV donor is not too extreme, this technique provides a useful estimate of 
the contamination level if the spectral type is approximately known, or the 
spectral type if the contamination level can be estimated, without recourse to 
synthetic or template spectra matching.

Thus, we can now state with confidence that magnesium and carbon deficiencies
are common among the donor stars in CVs. There are also systems
with enhanced levels of sodium. H16 discussed the possible mechanisms to
produce such systems. It is clear that long period CVs have a very wide range
of properties, from systems with completely normal abundance patterns, to highly
peculiar ones. It is interesting that the two longest period systems, GK Per and
V508 Dra, have subgiant donor stars that are relatively normal, with the only 
shared anomaly being a carbon deficit. While there are systems with orbital 
periods near 10 hrs that have significant hydrogen deficits. As explored in 
Kalomeni et al. (2016), there have to be a variety of paths for producing CVs.

\section{Conclusions}

Using new hydrogen deficient atmospheres, we have generated synthetic spectra 
to determine whether any CV donor stars show evidence for hydrogen 
deficiencies. We have found four systems that appear to have such deficits. We 
have also shown that when a contamination estimate is possible, one can use 
the EQW method to obtain both donor star temperatures and identify abundance 
anomalies without the need for model spectra. This includes application to
M-type donor stars, where it currently remains extremely difficult to construct
realistic synthetic spectra. Further refinement of the M star equivalent width 
technique used above, which would include obtaining additional spectra of 
M-type stars spanning a wider range in properties, is needed to enable its use 
for deriving accurate values for donor star temperatures, metallicities,
abundance anomalies, and contamination levels. The latter grows in 
importance in shorter period CV systems where the donor stars are much less 
luminous relative to the white dwarf and its accretion disk.

\acknowledgements T. E. H. was partially supported by a grant from the NSF 
(AST-1209451). The Gemini GNIRS data were acquired under the program 
GN-2017A-Q-91. This publication makes use of data products from the Wide-field
Infrared Survey Explorer, which was a joint project of the University of 
California, Los Angeles, and the Jet Propulsion Laboratory/California Institute 
of Technology, funded by the National Aeronautics and Space Administration. 
This publication makes use of data products from the Near-Earth Object 
Wide-field Infrared Survey Explorer (NEOWISE), which is a project of the Jet 
Propulsion Laboratory/California Institute of Technology. NEOWISE is funded by 
the National Aeronautics and Space Administration. We acknowledge with thanks 
the variable star observations from the AAVSO International Database contributed
by observers worldwide and used in this research. This publication makes use of 
data products from the 2MASS, which is a joint project of the University of 
Massachusetts and the Infrared Processing and Analysis Center/California 
Institute of Technology, funded by the National Aeronautics and Space 
Administration and the National Science Foundation.

\clearpage
\begin{center}
{\bf References}
\end{center}
Abada-Simon, M., Casares, J., Evans, A., Eyres, S., Fender, R., et al. 2005,
A\&A, 433, 1063\\
Ag\"{u}eros, M. A., Anderson, S. F., Covey, K. R., Hawley, S. L., Margon, B.,
et al. 2009, ApJS, 181, 444\\
Beuermann, K., Baraffe, I., Kolb, U., \& Weichhold, M. 1998, A\&A, 339, 518\\
Bitner, M. A., Robinson, E. L., \& Behr, B. B. 2007, ApJ, 662, 564\\
Bonnet-Bidaud, J. M., \& Mouchet, M. 1987, A\&A, 188, 89\\
Boyajian, T. S., von Braun, K., van Belle, G., McAlister, H. A., ten 
Brummelaar, T. A., et al. 2012, ApJ, 757, 112\\
Bruch, A., Vrielmann, S., Hessman, F. V., Kocksiek, A., \& Schimpke, T. 1997,
A\&A, 327, 1107\\
Casares, J., Mart\'{i}nez-Pais, I. G., \& Rodr\'{i}guez-Gil, P. 2009, MNRAS, 
399, 1534\\
Castelli, F. 2005, MSAIS, 8, 25\\
Covey, K. R., Lada, C. J., Roman-Zuniga, C. et al. 2010, ApJ, 722, 971\\
Cushing, M. C., Rayner, J. T., \& Vacca, W. D. 2005, ApJ , 623, 1115\\
Drew, J. E., Jones, D. H. P., \& Woods, J. A. 1993, MNRAS, 260, 803\\
Dubus, G., Campbell, R., Kern, B., Taam, R. E., \& Spruit, H. C. 2004, MNRAS,
349, 869\\
Dunford, A., Watson, C. A., Smith, R. C. 2012, MNRAS, 422, 3444\\
Echevarr\'{i}a, J., Connon Smith, R., Costero, R., Zharikov, S., \& Michel, R. 
2007, A\&A, 462, 1068\\
Friend, M. T., Martin, J. S., Connon Smith, R., \& Jones, D. H. P. 1990b, 
MNRAS, 246, 654\\
Friend, M. T., Martin, J. S., Connon Smith, R., \& Jones, D. H. P. 1990a, 
MNRAS, 246, 637\\
Gaia Collaboration: Brown, A. G. A., Vallenari, A., Prusti, T., de Bruijne,
J. H. J., Babusiaux, C., et al. 2018, arXiv: 1804.09365\\
G\"{a}nsicke, B. T., Szkody, P., de Martino, D., Beuermann, K., Long, K. S.,
et al. 2003, ApJ, 594, 443\\
Godon, P., Sion, E. M., Barrett, P. E., Hubeny, I.,Linnell, A. P., \&
Szkody, P. 2008, ApJ, 679, 1447\\
Goliasch, J., \& Nelson, L. 2015, ApJ, 809, 80\\
Harrison, T. E., \& Marra, R. E. 2017 (HM17), ApJ, 843, 152\\
Harrison, T. E. 2016 (H16), ApJ, 833, 14\\
Harrison, T. E., \& Hamilton, R. T. 2015, AJ, 150, 142\\
Harrison, T. E., Hamilton, R. T., Tappert, C., Hoffman, D. I., \& Campbell,
R. K. 2013a, ApJ, 145, 19\\
Harrison, T. E., Campbell, R. D., \& Lyke, J. E. 2013b, ApJ, 146, 37\\
Harrison, T. E., Bornak, J., Howell, S. B., Mason, E., Szkody, P., \& McGurk, 
R. 2009, AJ, 137, 4061\\
Harrison, T. E., Howell, S. B., Szkody, P., \& Cordova, F. A. 2007, ApJ, 133,
162\\
Harrison, T. E., Osborne, H. L., \& Howell, S. B., 2005, AJ, 129, 2400\\
Harrison, T. E., Osborne, H. L., \& Howell, S. B., 2004a, AJ, 127, 3493\\
Harrison, T. E., Johnson, J. J., McArthur, B. E., Benedict, G. F., Szkody, P.,
et al. 2004b, AJ, 127, 460\\
Hartley, L. E., Long, K. S., Froning, C. S., \& Drew, J. E. 2005, ApJ, 623,
425\\
Henry, T. J., \& McCarthy, D. W. 1993, AJ, 106, 773\\
Hill, C. A., Watson, C. A., Shahbaz, T., Steeghs, D., \& Dhillon, V. S. 2014,
MNRAS, 444, 192\\
Howell, S. B., Nelson, L. A., \& Rappaport, S. 2001, ApJ, 550, 897\\
Houk, N., Swift, C. M., Murray, C. A., Penston, M. J., \& Binney, J. J. 1997, 
in Proc. ESA Symp. Hipparcos — Venice, ESA SP-402, ed. M. A. C. Perryman \& 
P. L. Bernacca (Noordwijk: ESA), 279\\
Kaitchuck, R. R. 1989, PASP 101, 1129\\
Kalomeni, B., Nelson, L., Rappaport, S., et al. 2016, ApJ , 833, 83\\
Kato, T., Nogamin, D., Moilanen, M., \& Yamaoka, H. 2003, PASJ, 55, 989\\
King, A. R., \& Kolb, U. 1995, ApJ, 439, 330\\
Knigge, C., Baraffe, I., \& Patterson, J. 2011, ApJS, 194, 28\\
K\"{o}rding, E., Rupen, M., Knigge, C., et al. 2008, Sci, 320, 1318\\
Kraft, R. P., Krzeminski, W., Mumford, G. S. 1969, ApJ, 158, 589\\
Kraft, R. P. 1962, ApJ, 135, 408\\
Kurucz, R. L. 2005, MSAIS, 8, 14\\
Lindegren, L., Hernandez, J., Bombrun, A., Klioner, S., Bastian, U., et al.
2018, arXiv 1804,09366\\
Mainzer, A., Bauer, J., Grav, T., Masiero, J., Cutri, R. M., et al. 2011,
ApJ, 731, 53\\
Mauche, C. W., Lee, Y. P., \& Kallman, T. R. 1997, ApJ , 477, 832\\
Neustroev, V. V., \& Zharikov, S. 2008, MNRAS, 386, 1366\\
North, R. C., Marsh, T. R., Moran, C. K. J., Kolb, U., Smith, R. C., \& Stehle,
R. 2000, MNRAS, 313, 383\\
North, R. C., Marsh, T. R., Kolb, U., Dhillon, V. S., \& Moran, C. K. J. 2002,
MNRAS, 337, 1215\\
Pecaut, M. J., \& Mamajek, E. E. 2013, ApJS, 208, 9\\
Podsiadlowski, Ph., Han, Z., \& Rappaport, S. 2003, MNRAS, 340, 1214\\
Politano, M., \& Weiler, K. P. 2007, ApJ, 655, 663\\
Ritter, H. 2012, MmSAI, 83, 505\\
Ritter H., \& Kolb U. 2003, A\&A, 404, 301 \\
Rodriguez-Gil. P., Torres, M. A. P., Gansicke, B. T., Munoz-Darius, T., 
Steeghs, D., Schwarz, R., Rau, A., Hagen, H. J. 2009, MNRAS, 395, 973\\
Salaris, M., Serenelli, A., Weiss, A., \& Bertolami, M. M. 2009, ApJ, 692, 
1013\\
Schreiber, M. R., G\"{a}nsicke, B. T., \& Mattei, J. A., 2002, A\&A, 384, 6\\
Sepinsky, J. F., Sion, E. M., Szkody, P., \& G\"{a}nsicke, B. T. 2002, ApJ,
574, 937\\
Shafter, A. W. 1983, PhD thesis, University of California, Los Angeles\\
Shenker, K., King, A. R., Kolb, U., Wynn, G. A., \& Zhang, Z. 2002, MNRAS, 337,
1105\\
Simonsen, M., Boyd, D., Goff, W., Krajci, T., Menzies, K., et al. 2017, JAAVSO,
42, 177\\
Sion, E. M., Winter, L., Urban, J. A., Tovmassian, G. H., Zharikov, S.
et al. 2004, AJ, 128, 1795 \\
Sion, E. M., \& Urban, J. A. 2002, ApJ, 572, 456\\
Soubiran, C., Le Campion, J. -F., Brouillet, N., \& Chemin, L. 2016, A\&A,
591, 118\\
Szkody, P., \& Silber, A. 1996, AJ , 112, 289\\
Tanzi, E. G., Chincarini, G., \& Tarenghi, M. 1981, PASP, 93, 68\\
Thorstensen, J. R., Fenton, W. H., Patterson, J., et al. 2002, PASP, 114, 1117\\
Thorstensen, J. R., \& Ringwald, F. A. 1995, IBVS, 4249\\
Urban, J. A., \& Sion, E. M. 2006, ApJ, 642, 1029\\
Vollmann, K., \& Eversberg, T. 2006, AN, 327, 862\\
Wade, R. A. 1980, PhD thesis, California Institute of Technology\\
Warner, B. 1995, in ``Cataclysmic Variable Stars'', (Cambridge University 
Press:Cambridge), p495\\
Welsh, W. F., Horne, K., \& Gomer, R. 1995, MNRAS, 275, 649\\
Williams, R. E., \& Ferguson, D. H. 1982, ApJ, 257, 672\\
Wils, P. 2011, JAAVSO, 39, 60\\
Wright, E. L., Eisenhardt, P. R. M., Mainzer, A. K., et al. 2010,
AJ, 140, 1868\\

\begin{deluxetable}{lcccc}
\tablecolumns{5}
\tablewidth{0pt}
\centering
\tablecaption{Observation Log}
\tablehead{
\colhead{Object} & \colhead{Date Observed} & \colhead{Number Exps.} & \colhead{Orbital Phase}& \colhead{Instrument}\\
\colhead{ } & \colhead{(UT)} & \colhead{$\times$ Exp. Time} & \colhead{(mean)} &
\colhead{ }\\
}
\startdata
DX And & 2017 07 10 & 10 $\times$ 340 s & 0.00 & GNIRS\\
Z  Cam & 2018 02 02 &  8 $\times$ 300 s & 0.27 & TSPEC\\ 
SY Cnc & 2018 02 02 &  4 $\times$ 300 s & 0.68 & TSPEC\\
EM Cyg & 2017 06 02 & 10 $\times$ 300 s & 0.28 & GNIRS\\
V508 Dra& 2017 07 18 & 12 $\times$ 240 s & 0.17 & TSPEC\\
AH Her & 2018 02 02 &  8 $\times$ 300 s & 0.57 & TSPEC\\
\enddata
\label{obslog}
\end{deluxetable}

\begin{deluxetable}{lccc}
\tablecolumns{4}
\tablewidth{0pt}
\centering
\tablecaption{Equivalent Width Measurement Bandpasses}
\tablehead{
\colhead{Element} & \colhead{Blue Continuum} & \colhead{Line Bandpass} & \colhead{Red Continuum}\\
\colhead{  } & \colhead{Bandpass (\AA)}& \colhead{(\AA)} & \colhead{Bandpass (\AA)}}
\startdata
Na I & 11250 --- 11360 & 11372 --- 11418 & 11450 --- 11570\\
Si I & 11450 --- 11570 & 11585 --- 11630 & 11709 --- 11732\\
K I  & 11450 --- 11570 & 11682 --- 11709 & 11709 --- 11732\\
Mg I & 11785 --- 11814 & 11822 --- 11853 & 11853 --- 11868\\
Fe I & 11853 --- 11868 & 11875 --- 11908 & 11908 --- 11940\\
Si I & 11905 --- 11940 & 12090 --- 12115 & 12135 --- 12235\\
Si I & 12135 --- 12235 & 12265 --- 12289 & 12290 --- 12365\\
K I  & 12450 --- 12495 & 12515 --- 12545 & 12580 --- 12655\\
Ni I & 12907 --- 12923 & 12933 --- 12955 & 12955 --- 12958\\
Mn I & 12955 --- 12958 & 12970 --- 12995 & 12995 --- 13016\\
Si I & 12995 --- 13016 & 13028 --- 13055 & 13055 --- 13105\\
Al I & 13055 --- 13105 & 13119 --- 13174 & 13174 --- 13255\\
Mn I & 13174 --- 13255 & 13275 --- 13342 & 13342 --- 13365\\
Mg I & 14730 --- 14840 & 14850 --- 14900 & 14910 --- 14994\\
Mg I & 14910 --- 14994 & 15000 --- 15070 & 15100 --- 15187\\
Fe I & 15267 --- 15280 & 15283 --- 15302 & 15340 --- 15365\\
Na I & 16245 --- 16295 & 16369 --- 16410 & 16460 --- 16630\\
Al I & 16560 --- 16645 & 16732 --- 16765 & 16885 --- 16990\\
Mg I & 17030 --- 17080 & 17081 --- 17140 & 17150 --- 17290\\
Al I & 20990 --- 21034 & 21079 --- 21185 & 21280 --- 21330\\
Si I & 21280 --- 21330 & 21335 --- 21370 & 21380 --- 21415\\
Na I & 21940 --- 21990 & 22030 --- 22119 & 22190 --- 22245\\
Fe I & 22190 --- 22245 & 22360 --- 22405 & 22550 --- 22590\\
Ca I & 22550 --- 22590 & 22591 --- 22685 & 22690 --- 22780\\
Mg I & 22690 --- 22780 & 22781 --- 22840 & 22842 --- 22900\\
\enddata
\label{eqwwaves}
\end{deluxetable}

\begin{deluxetable}{lccc}
\tablecolumns{4}
\tablewidth{0pt}
\centering
\tablecaption{Spectral Type Template Parameters}
\tablehead{
\colhead{Name} & \colhead{Spectral Type} & \colhead{$\langle$T$_{\rm eff} \rangle$}& \colhead{$\langle$[Fe/H]$\rangle$}\\
\colhead{}&\colhead{}&\colhead{(K)}&\colhead{}}
\startdata
HD145675 & K0V & 5320 $\pm$ 114 & $+$0.41 $\pm$ 0.12\\
HD10476  & K1V & 5189 $\pm$ 53  & $-$0.06 $\pm$ 0.06\\
HD3765   & K2V & 5023 $\pm$ 66  & $+$0.05 $\pm$ 0.10\\
HD219134 & K3V & 4837 $\pm$ 138 & $+$0.05 $\pm$ 0.10\\
HD45977  & K4V & 4689 $\pm$ 174 & $+$0.03 $\pm$ 0.18\\
HD36003  & K5V & 4615 $\pm$ 29  & $-$0.14 $\pm$ 0.08\\
61 Cyg A & K5V & 4394 $\pm$ 150 & $-$0.18 $\pm$ 0.15\\
HD237903 & K7V & 4110 $\pm$ 102 & $-$0.21 $\pm$ 0.10\\
\enddata
\label{mkdata}
\end{deluxetable}

\begin{deluxetable}{lccccc}
\small
\tablecolumns{6}
\tablewidth{0pt}
\centering
\tablecaption{Results for the Program CVs}
\tablehead{
\colhead{Name}&\colhead{T$_{\rm eff}$}&\colhead{log$g$}&\colhead{[Fe/H]}&
\colhead{Hydrogen}&\colhead{Abundance Notes}\\
\colhead{}&\colhead{(K)}&\colhead{}&\colhead{}&\colhead{Deficit}&\colhead{}
}
\startdata
DX And &5000&4.5&0.0&25\% &[C/Fe] = $-$0.5, [Mg/Fe] = $-$0.3, [Na/Fe] = $+$0.3\\
DX And &5000&4.0&0.0&25\% &[C/Fe] = $-$0.7, [Mg/Fe] = $-$0.2, [Na/Fe] = $+$0.5\\
RX And &3500&4.5&$+$0.07& 0\% &[C/Fe] $\leq$ $-$0.3, [Mg/Fe] $\leq$ $-$0.2\\
AE Aqr &5000&4.5&0.0&25\% & see HM17\\
AE Aqr &5000&4.0&0.0&25\% & see HM17\\
Z Cam  &3575&4.5&$+$0.03& 0\% & solar\\
SY Cnc &5250&4.5&0.0& 0\%& solar\\
SS Cyg &4750&4.5&0.0& 0\%& HM17, except [Na/Fe] = $+$0.2\\
EM Cyg &4500&4.5&0.0& 0\%& [C/Fe] $\leq$ $-$0.5, [Mg/Fe] $\leq$ $-$0.5\\
EY Cyg &5250&4.5&0.0&25\%& see H16\\
EY Cyg &5250&4.0&0.0&50\%& see H16\\
V508 Dra&4500&4.0&0.0&0 \%& [C/Fe] = $-$0.3\\
AH Her &3600&4.5&0.12& 0\%& solar\\
RU Peg &5000&4.5&0.0& 0\%& see HM17\\
RU Peg &5000&4.0&0.0& 0\%& see HM17\\
GK Per &5000&4.0&$-$0.3& 0\%& see H16\\
QZ Ser &4500&4.5&0.0&90\%& H16, except [Ti/Fe] = $+$0.5?\\
\enddata
\label{results}
\end{deluxetable}

\renewcommand{\thefigure}{1}
\begin{figure}[htb]
\centerline{{\includegraphics[width=15cm]{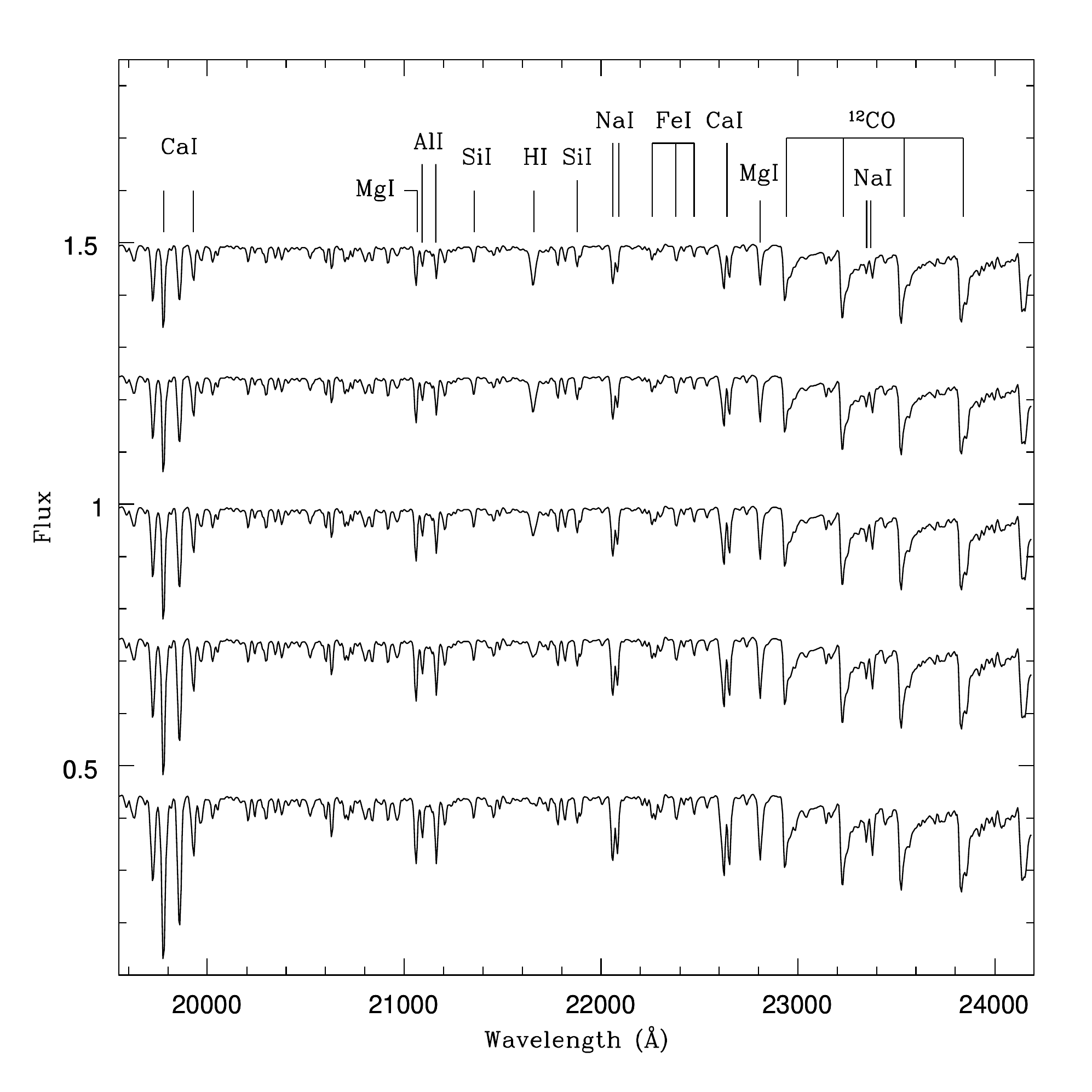}}}
\caption{Synthetic $K$-band spectra for T$_{\rm eff}$ = 5000 K and [Fe/H] = 0.0,
with five different values for the hydrogen deficit. The model at top has Hd 
= 0\%, while that at the bottom has Hd = 90\%. The metal lines (the strongest
of which are identified at the top of the plot) all get stronger as hydrogen 
becomes depleted, while the H I Br$\gamma$ line at 21600 \AA ~slowly 
disappears.}
\label{speccomp}
\end{figure}

\renewcommand{\thefigure}{2}
\begin{figure}[htb]
\centerline{{\includegraphics[width=15cm]{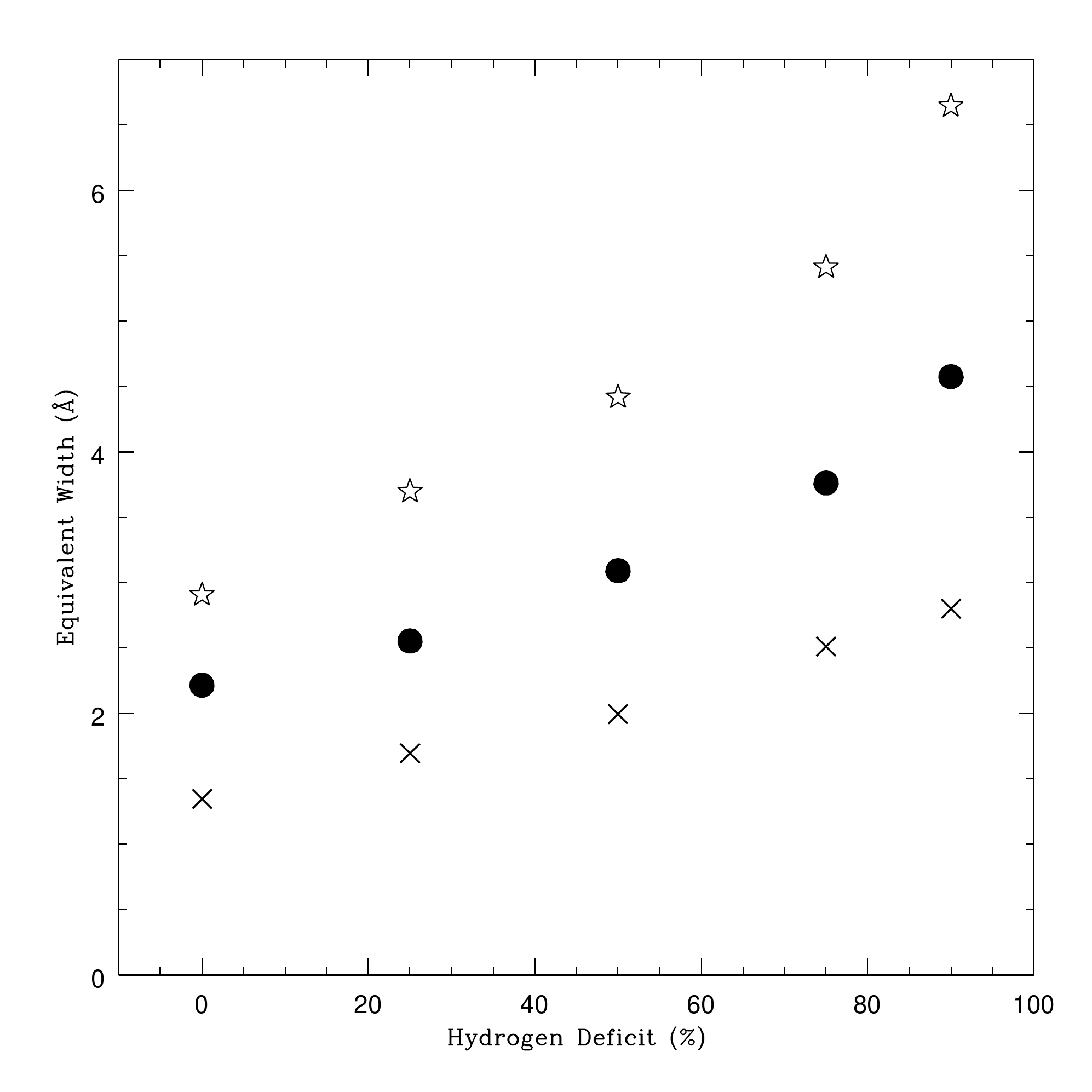}}}
\caption{Equivalent width measurements from the synthetic $K$-band spectra
presented in Fig. \ref{speccomp}. The star-shaped symbols are for the Na I
doublet at 22000 \AA, the filled circles are for the Ca I triplet at
22632 \AA, and the crosses are for the Mg I feature 22808 \AA.}
\label{eqwbands}
\end{figure}

\renewcommand{\thefigure}{3abc}
\begin{figure}[htb]
\centerline{{\includegraphics[width=15cm]{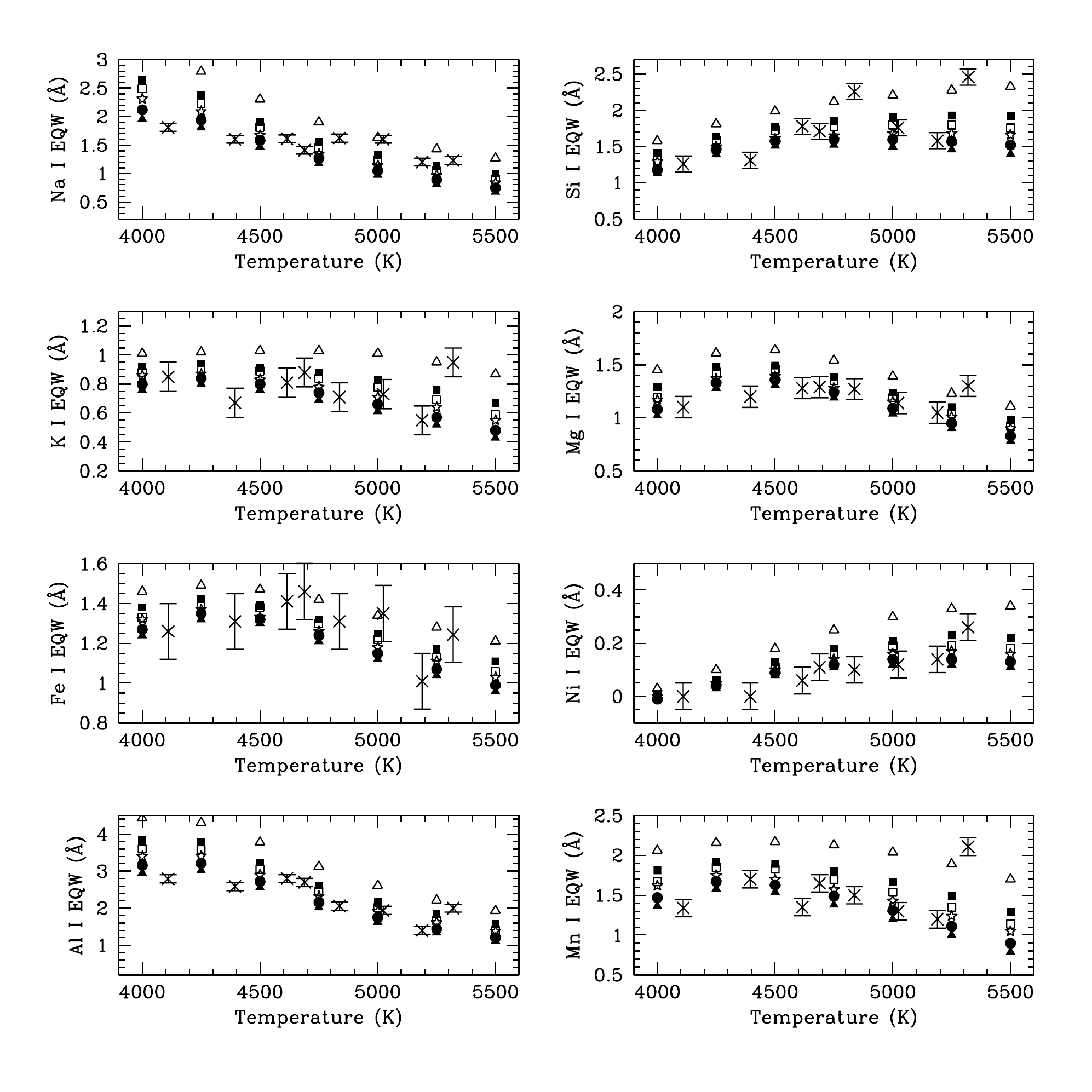}}}
\caption{The equivalent width measurements (crosses with error bars) in the
$J$-band for the K type dwarfs listed in Table \ref{mkdata}, plotted against 
those measured from the synthetic spectra for five metallicities: [Fe/H] = 0.0 
(stars), [Fe/H] = $+$0.1 (open squares), [Fe/H] = $+$0.2 (filled squares), 
[Fe/H] = $+$0.5 (open triangles), [Fe/H] = $-$0.1 (filled circles), and
[Fe/H] = $-$0.2 (filled triangles).}
\label{mkeqw}
\end{figure}

\renewcommand{\thefigure}{3b}
\begin{figure}[htb]
\centerline{{\includegraphics[width=15cm]{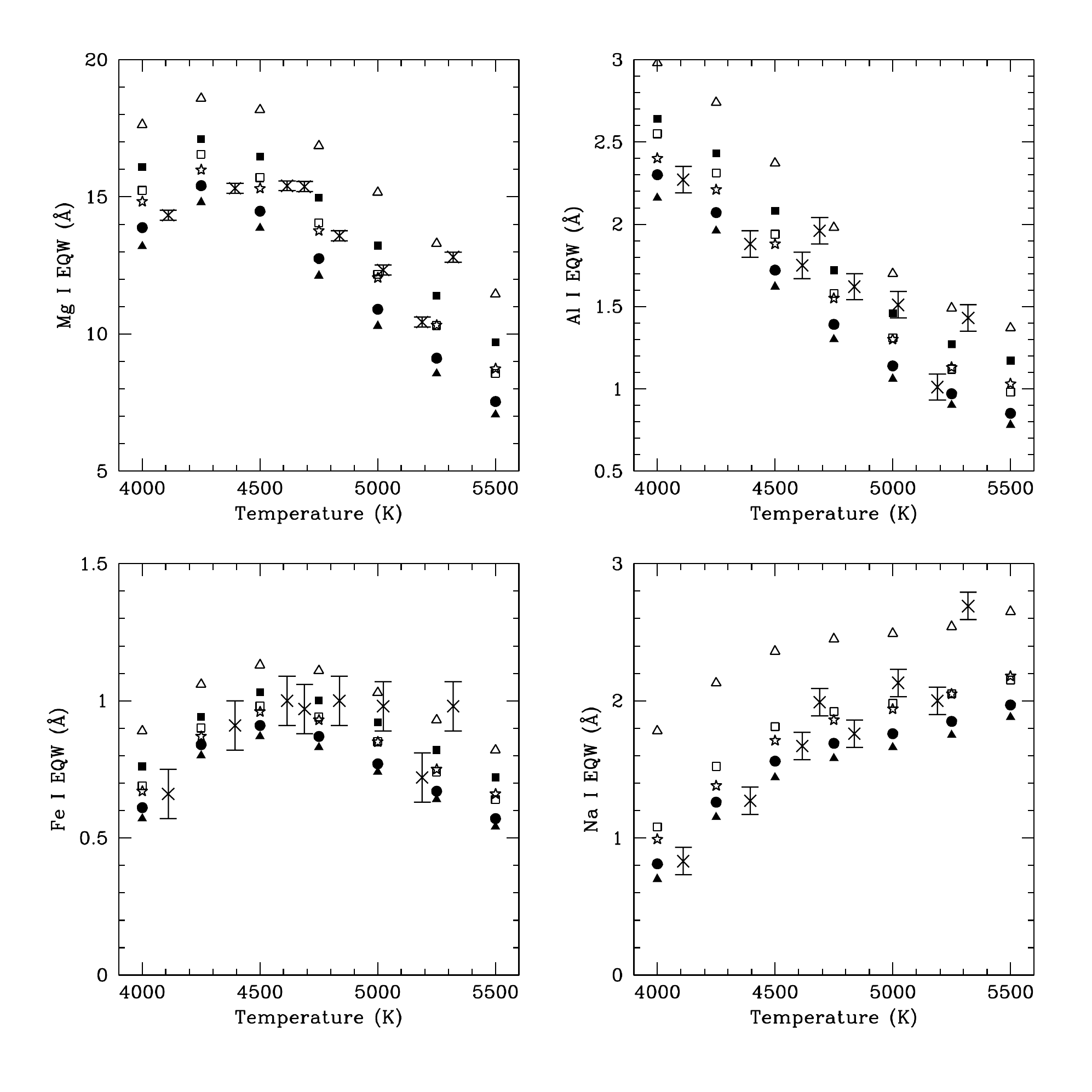}}}
\caption{Same as 3a, but for the $H$-band.}
\end{figure}

\renewcommand{\thefigure}{3c}
\begin{figure}[htb]
\centerline{{\includegraphics[width=15cm]{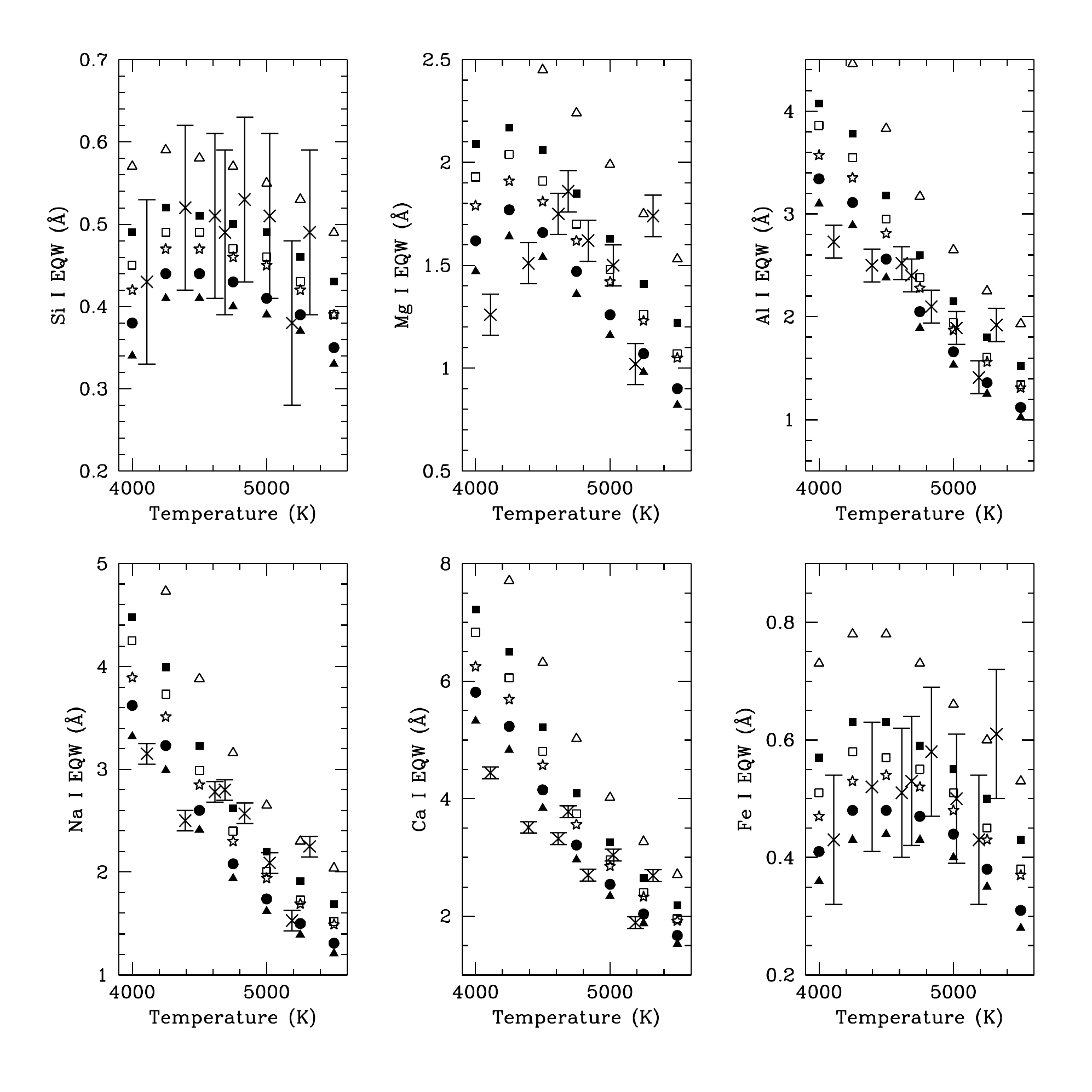}}}
\caption{Same as 3a, but for the $K$-band.}
\end{figure}

\renewcommand{\thefigure}{4abc}
\begin{figure}[htb]
\centerline{{\includegraphics[width=15cm]{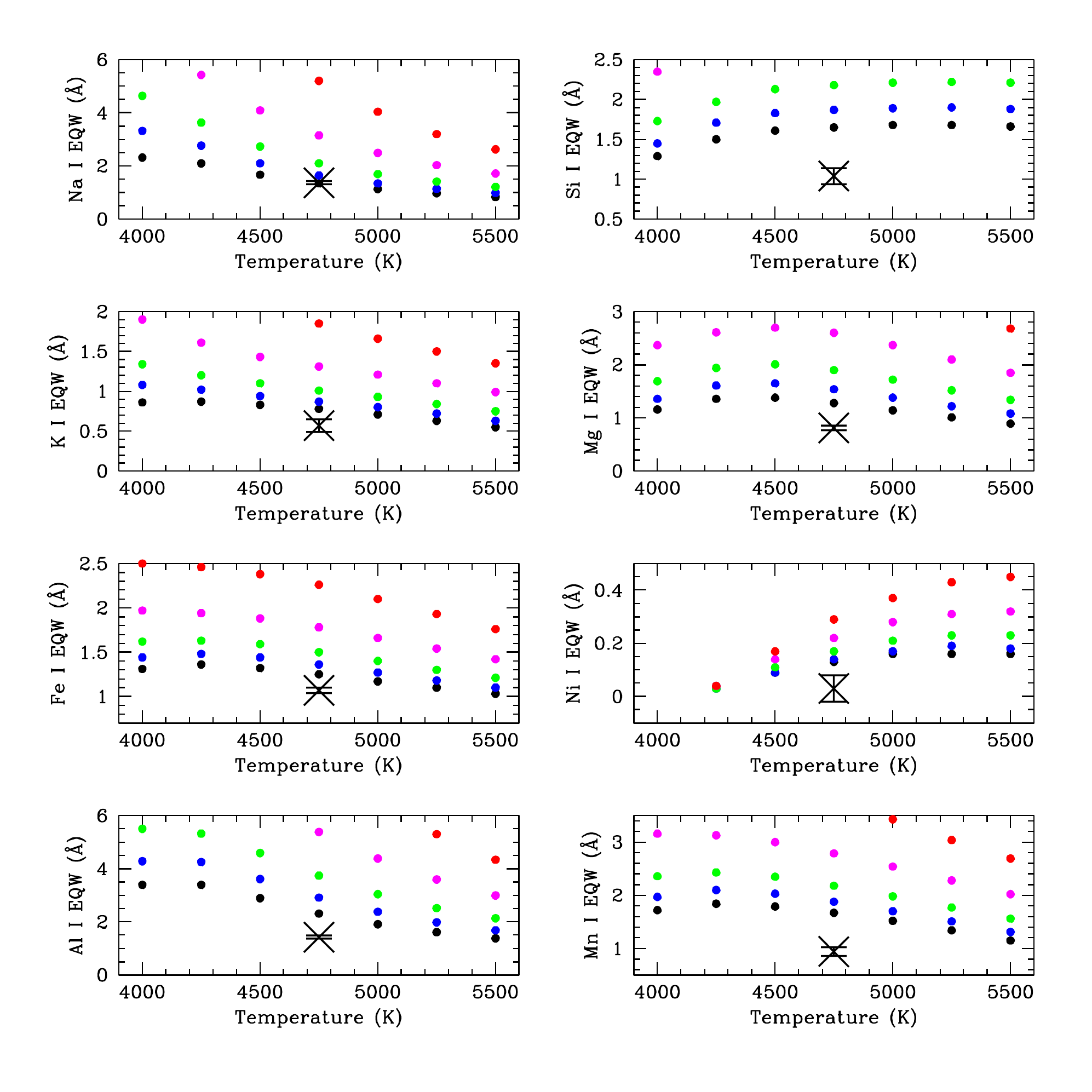}}}
\caption{The fit of the $J$-band EQW measurements for SS Cyg to those for the
model synthetic spectra with log$g$ = 4.5, and [Fe/H] = 0.0. The filled
black circles are for no hydrogen deficit, blue circles are for a hydrogen
deficit of 25\%, green is for a deficit of 50\%, magenta for a deficit of 
75\%, and red for a deficit of 90\%.}
\label{sseqwsolar}
\end{figure}

\renewcommand{\thefigure}{4b}
\begin{figure}[htb]
\centerline{{\includegraphics[width=15cm]{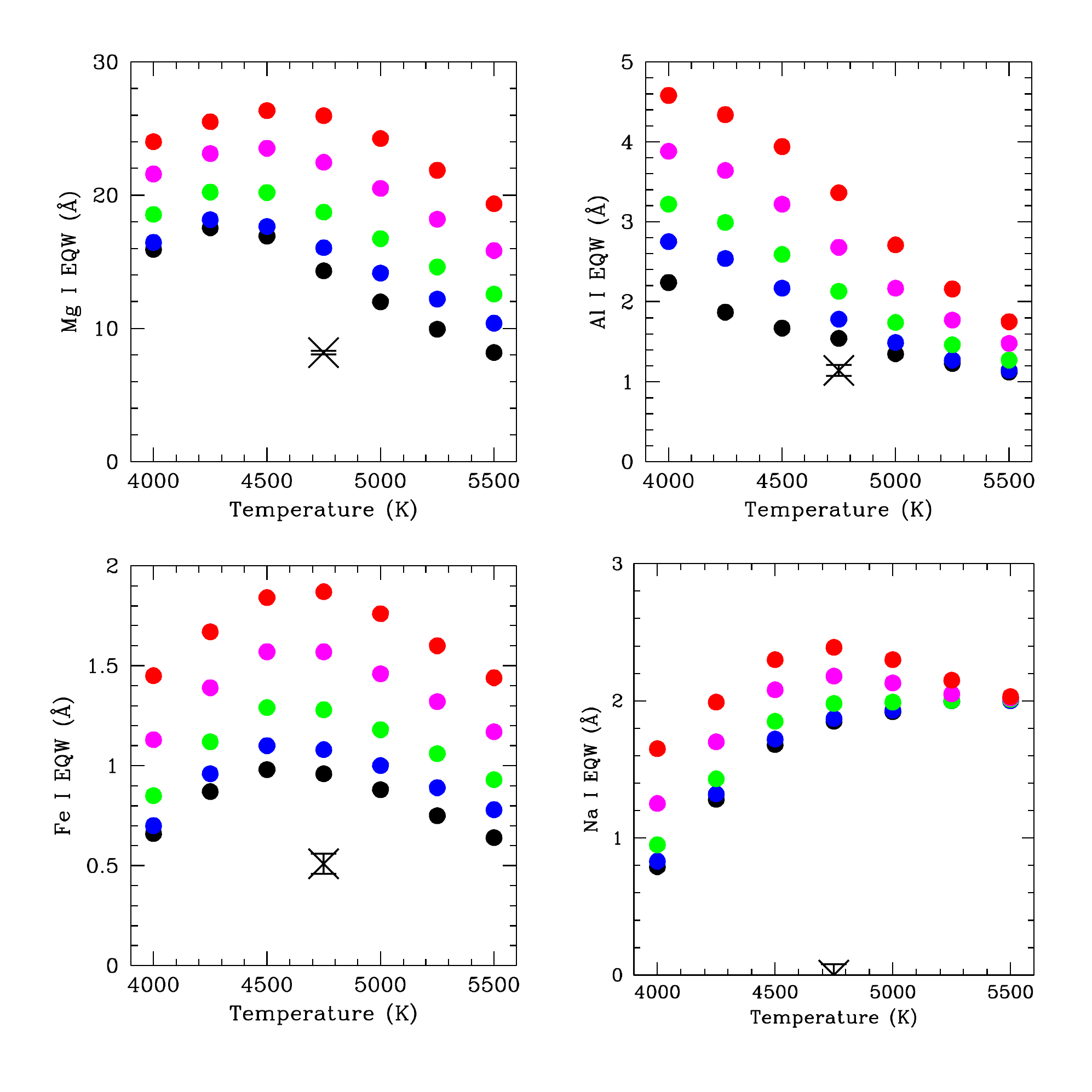}}}
\caption{The same as panel $a$, but for the $H$-band.}
\label{sseqwh}
\end{figure}

\renewcommand{\thefigure}{4c}
\begin{figure}[htb]
\centerline{{\includegraphics[width=15cm]{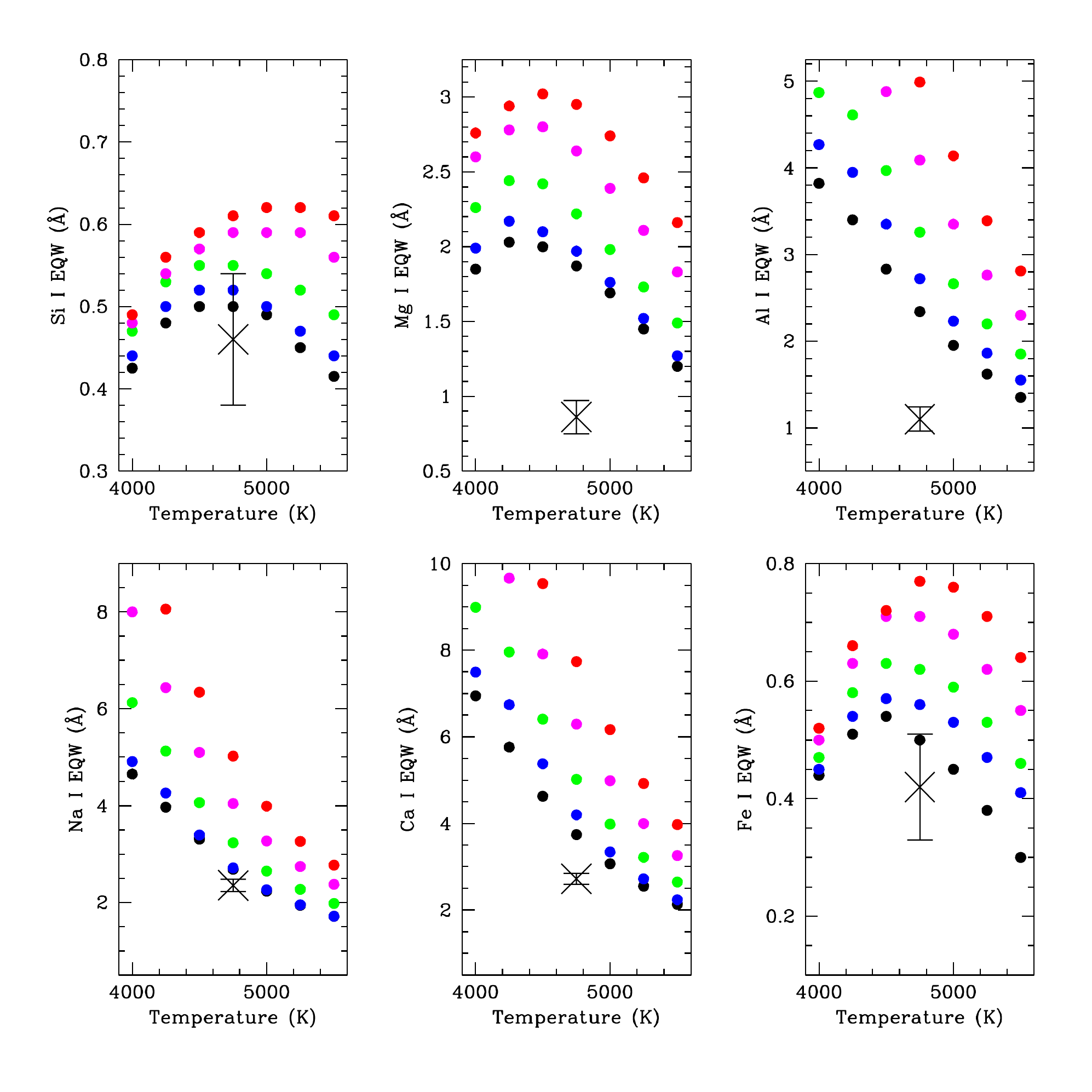}}}
\caption{The same as panel $a$, but for the $K$-band.}
\label{sseqwk}
\end{figure}

\renewcommand{\thefigure}{5abc}
\begin{figure}[htb]
\centerline{{\includegraphics[width=15cm]{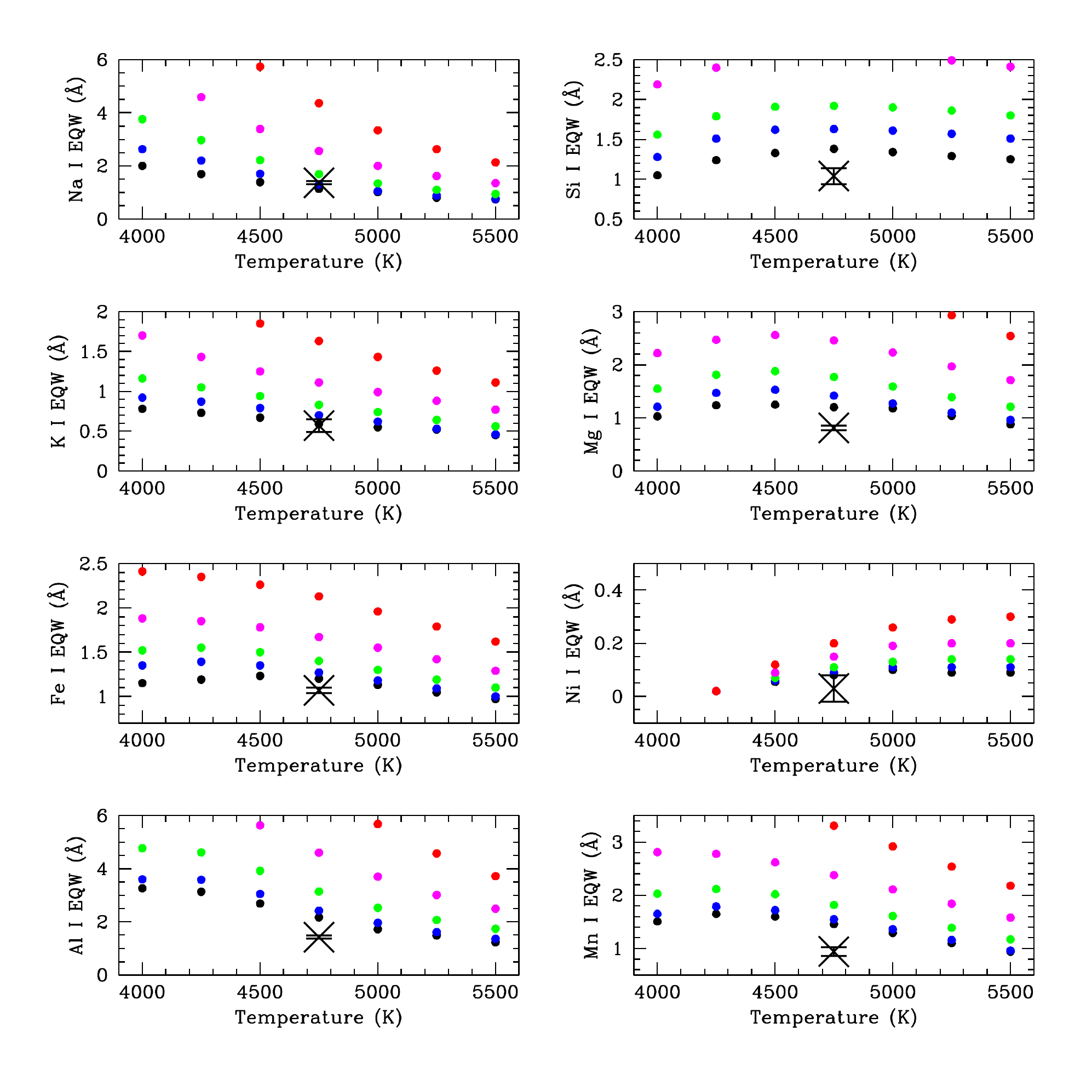}}}
\caption{As in Fig. \ref{sseqwsolar}, but for [Fe/H] = $-$0.3.}
\label{sseqwnonsolar}
\end{figure}

\renewcommand{\thefigure}{5b}
\begin{figure}[htb]
\centerline{{\includegraphics[width=15cm]{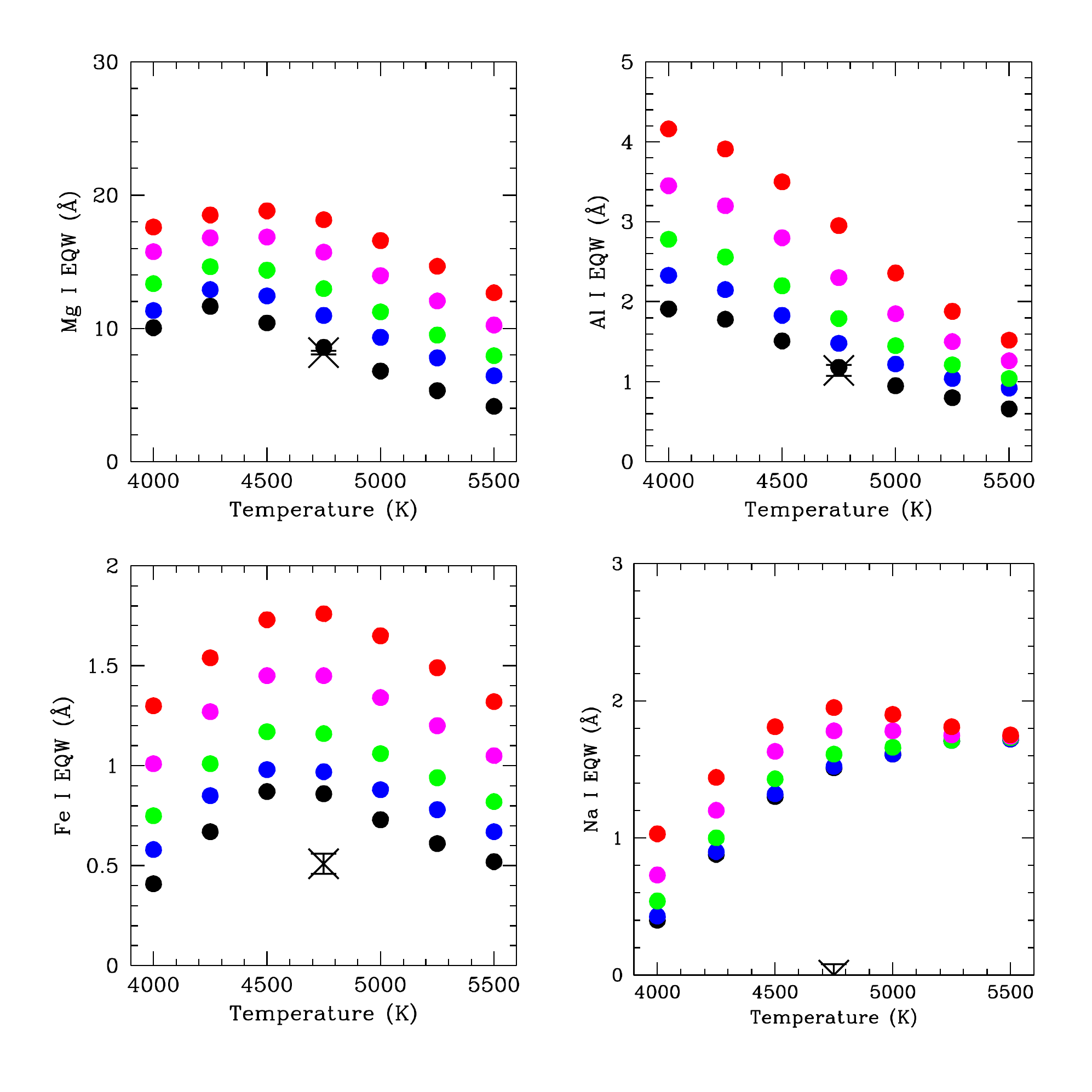}}}
\caption{Same as panel $a$, but for the $H$-band.}
\end{figure}

\renewcommand{\thefigure}{5c}
\begin{figure}[htb]
\centerline{{\includegraphics[width=15cm]{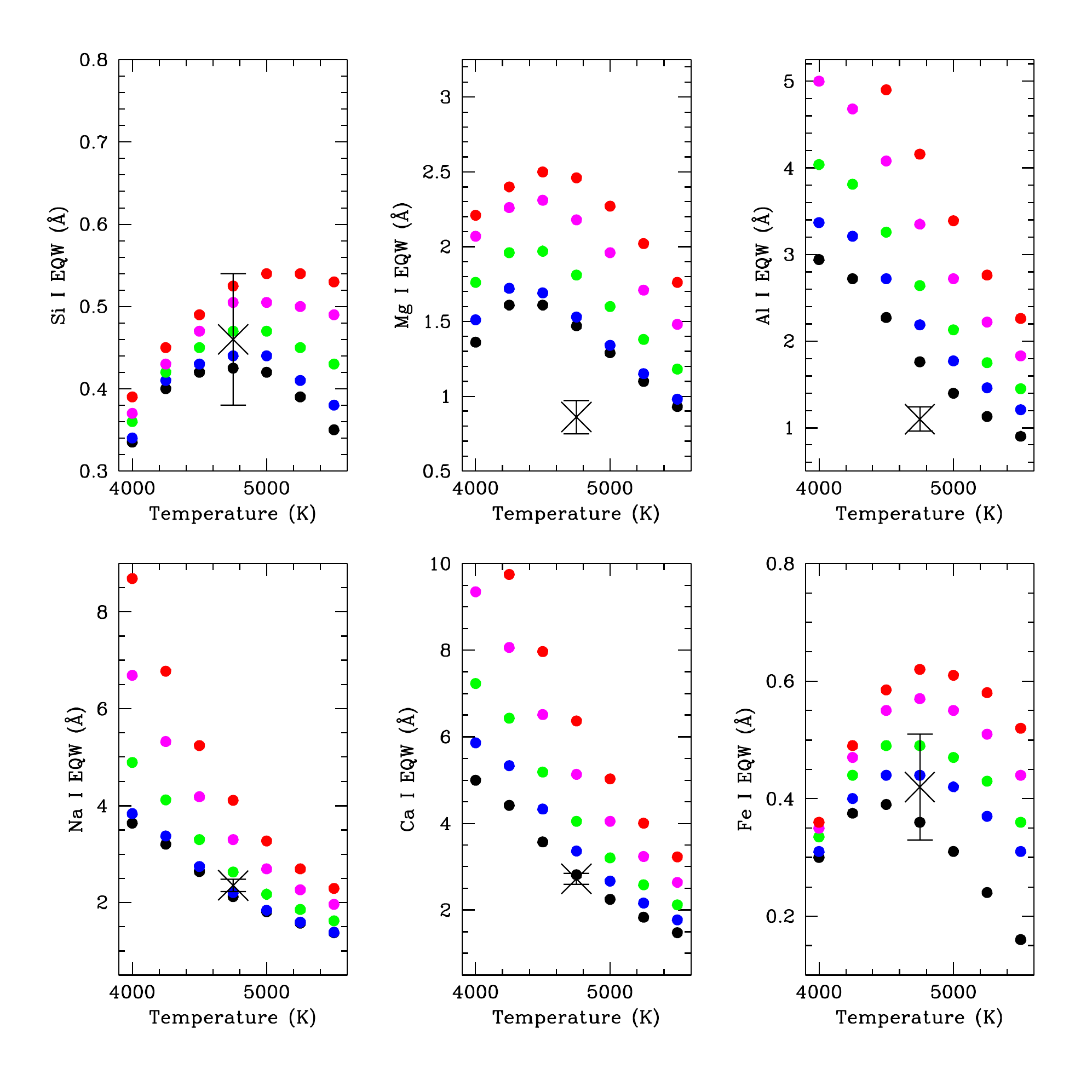}}}
\caption{The same as panel $a$, but for the $K$-band.}
\end{figure}

\renewcommand{\thefigure}{6}
\begin{figure}[htb]
\centerline{{\includegraphics[width=14cm]{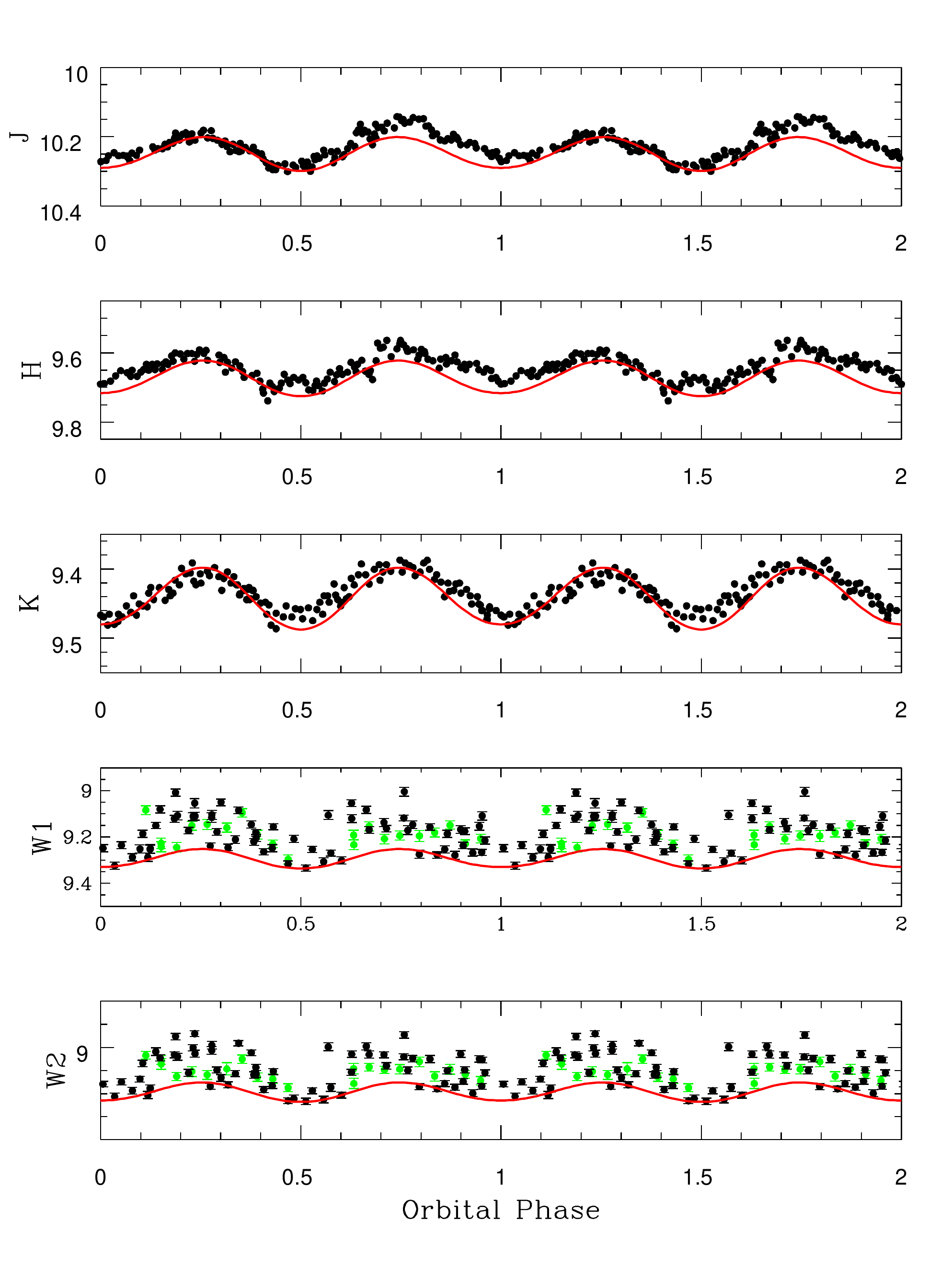}}}
\caption{The $JHK$ and $WISE$/$NEOWISE$ light curves of SS Cyg. In the
$W1$ (3.4 $\mu$m) and $W2$ (4.6 $\mu$m) panels, the $WISE$ light curve
data is plotted in green, and the $NEOWISE$ data is plotted in black. The
model light curve is in red and has T$_{\rm eff_{1}}$ = 20,000 K, 
T$_{\rm eff_{\rm 2}}$ = 4750 K, $i$ = 45$^{\circ}$, with a contamination level 
of 25\% from the white dwarf, accretion disk and its hot spot.}
\label{sscyglc}
\end{figure}

\renewcommand{\thefigure}{7abc}
\begin{figure}[htb]
\centerline{{\includegraphics[width=15cm]{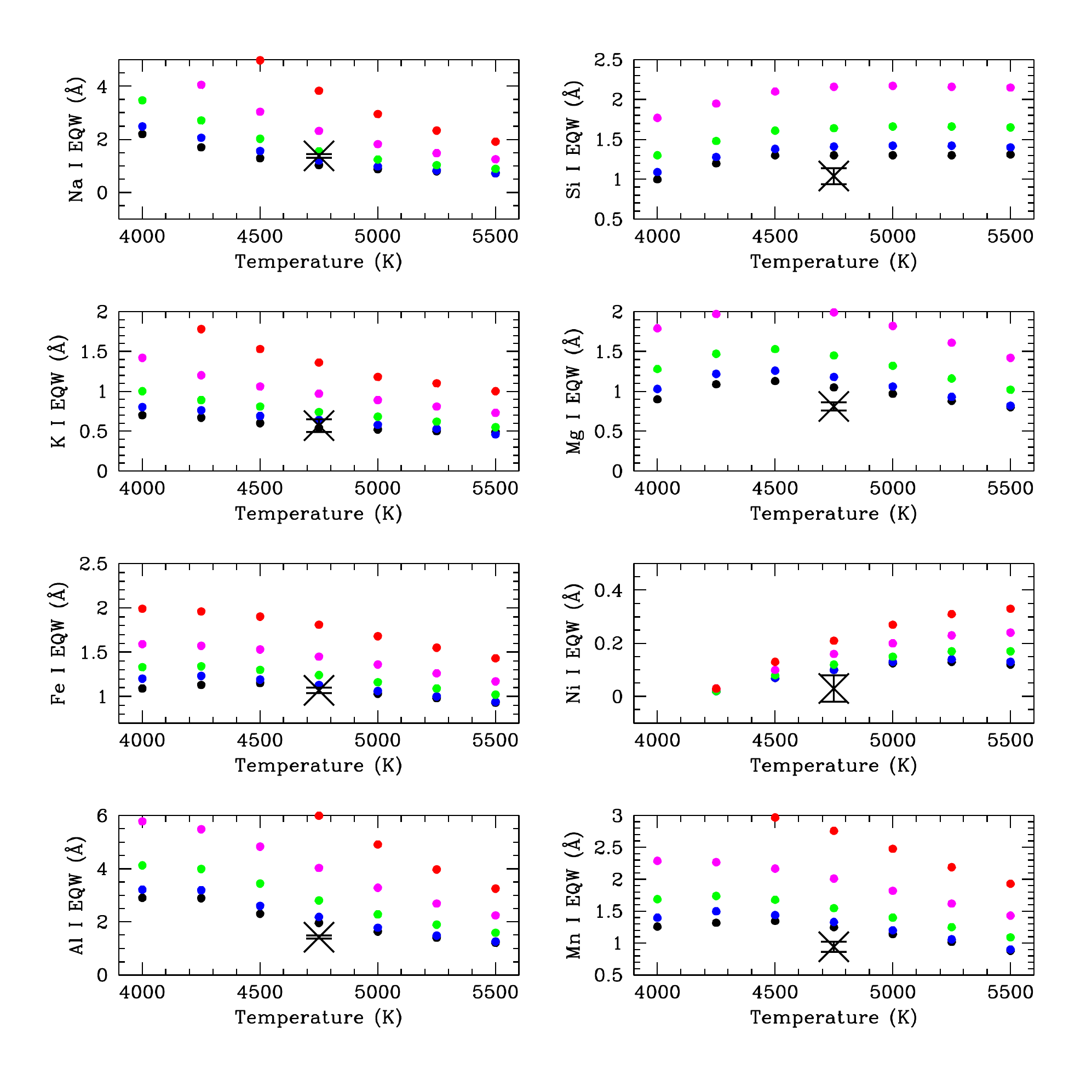}}}
\caption{The fit of the $J$-band EQW measurements for SS Cyg to those for the
model synthetic spectra with [Fe/H] = 0.0 and with the donor star supplying 
75\% of the broadband flux.}
\label{sscygcont}
\end{figure}

\renewcommand{\thefigure}{7b}
\begin{figure}[htb]
\centerline{{\includegraphics[width=15cm]{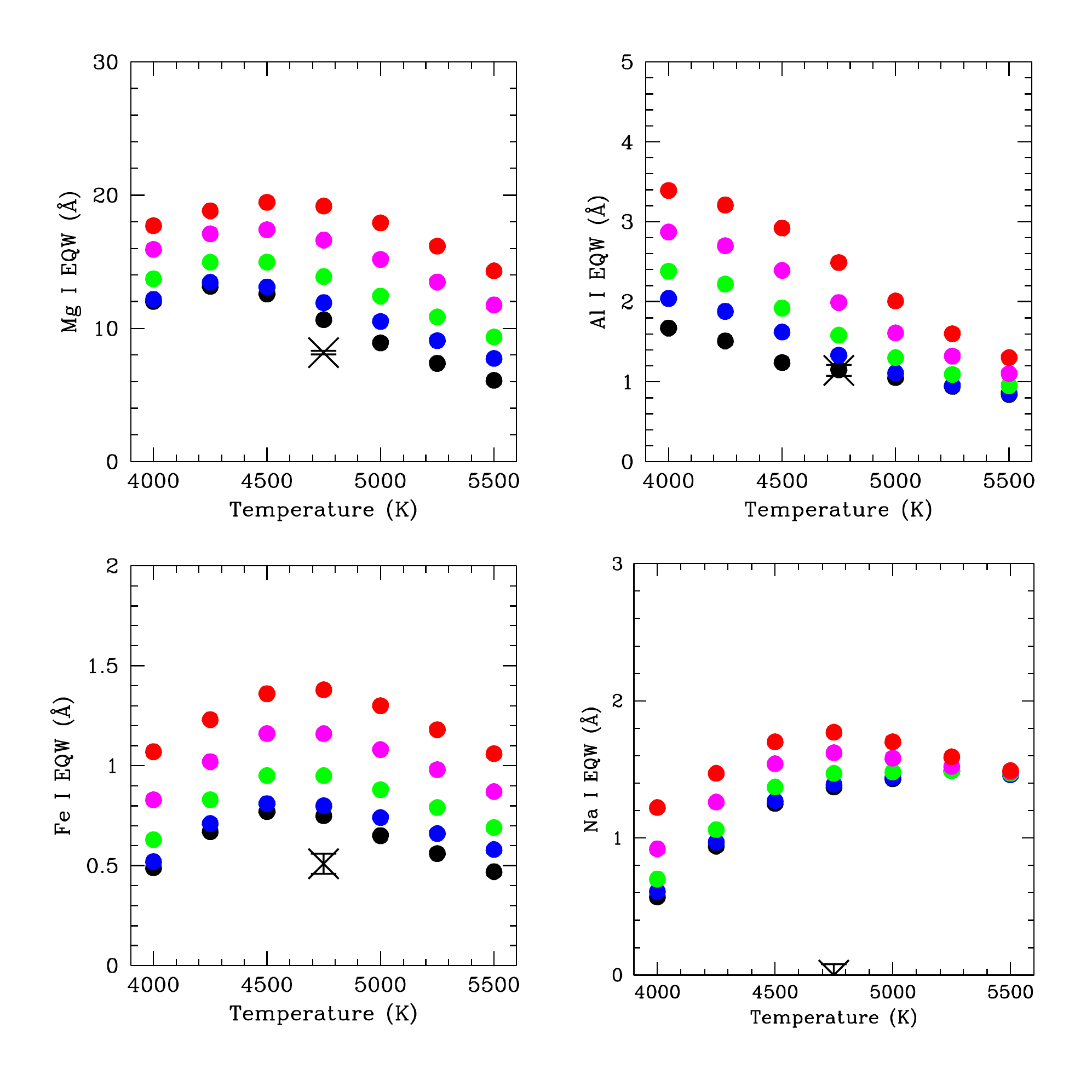}}}
\caption{The same as in panel $a$, but for the $H$-band.}
\end{figure}

\renewcommand{\thefigure}{7c}
\begin{figure}[htb]
\centerline{{\includegraphics[width=15cm]{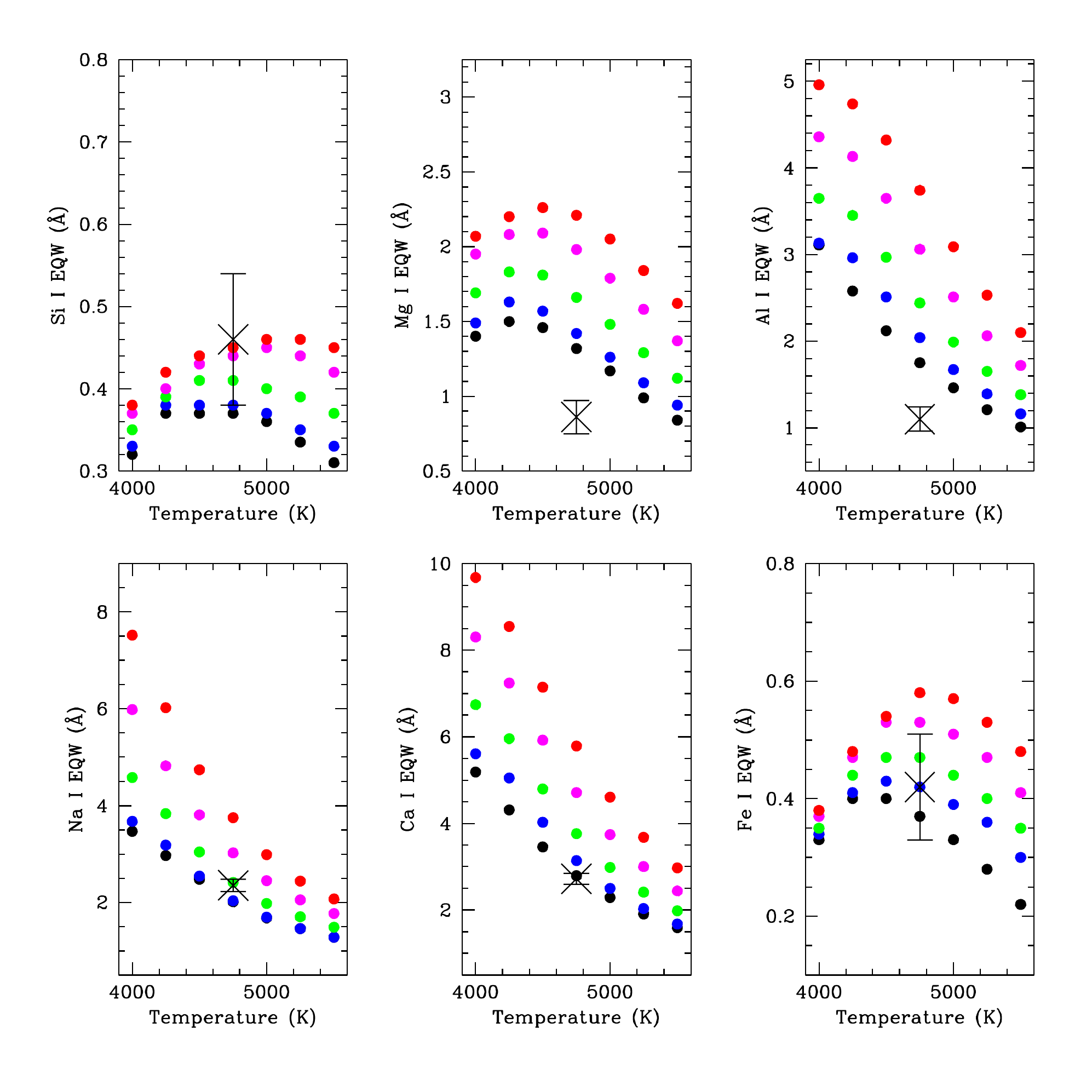}}}
\caption{The same as as in panel $a$, but for the $K$-band.}
\end{figure}

\renewcommand{\thefigure}{8}
\begin{figure}[htb]
\centerline{{\includegraphics[width=15cm]{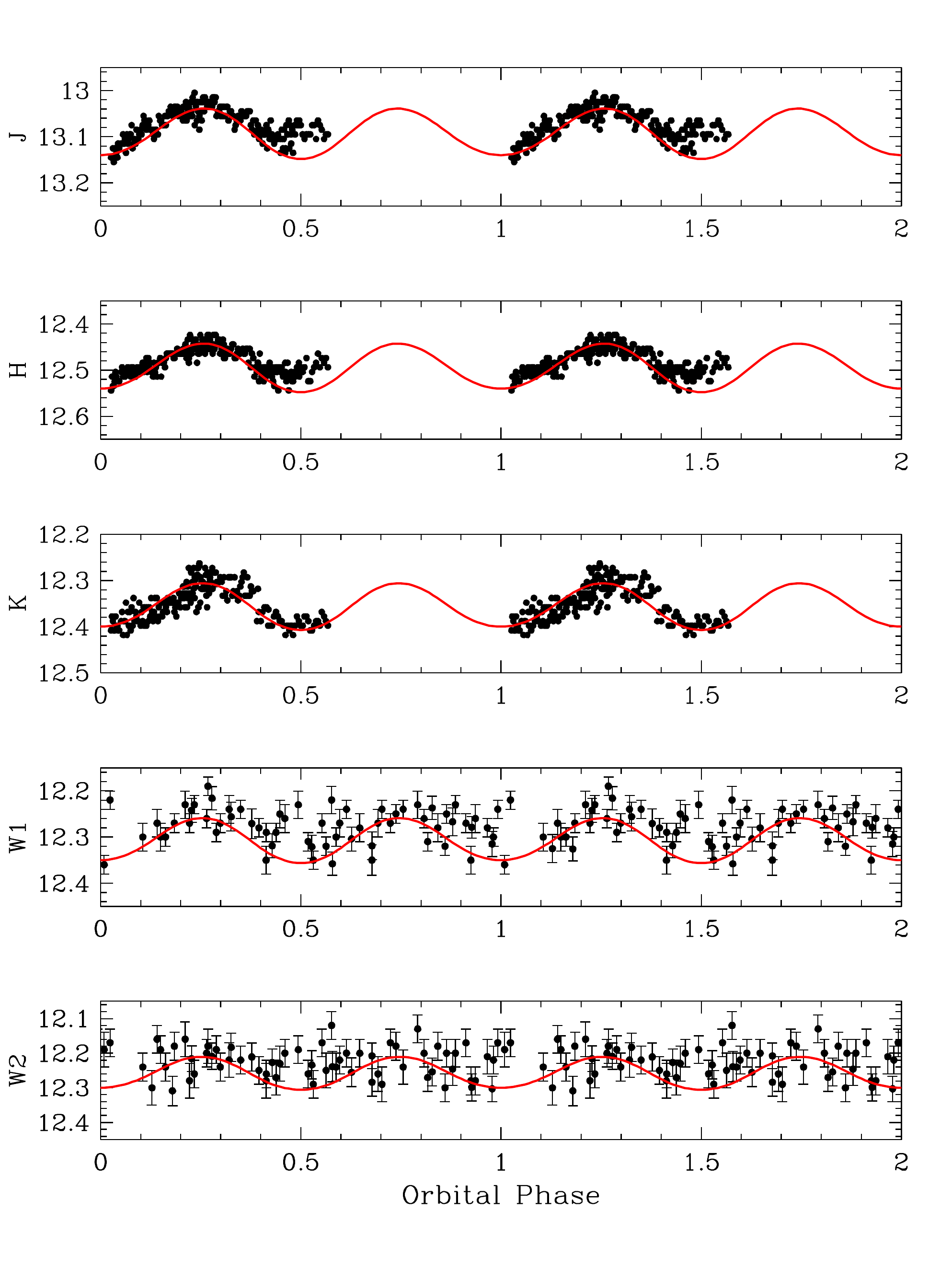}}}
\caption{The $JHK$ and $WISE$/$NEOWISE$ light curves of DX And. Using
the AAVSO data base, DX And was in a quiescent state for all epochs of
observation. The light curve model (red) has T$_{\rm eff_{\rm 1}}$ = 25,000 K,
T$_{\rm eff_{\rm 2}}$ = 5,000 K, $q$ = 0.96, and $i$ = 45$^{\circ}$.}
\label{dxandlc}
\end{figure}

\renewcommand{\thefigure}{9abc}
\begin{figure}[htb]
\centerline{{\includegraphics[width=15cm]{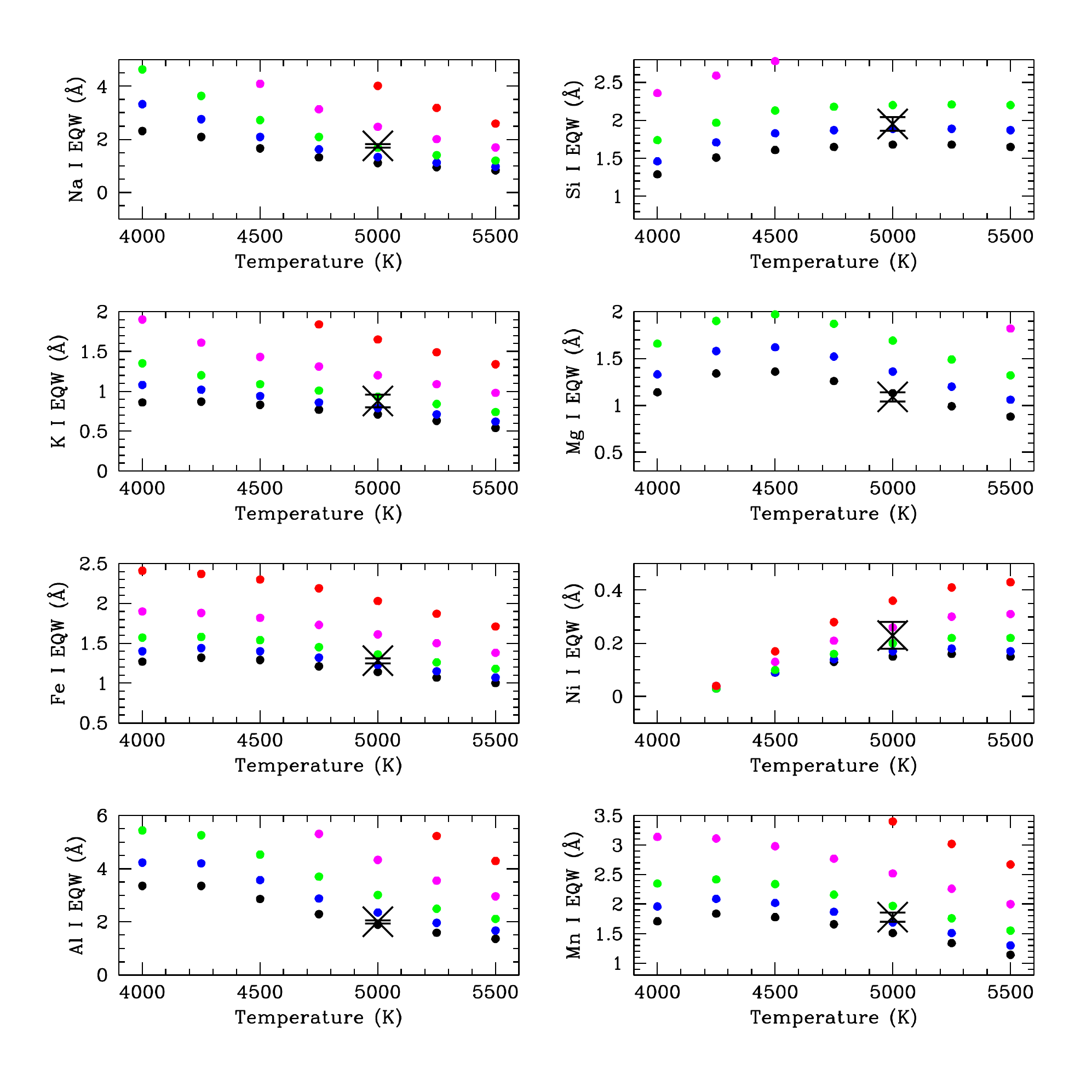}}}
\caption{The fit of the $J$-band EQW measurements for DX And to those for the
synthetic spectra with log$g$ = 4.5, and [Fe/H] = 0.0.}
\label{dxandeqw}
\end{figure}

\renewcommand{\thefigure}{9b}
\begin{figure}[htb]
\centerline{{\includegraphics[width=15cm]{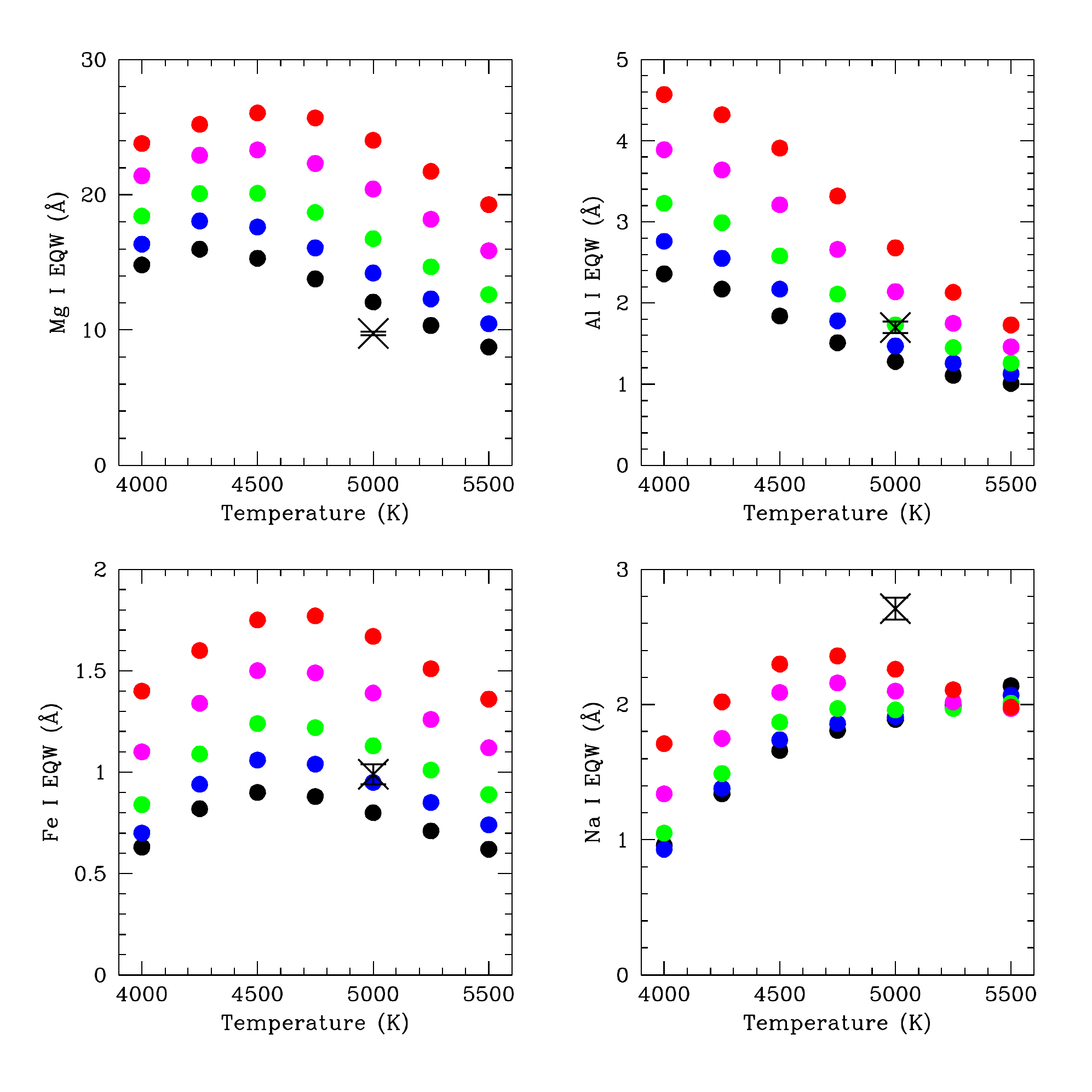}}}
\caption{The same as panel $a$, but for the $H$-band.}
\end{figure}

\renewcommand{\thefigure}{9c}
\begin{figure}[htb]
\centerline{{\includegraphics[width=15cm]{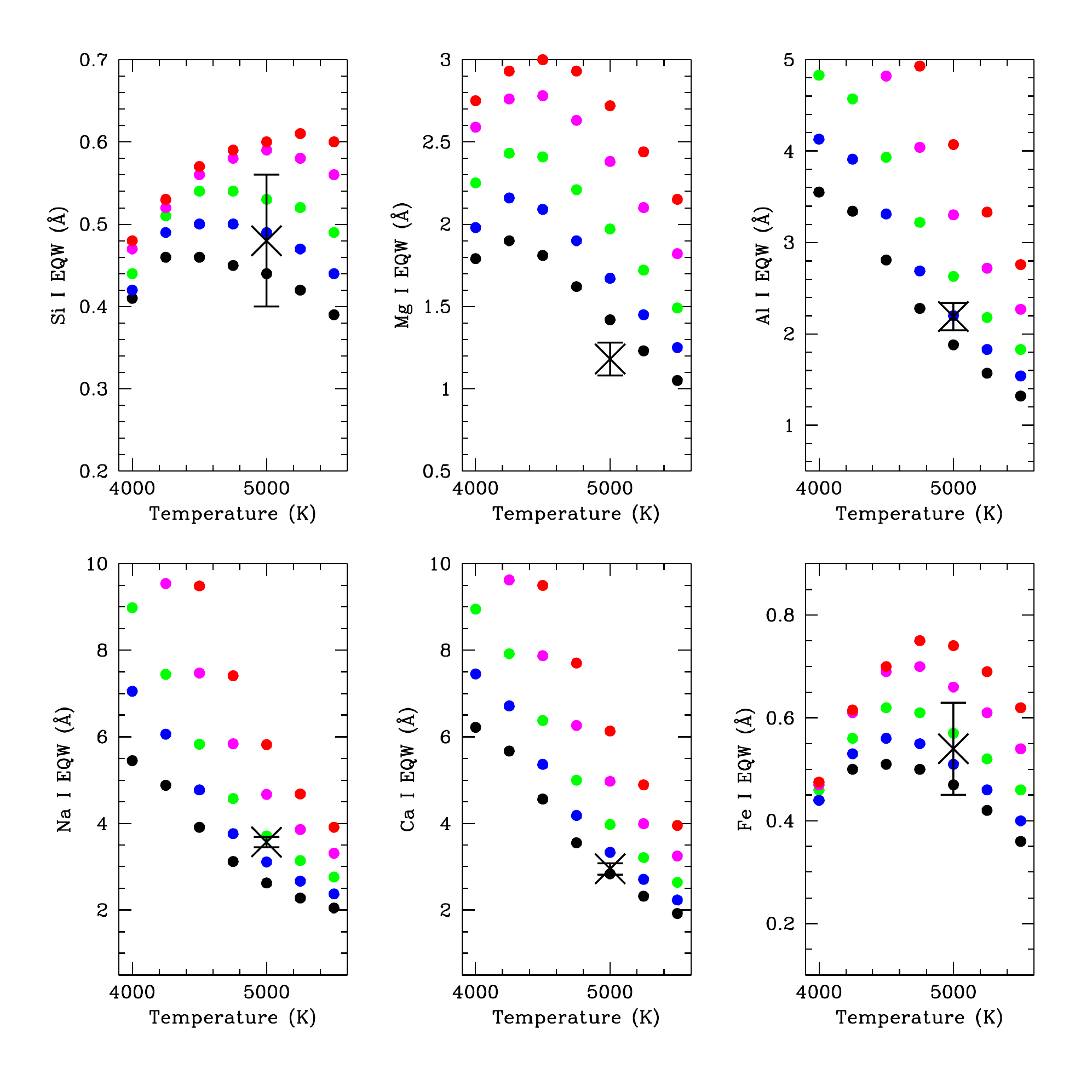}}}
\caption{The same as panel $a$, but for the $K$-band.}
\end{figure}

\renewcommand{\thefigure}{10abc}
\begin{figure}[htb]
\centerline{{\includegraphics[width=15cm]{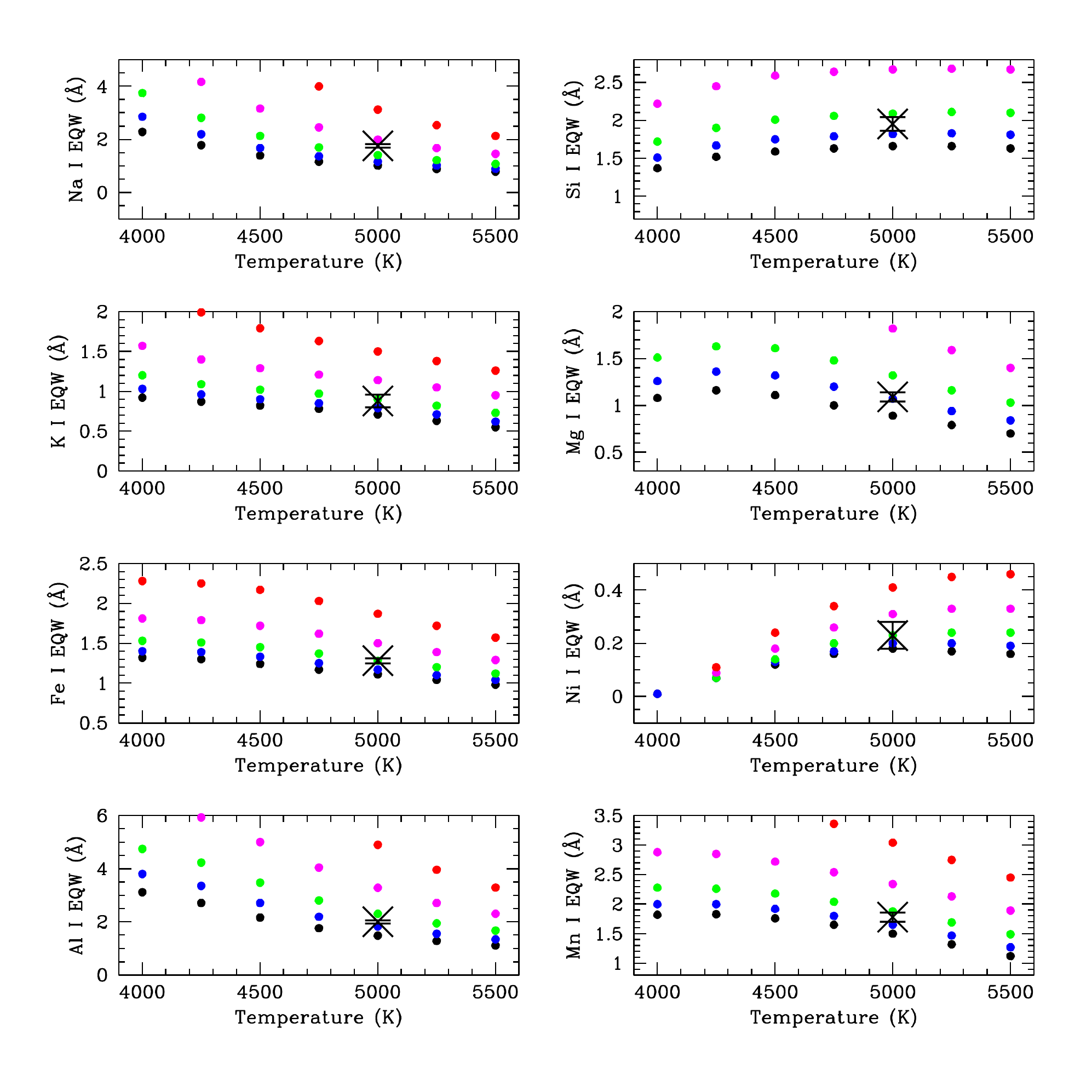}}}
\caption{The fit of the $J$-band EQW measurements for DX And to those for the
synthetic spectra with log$g$ = 4.0, and [Fe/H] = 0.0.}
\label{dxand}
\end{figure}

\renewcommand{\thefigure}{10b}
\begin{figure}[htb]
\centerline{{\includegraphics[width=15cm]{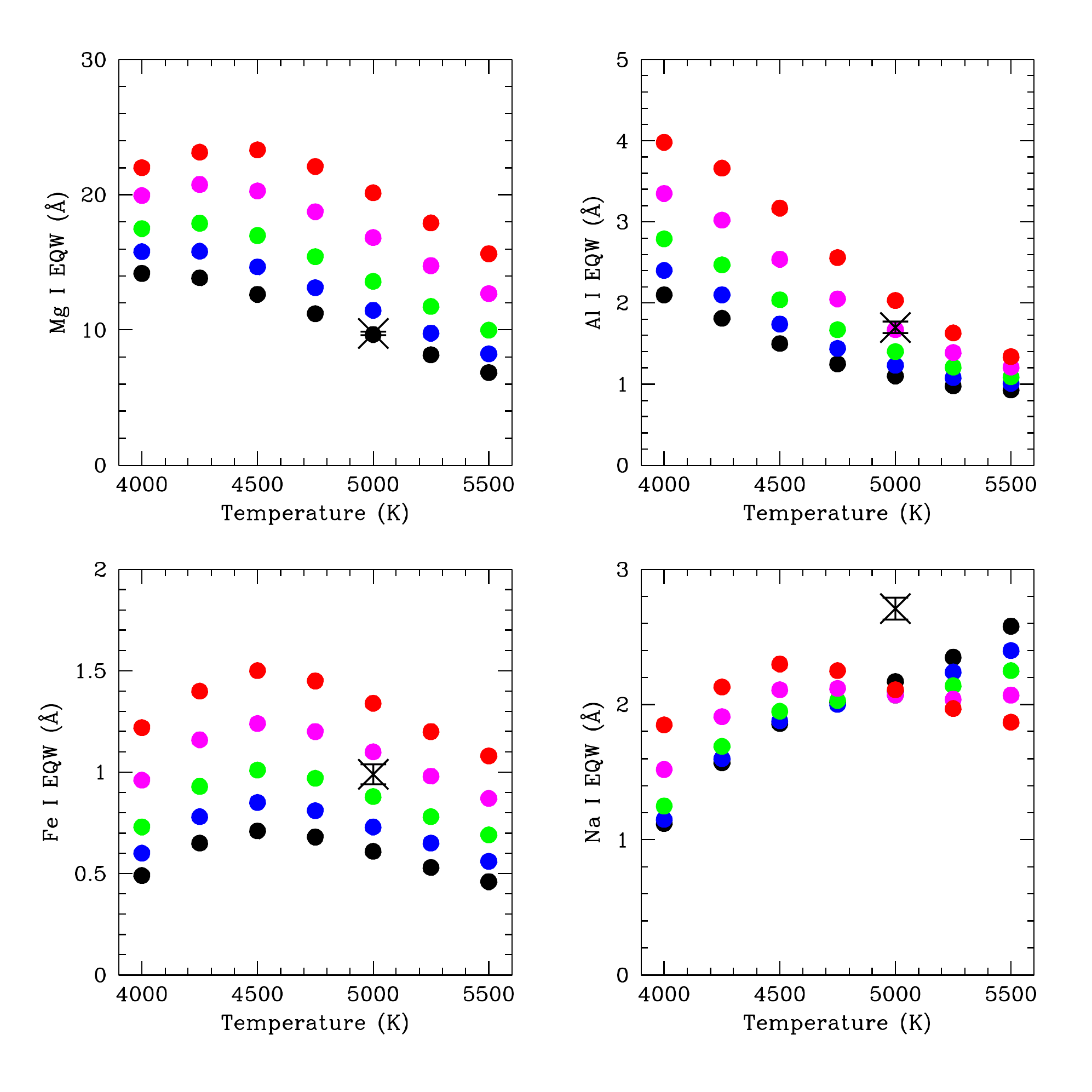}}}
\caption{The same as panel $a$, but for the $H$-band.}
\end{figure}

\renewcommand{\thefigure}{10c}
\begin{figure}[htb]
\centerline{{\includegraphics[width=15cm]{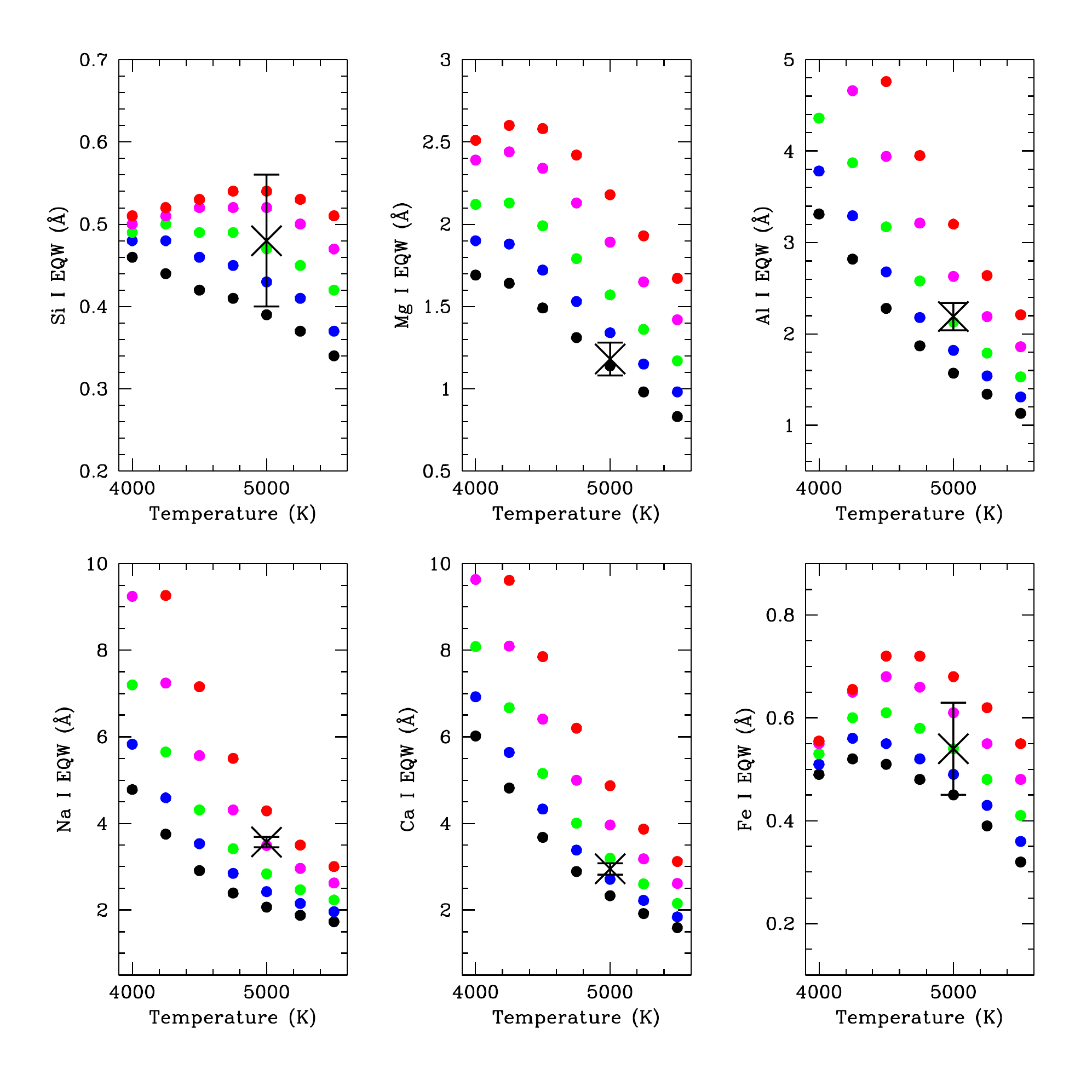}}}
\caption{The same as panel $a$, but for the $K$-band.}
\end{figure}

\renewcommand{\thefigure}{11}
\begin{figure}[htb]
\centerline{{\includegraphics[width=15cm]{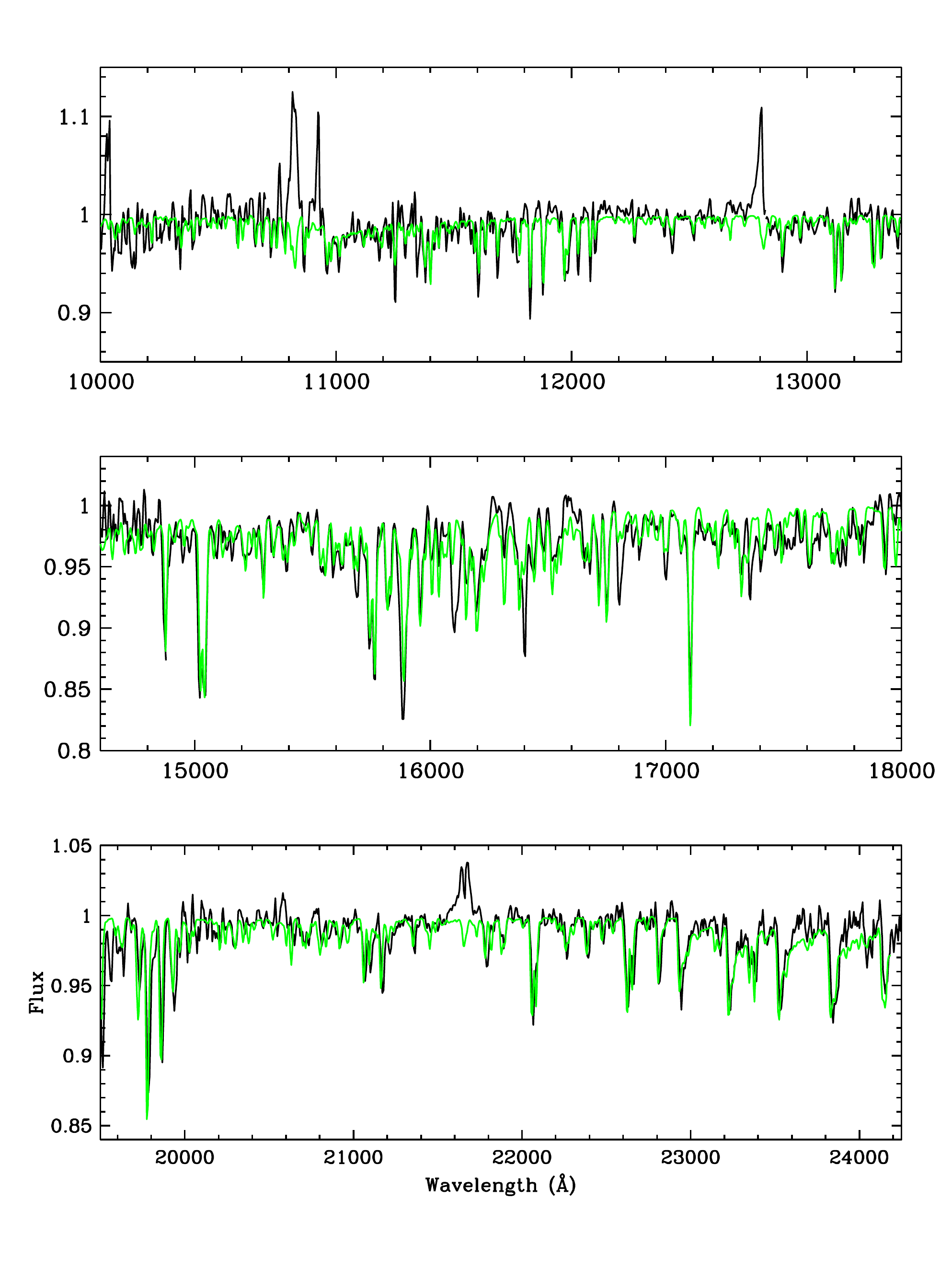}}}
\caption{The Gemini GNIRS spectrum (black) of DX And compared to a 
synthetic spectrum (green) with T$_{\rm eff}$ = 5000 K, log$g$ = 4.0, [Fe/H] = 
0.0, Hd = 50\%, [C/Fe] = $-$0.7, [Mg/Fe] = $-$0.2, and [Na/Fe] = $+$0.5.}
\label{dxspec}
\end{figure}

\renewcommand{\thefigure}{12abc}
\begin{figure}[htb]
\centerline{{\includegraphics[width=15cm]{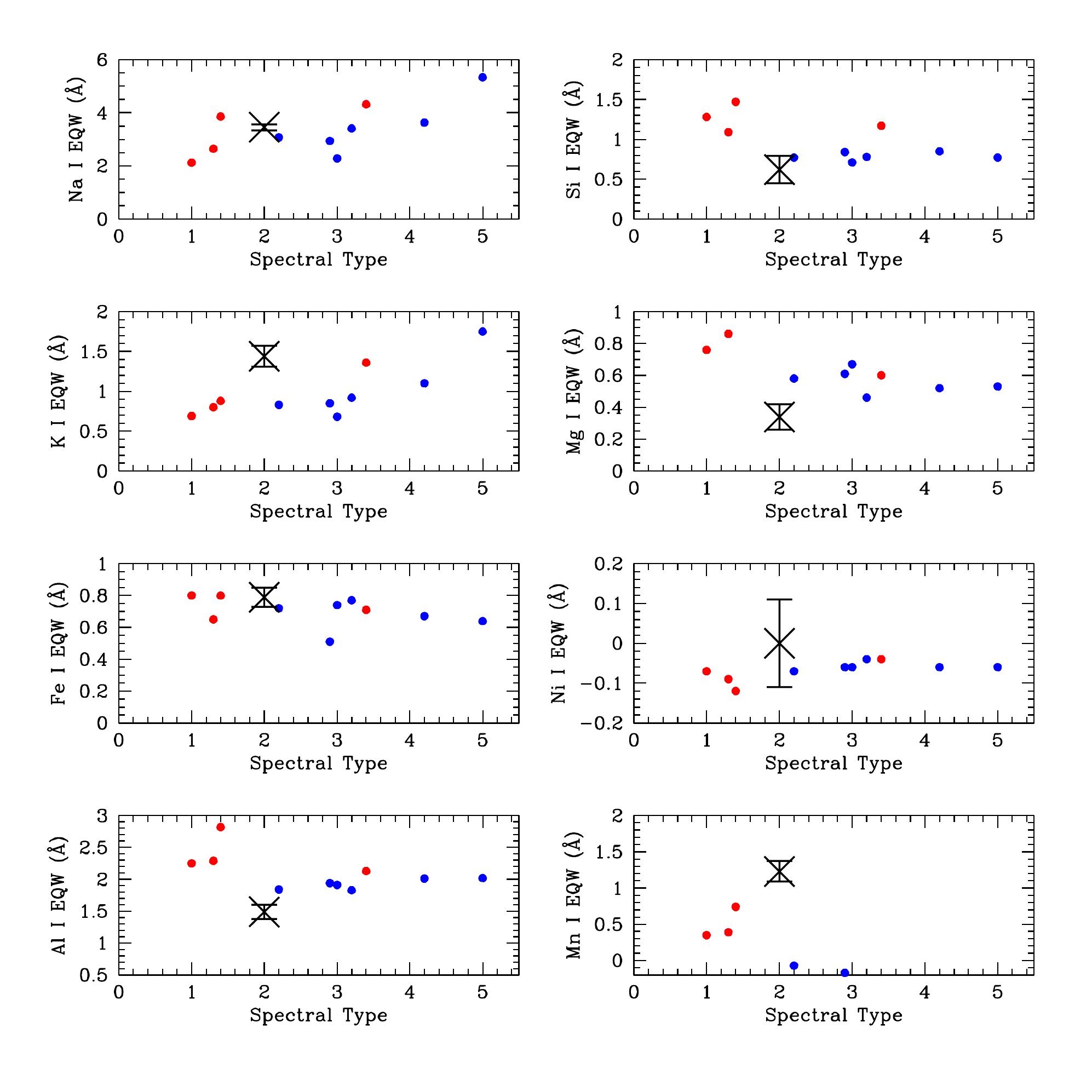}}}
\caption{The plot of the EQW measurements in the $J$-band of RX And versus 
those for the M dwarfs in the IRTF Spectral Library (Cushing et al. 2005) for 
the same set of elements used above. The abundances and spectral types of the 
M dwarfs were generated using the techniques discussed in H16. The x-axis is 
spectral type, ranging from M0V to M5V. The data plotted in blue are the EQW 
measures for M dwarfs with [Fe/H] $<$ 0, while those in red are for [Fe/H] 
$>$ 0.}
\label{rxmdwf}
\end{figure}

\renewcommand{\thefigure}{12b}
\begin{figure}[htb]
\centerline{{\includegraphics[width=15cm]{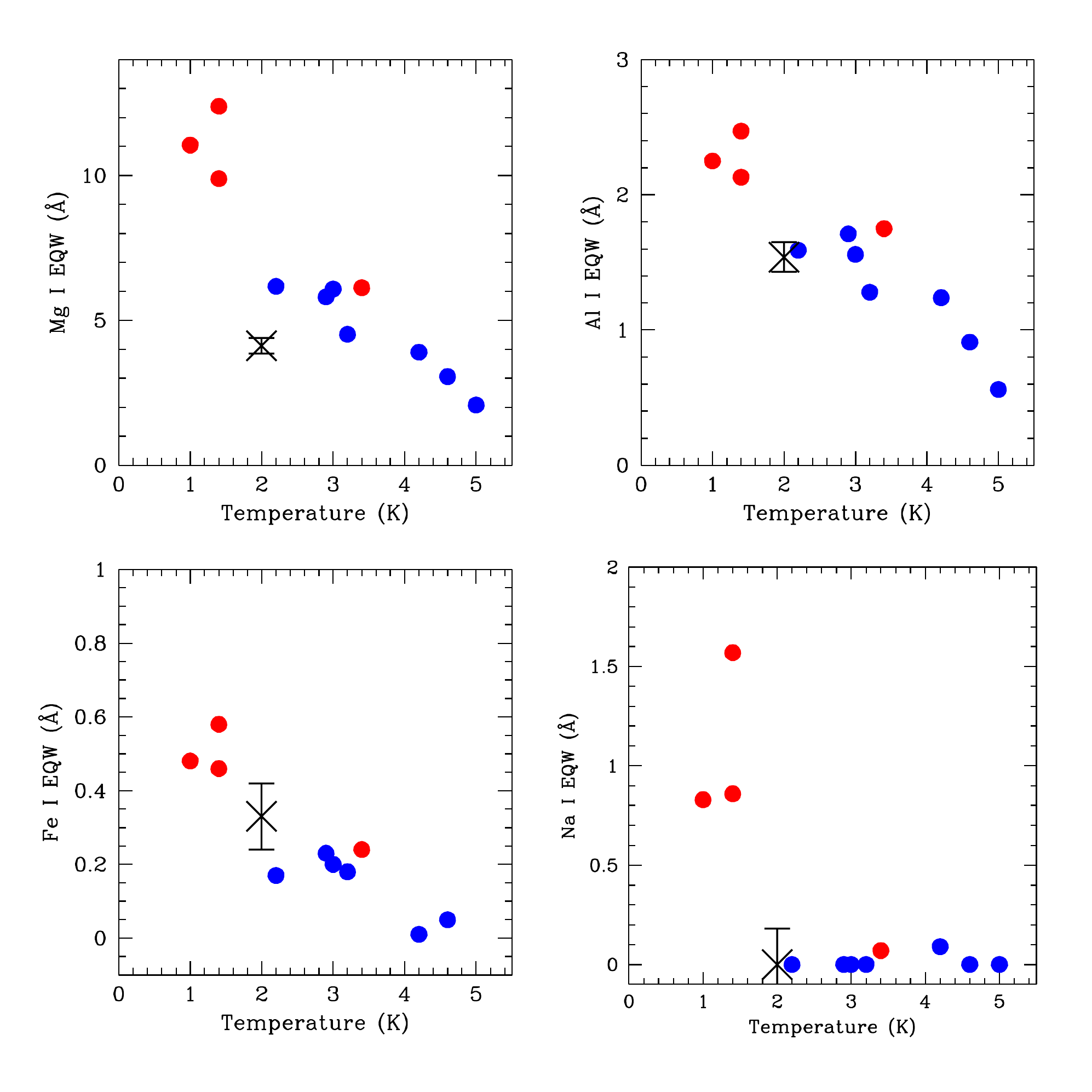}}}
\caption{The same as panel $a$, but for the $H$-band.}
\end{figure}

\renewcommand{\thefigure}{11c}
\begin{figure}[htb]
\centerline{{\includegraphics[width=15cm]{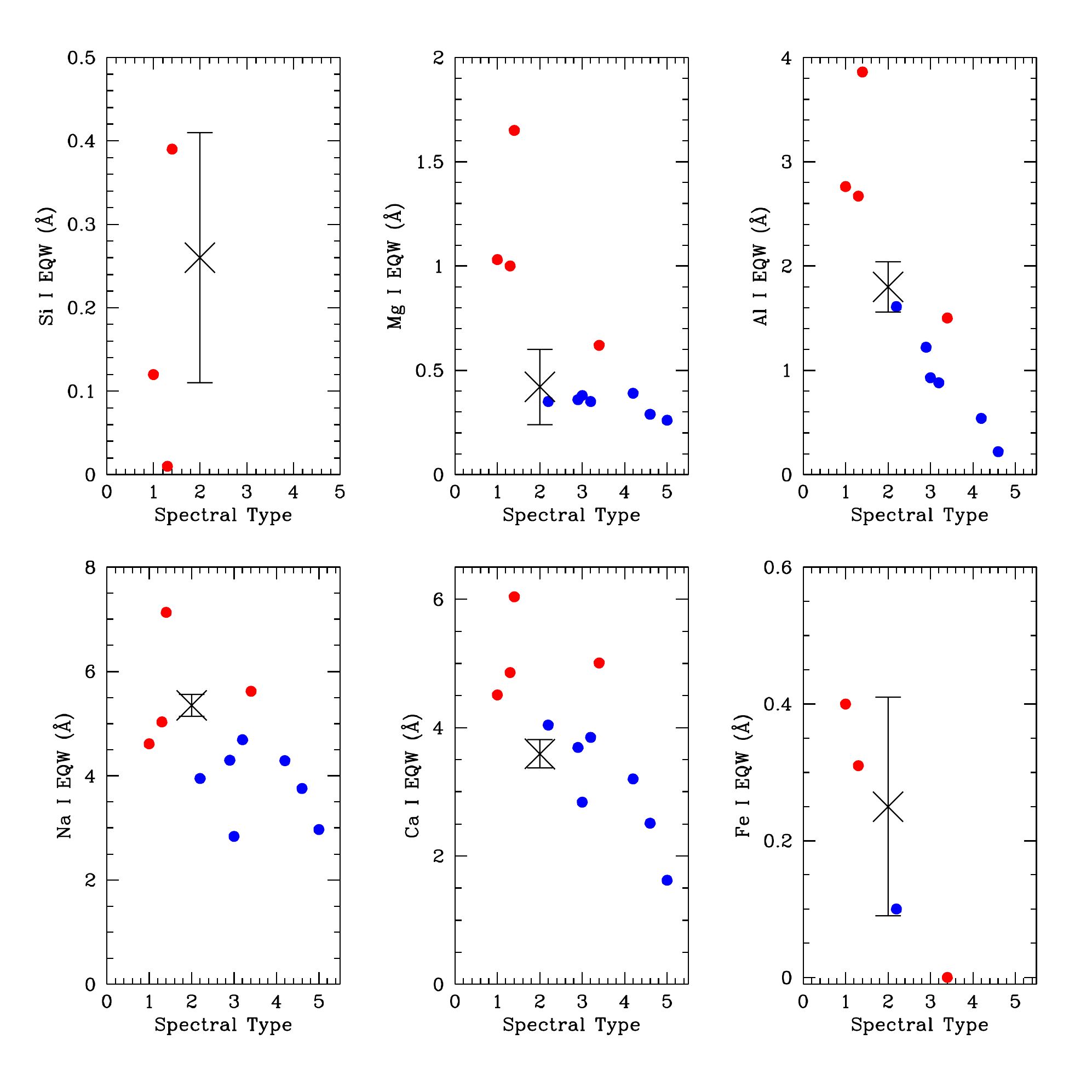}}}
\caption{The same as panel $a$, but for the $K$-band.}
\end{figure}

\renewcommand{\thefigure}{13abc}
\begin{figure}[htb]
\centerline{{\includegraphics[width=15cm]{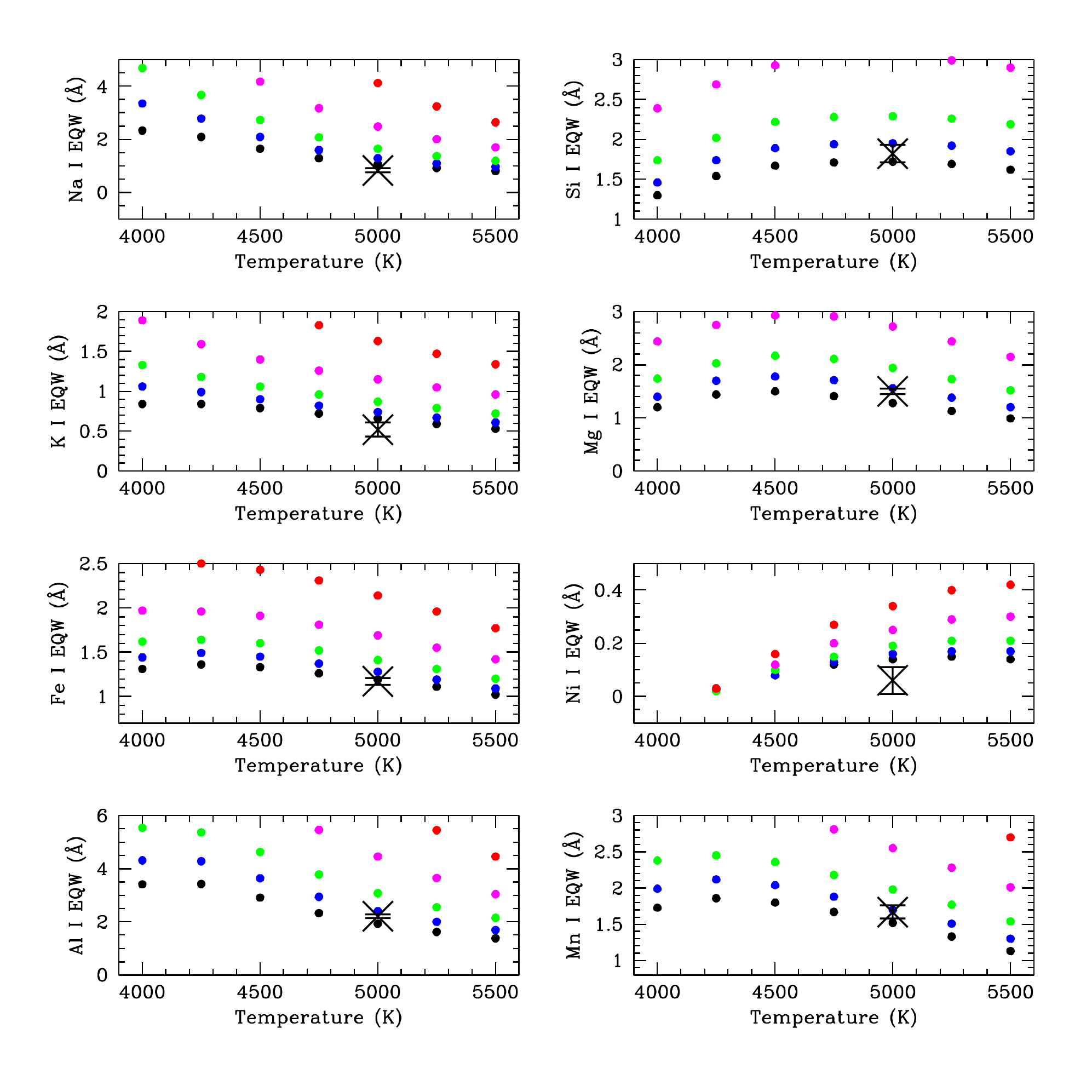}}}
\caption{The fit of the $J$-band EQW measurements for AE Aqr to those for the
model synthetic spectra with log$g$ = 4.5, and [Fe/H] = 0.0.}
\label{aeaqrEQWdw}
\end{figure}

\renewcommand{\thefigure}{13b}
\begin{figure}[htb]
\centerline{{\includegraphics[width=15cm]{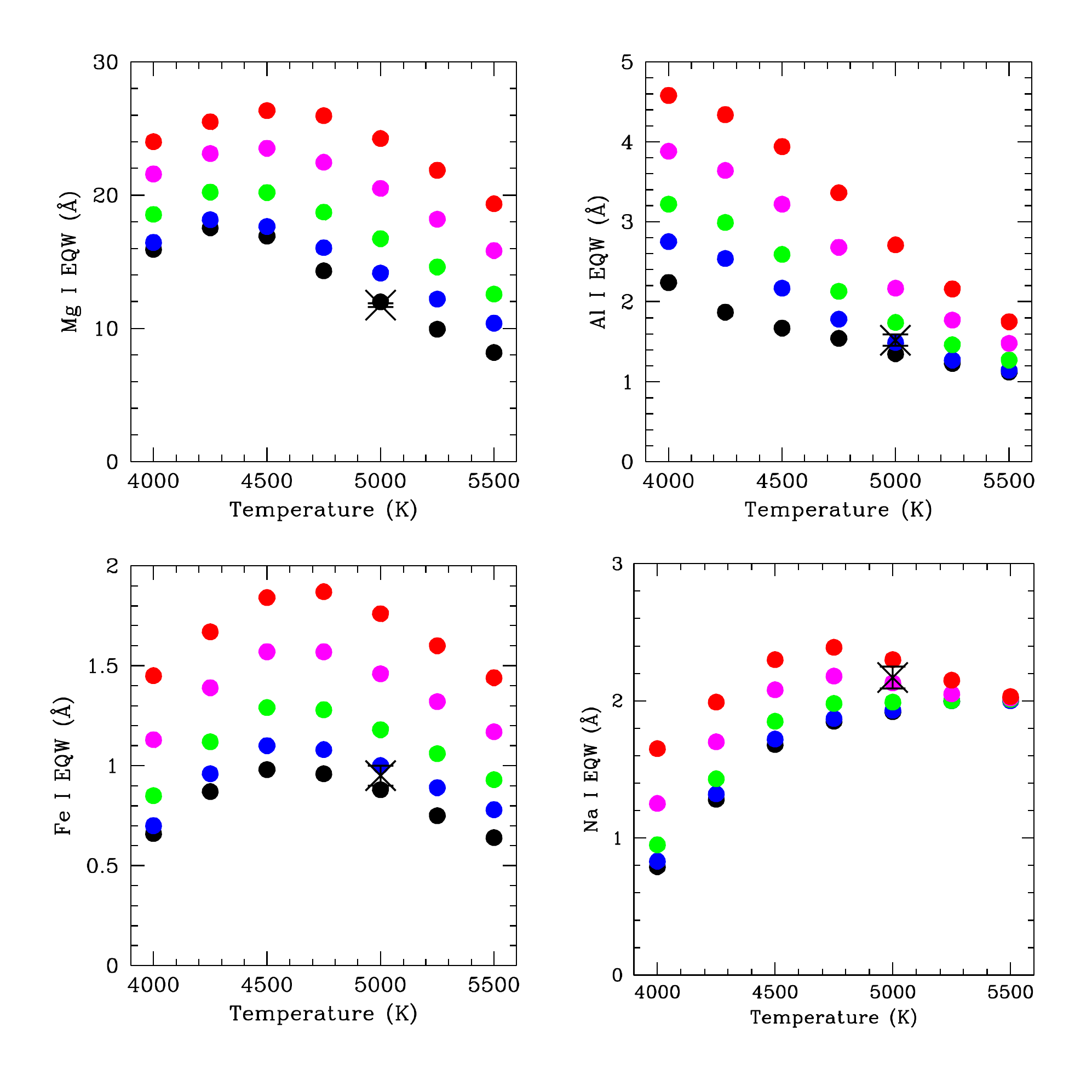}}}
\caption{The same as panel $a$, but for the $H$-band.}
\end{figure}

\renewcommand{\thefigure}{13c}
\begin{figure}[htb]
\centerline{{\includegraphics[width=15cm]{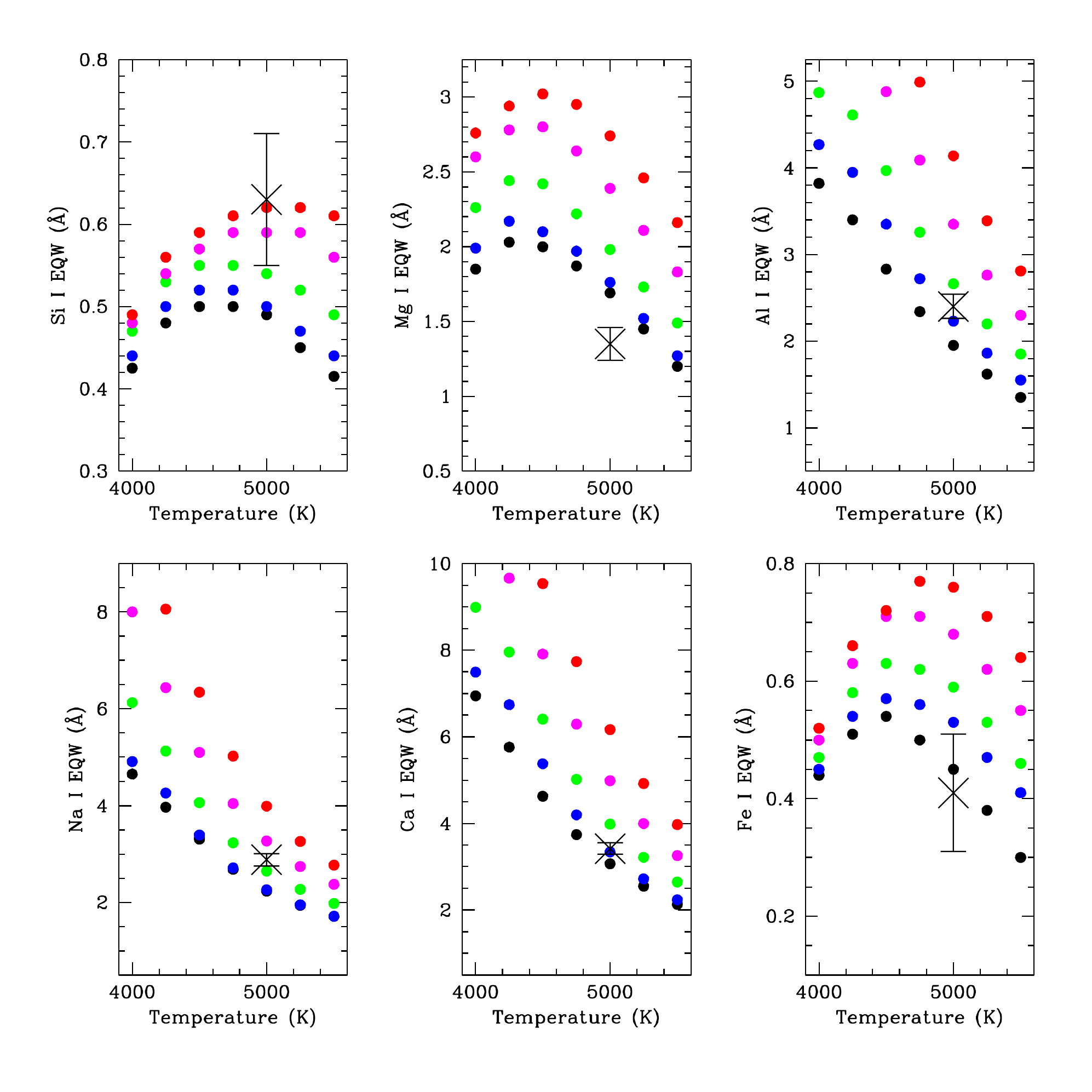}}}
\caption{The same as panel $a$, but for the $K$-band.}
\end{figure}

\renewcommand{\thefigure}{14abc}
\begin{figure}[htb]
\centerline{{\includegraphics[width=15cm]{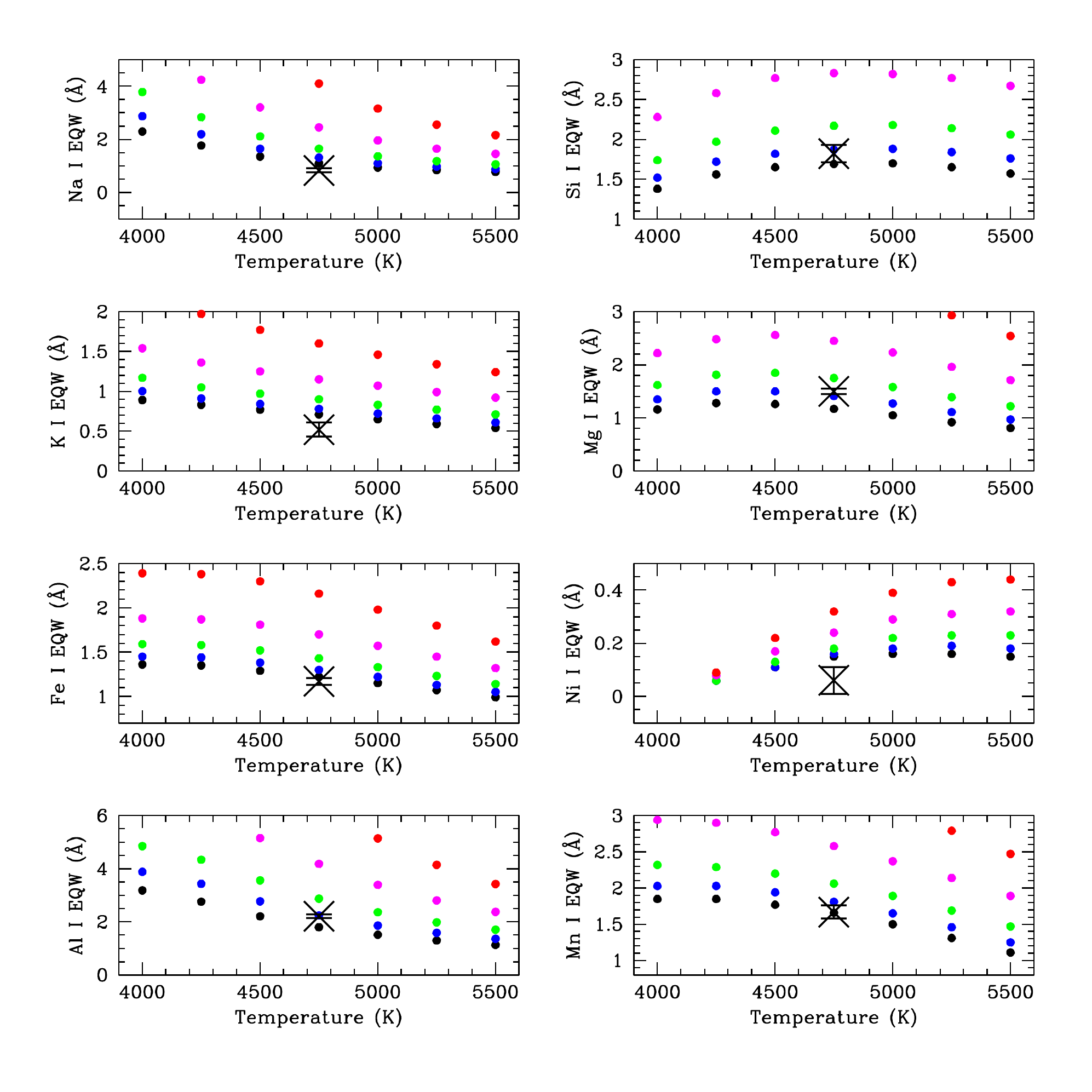}}}
\caption{The fit of the $J$-band EQW measurements for AE Aqr to those for the 
synthetic spectra with log$g$ = 4.0, and [Fe/H] = 0.0.}
\label{aeaqrEQWsg}
\end{figure}

\renewcommand{\thefigure}{14b}
\begin{figure}[htb]
\centerline{{\includegraphics[width=15cm]{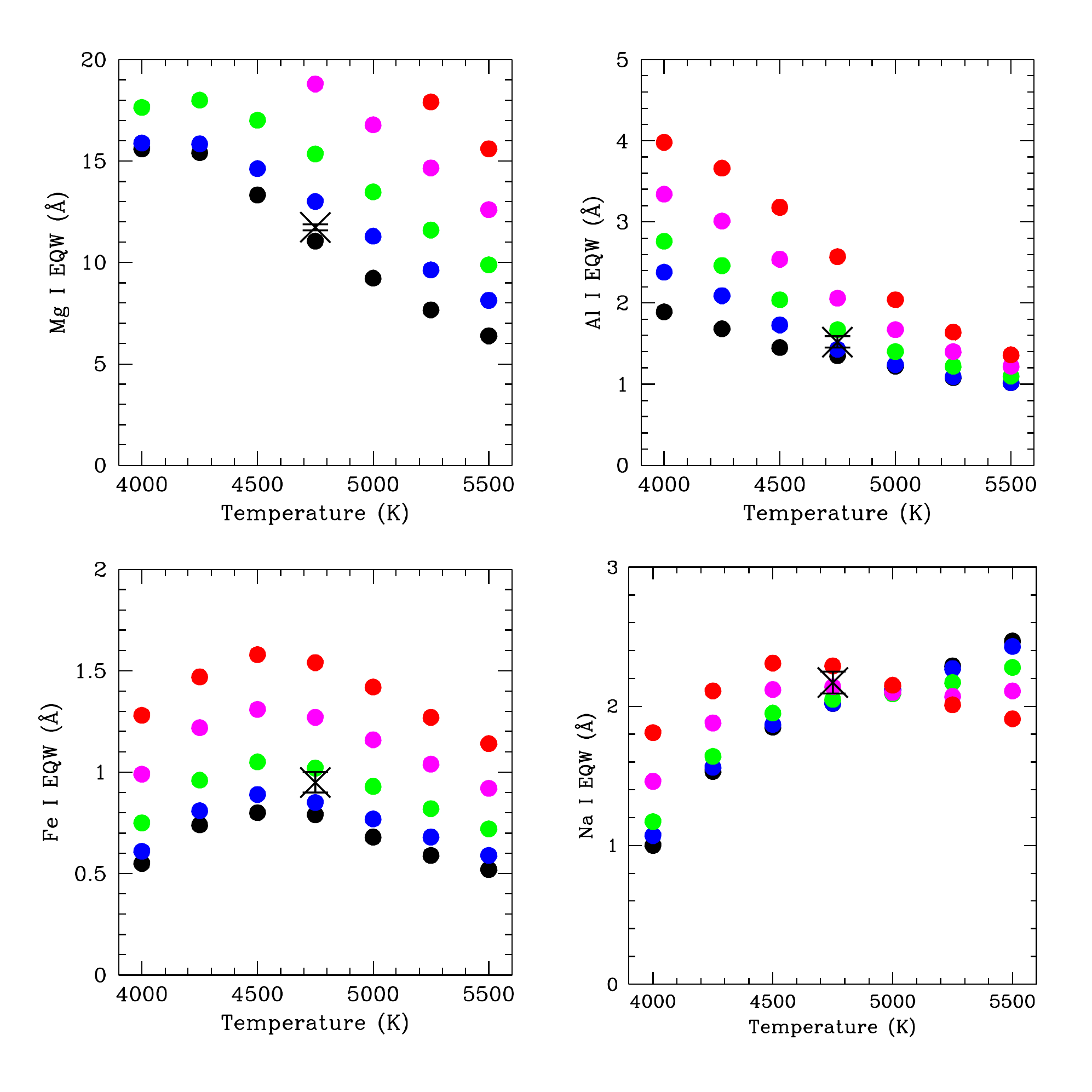}}}
\caption{The same as panel $a$, but for the $H$-band.}
\end{figure}

\renewcommand{\thefigure}{14c}
\begin{figure}[htb]
\centerline{{\includegraphics[width=15cm]{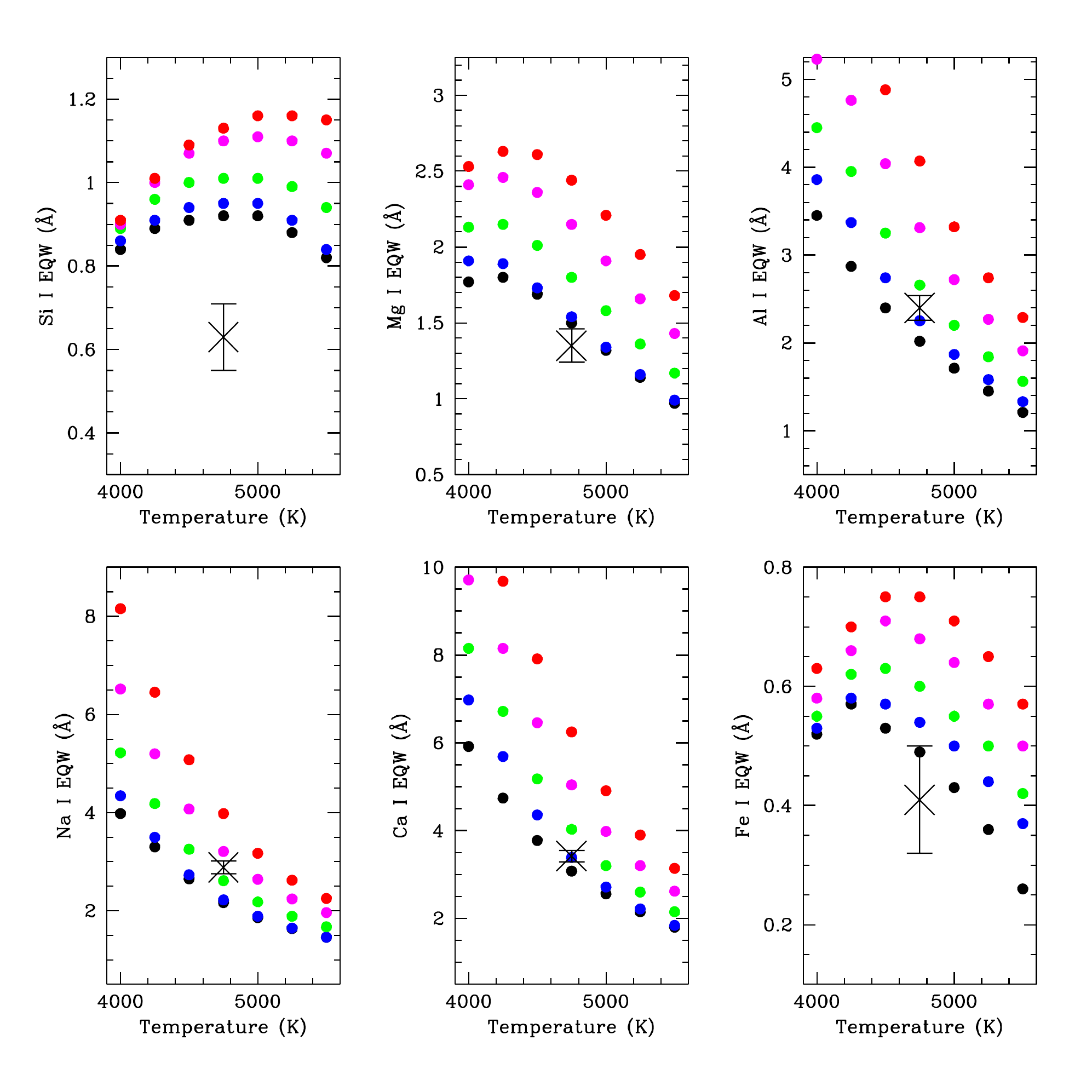}}}
\caption{The same as panel $a$, but for the $K$-band.}
\end{figure}

\renewcommand{\thefigure}{15}
\begin{figure}[htb]
\centerline{{\includegraphics[width=15cm]{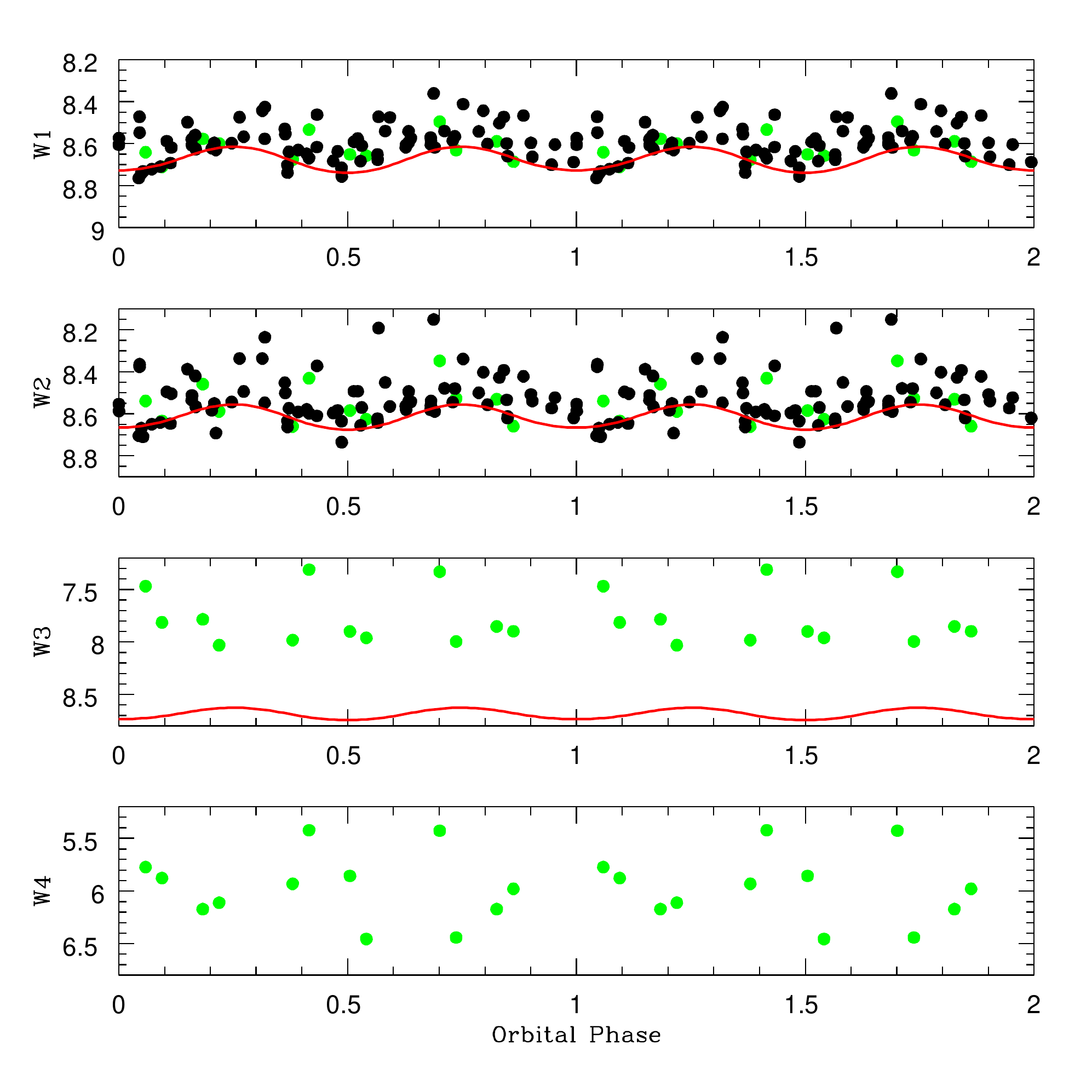}}}
\caption{The $WISE$ (green) and $NEOWISE$ (black) light curves of AE Aqr.}
\label{aeaqrlc}
\end{figure}

\renewcommand{\thefigure}{16}
\begin{figure}[htb]
\centerline{{\includegraphics[width=15cm]{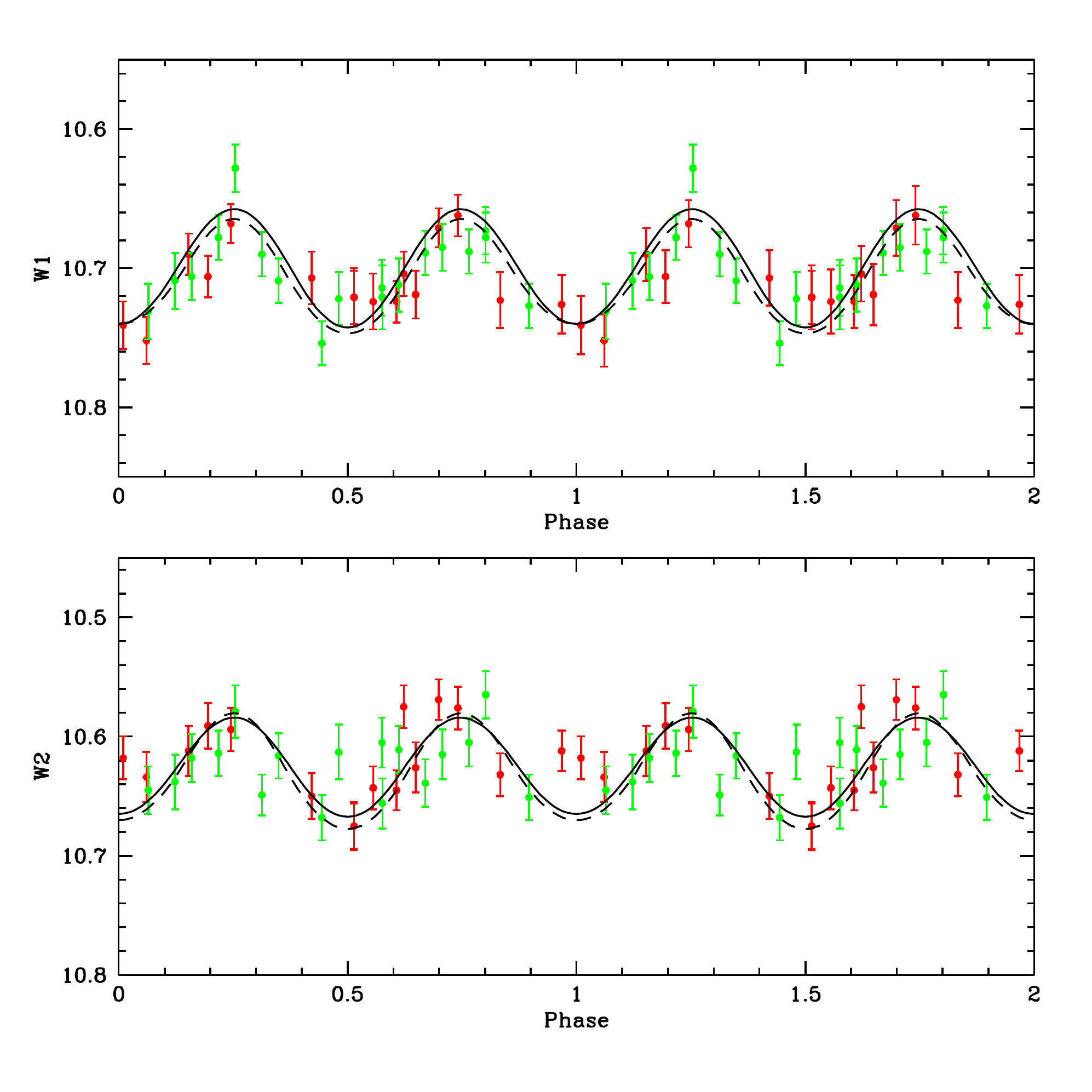}}}
\caption{The $NEOWISE$ light curves of Z Cam. The data for the 2014 October
10 epoch are plotted in red, and those for 2015 October 10 epoch are plotted 
in green. The photometry for 2014 has been offset by $\Delta$ = $+$0.17
mag so as to agree with the 2015 data. The solid black line is a light curve 
model with $i$ = 40$^{\circ}$, while the dashed line has $i$ = 57$^{\circ}$, 
and a contamination level of 32\% in the $W1$ band, and 28\% in the $W2$ band.}
\label{zcamlc}
\end{figure}
\clearpage

\renewcommand{\thefigure}{17abc}
\begin{figure}[htb]
\centerline{{\includegraphics[width=15cm]{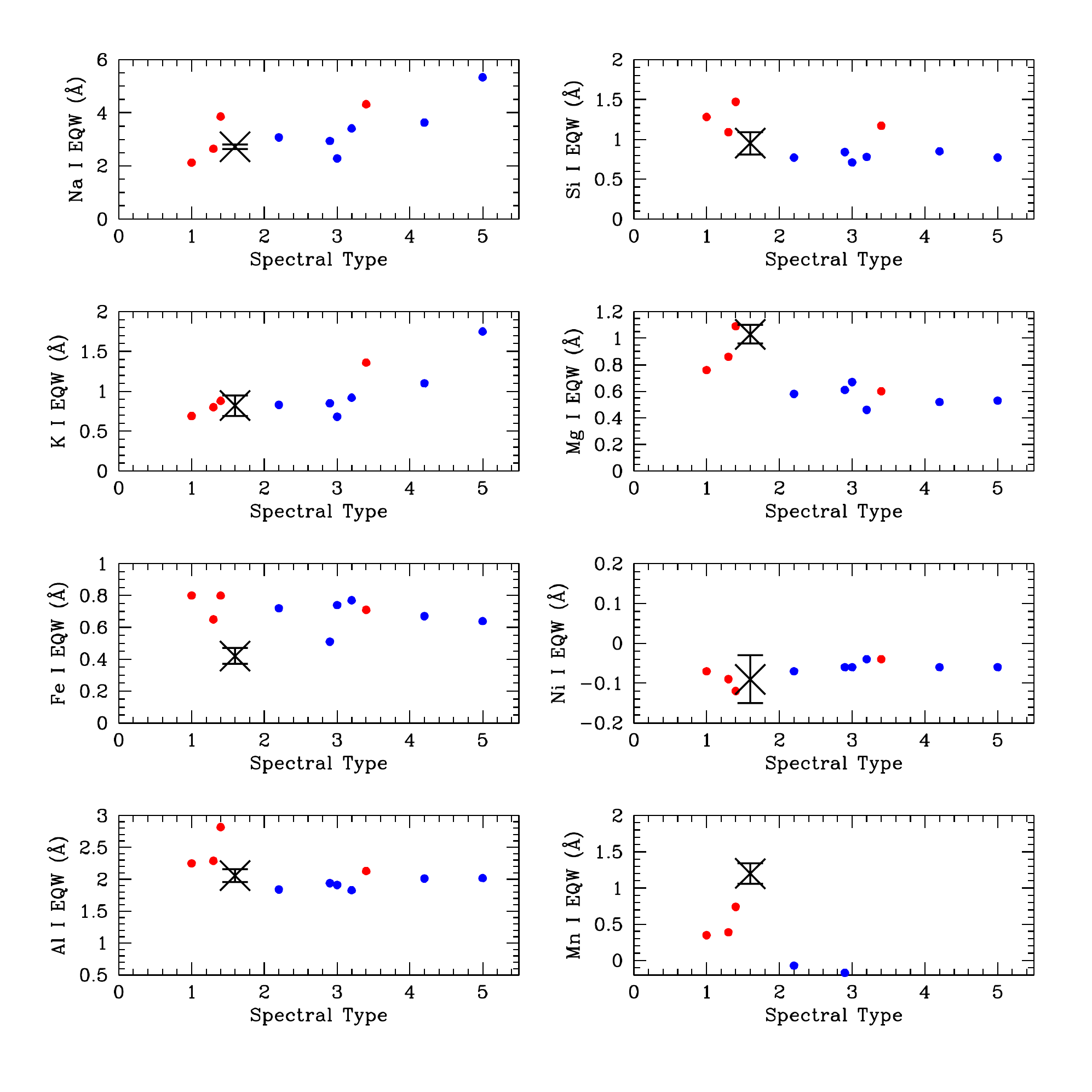}}}
\caption{As in Fig. \ref{rxmdwf}, we compare the EQW measurements of Z Cam
in the $J$-band to those of M dwarfs.}
\label{zcammdwarfs}
\end{figure}

\renewcommand{\thefigure}{17b}
\begin{figure}[htb]
\centerline{{\includegraphics[width=15cm]{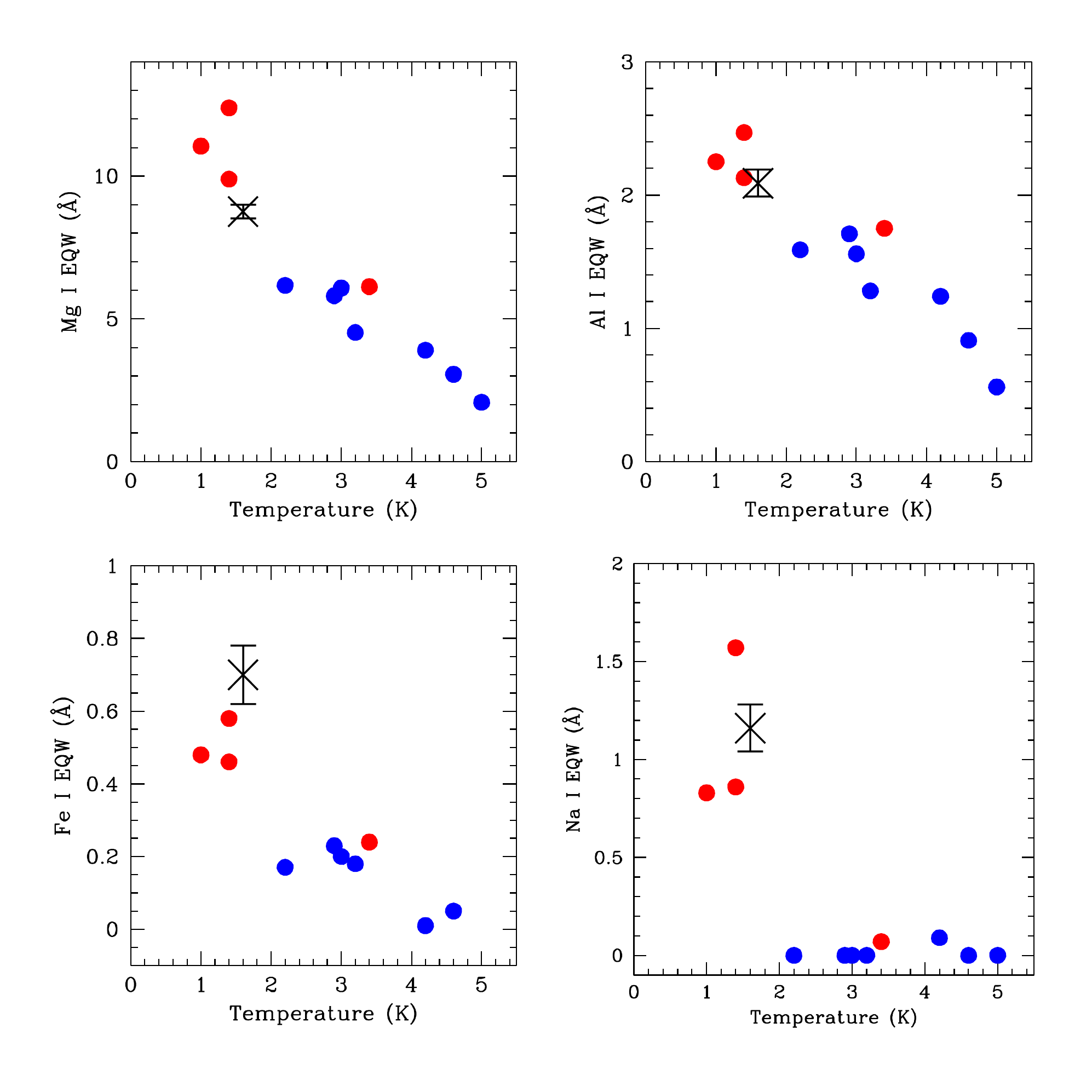}}}
\caption{As in Fig. \ref{zcammdwarfs}, but for the $H$-band.}
\label{zcammdwarfsh}
\end{figure}

\renewcommand{\thefigure}{17c}
\begin{figure}[htb]
\centerline{{\includegraphics[width=15cm]{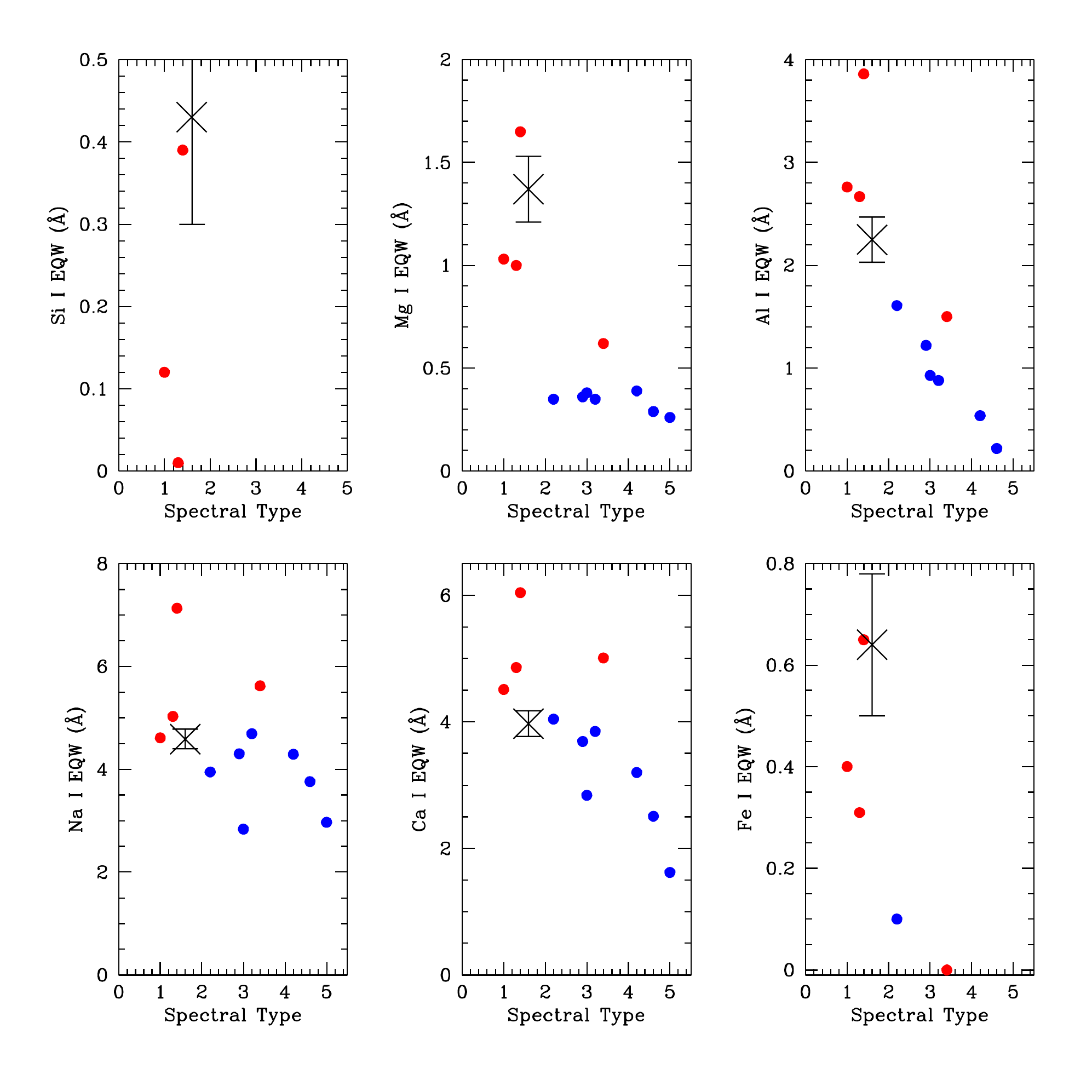}}}
\caption{As in Fig. \ref{zcammdwarfs}, but for the $K$-band.}
\label{zcammdwarfsk}
\end{figure}

\renewcommand{\thefigure}{18}
\begin{figure}[htb]
\centerline{{\includegraphics[width=15cm]{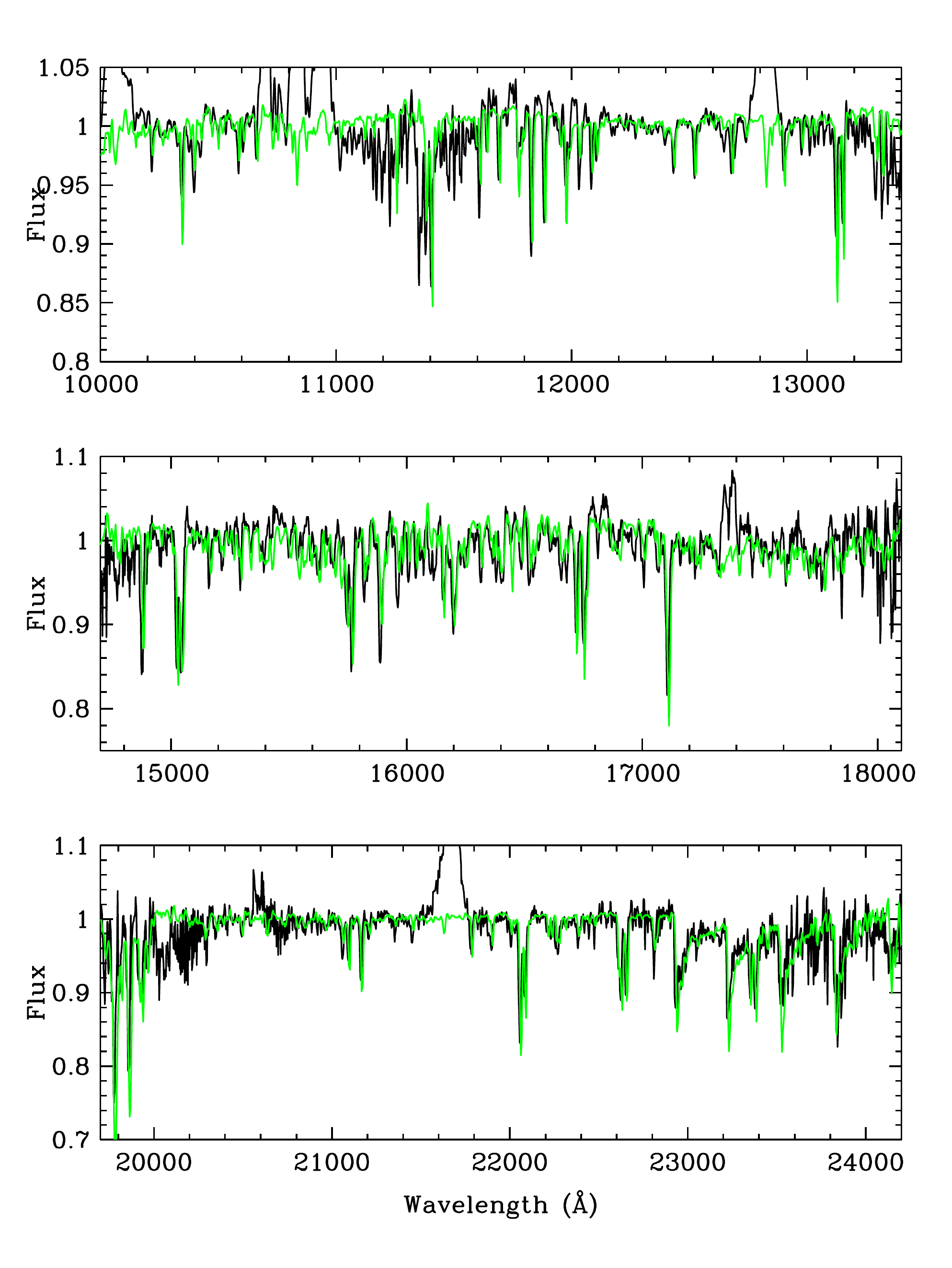}}}
\caption{The $JHK$ spectra of Z Cam (black) compared to to that of HD 42581,
an M1.3V (green).}
\label{zcamjhkspec}
\end{figure}

\renewcommand{\thefigure}{19}
\begin{figure}[htb]
\centerline{{\includegraphics[width=15cm]{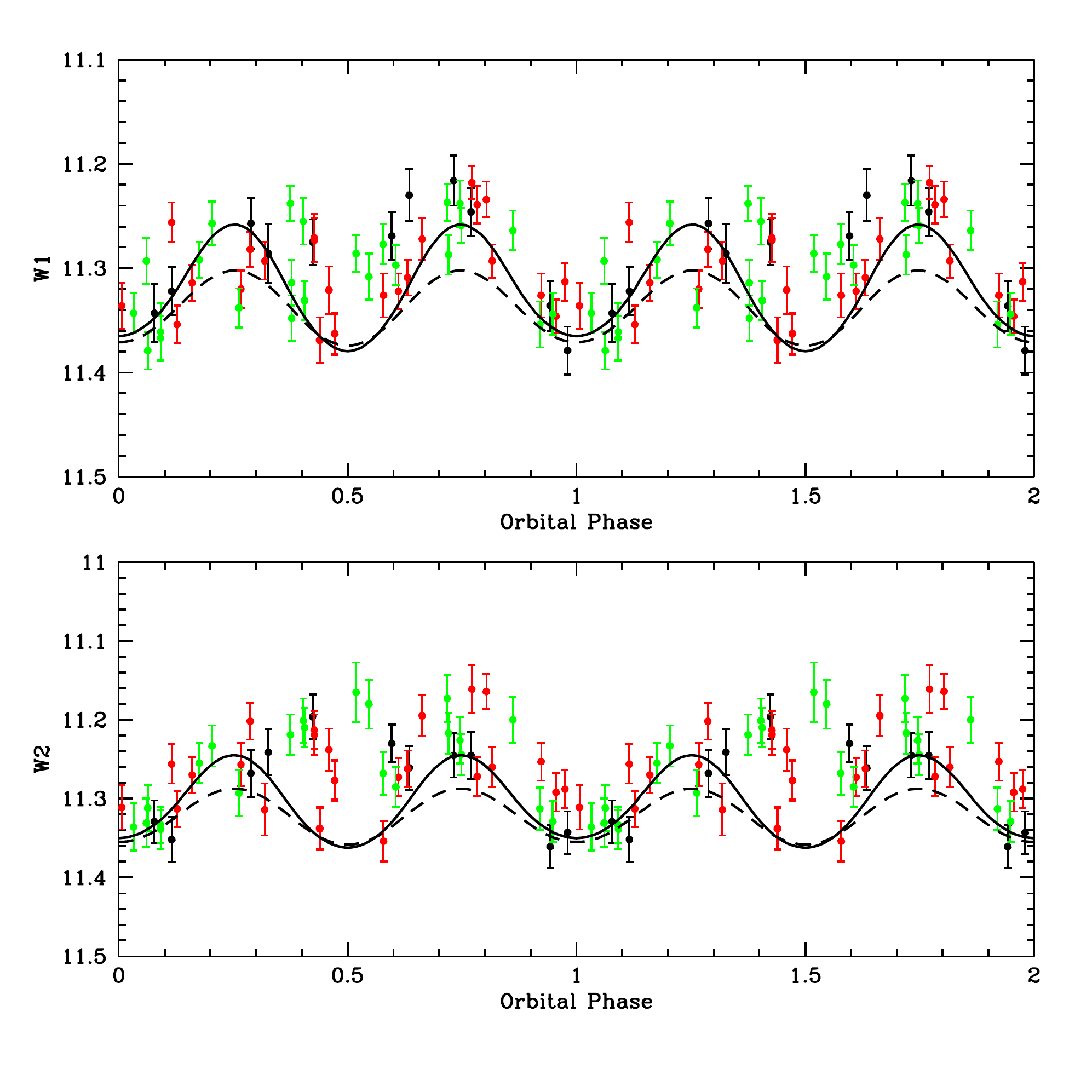}}}
\caption{The $WISE$ (black) and $NEOWISE$, 2016 April (red), and 2016
November (green) data for SY Cnc. The solid black line is a light curve model 
with $i$ = 55$^{\circ}$, while the dashed curve is for $i$ = 38$^{\circ}$. }
\label{sycnclc}
\end{figure}

\renewcommand{\thefigure}{20abc}
\begin{figure}[htb]
\centerline{{\includegraphics[width=15cm]{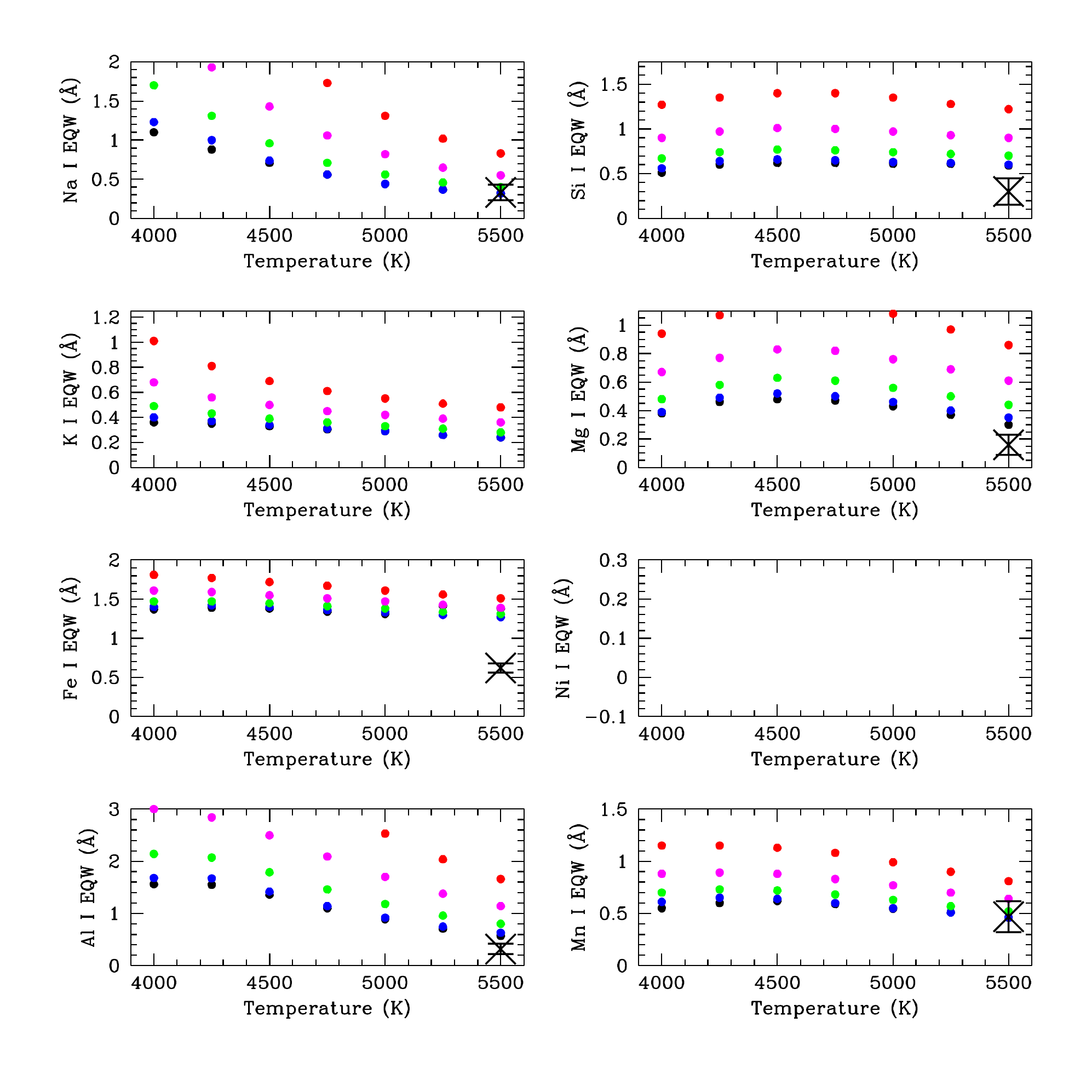}}}
\caption{The EQW measurements for SY Cnc in the $J$-band. Before the analysis,
both the source and synthetic spectra were boxcar smoothed by 5 pixels.}
\label{sycnceqw}
\end{figure}

\renewcommand{\thefigure}{20b}
\begin{figure}[htb]
\centerline{{\includegraphics[width=15cm]{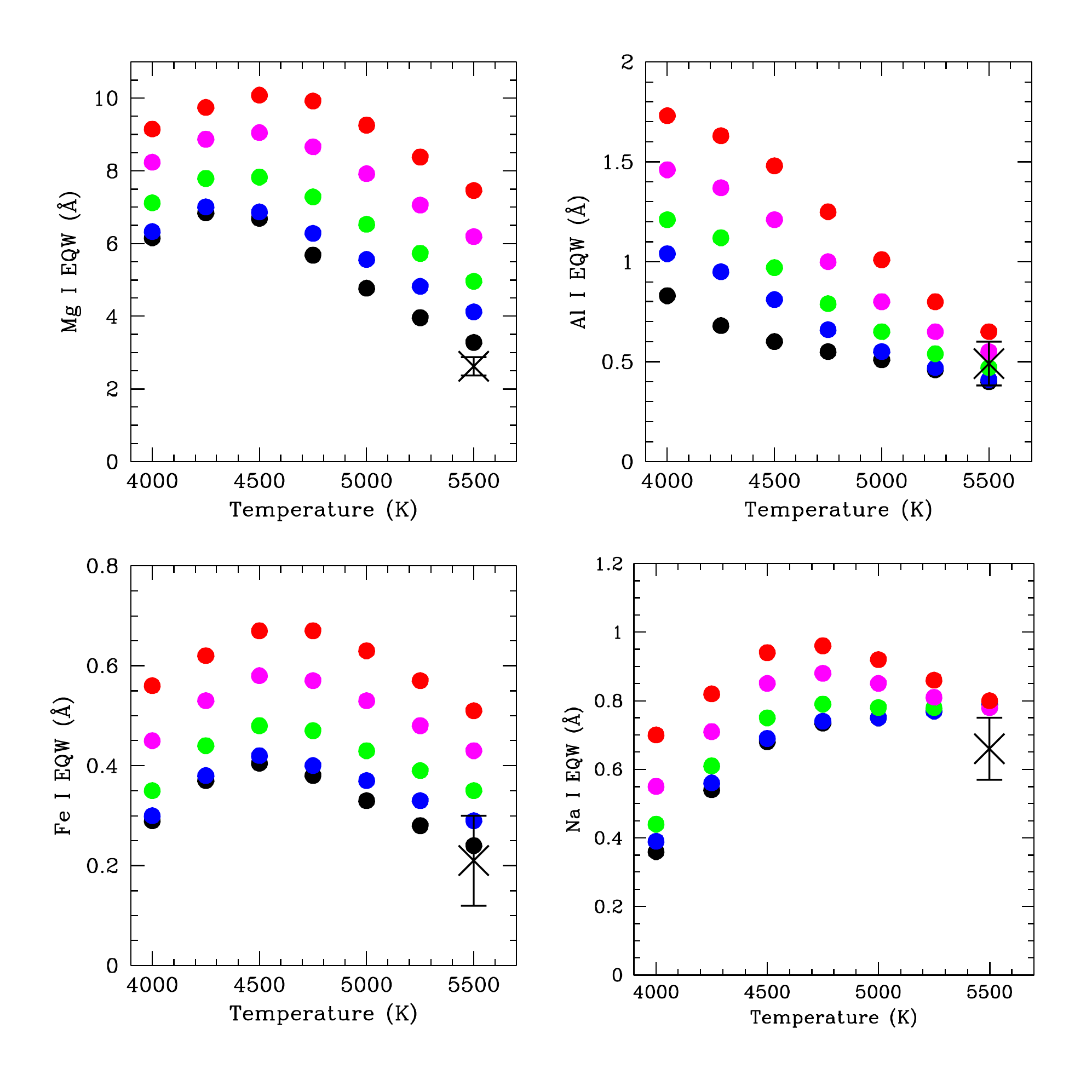}}}
\caption{The same as panel $a$, but for the $H$-band.}
\label{sycnceqwH}
\end{figure}

\renewcommand{\thefigure}{20c}
\begin{figure}[htb]
\centerline{{\includegraphics[width=15cm]{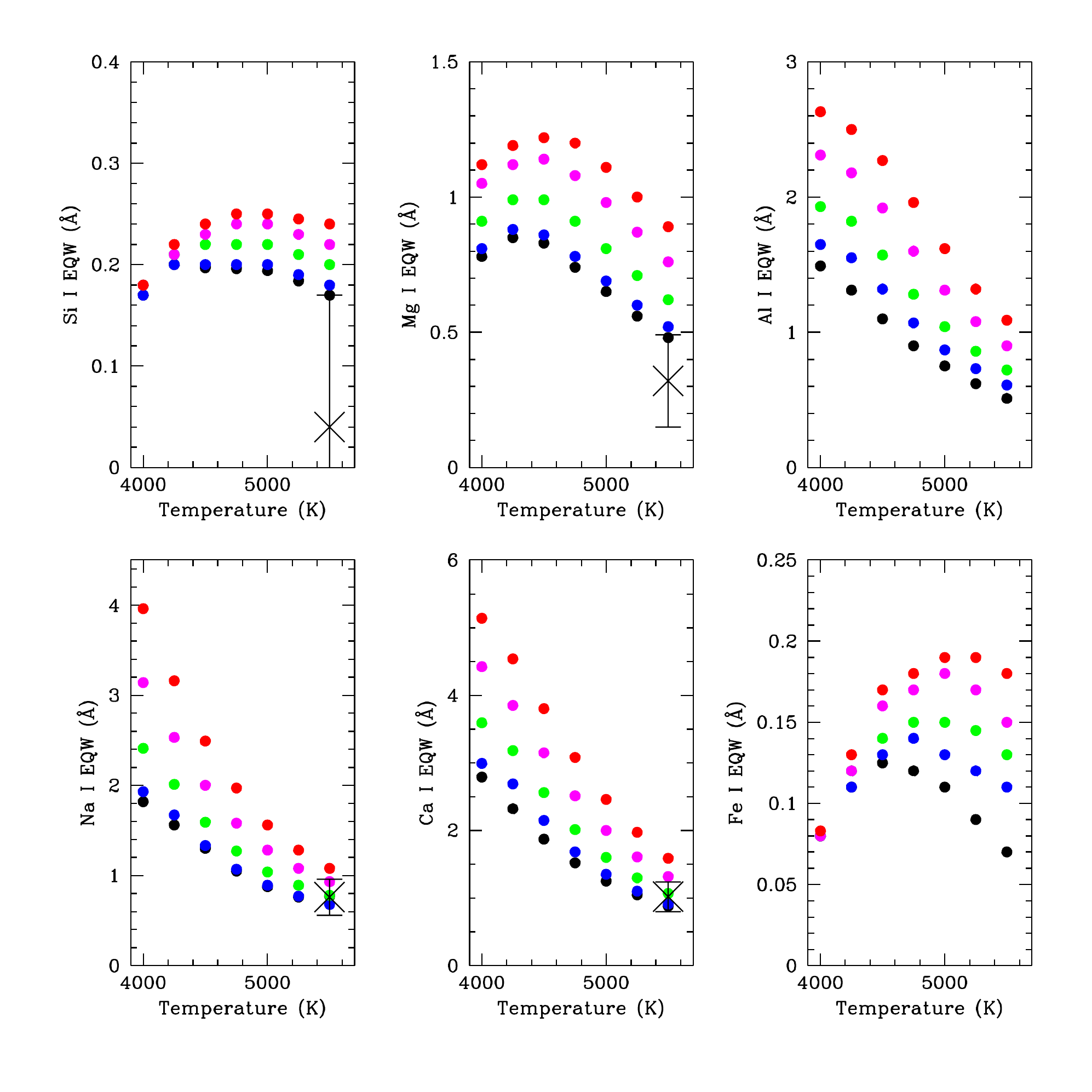}}}
\caption{The same as panel $a$, but for the $K$-band.}
\label{sycnceqwK}
\end{figure}

\renewcommand{\thefigure}{21}
\begin{figure}[htb]
\centerline{{\includegraphics[width=15cm]{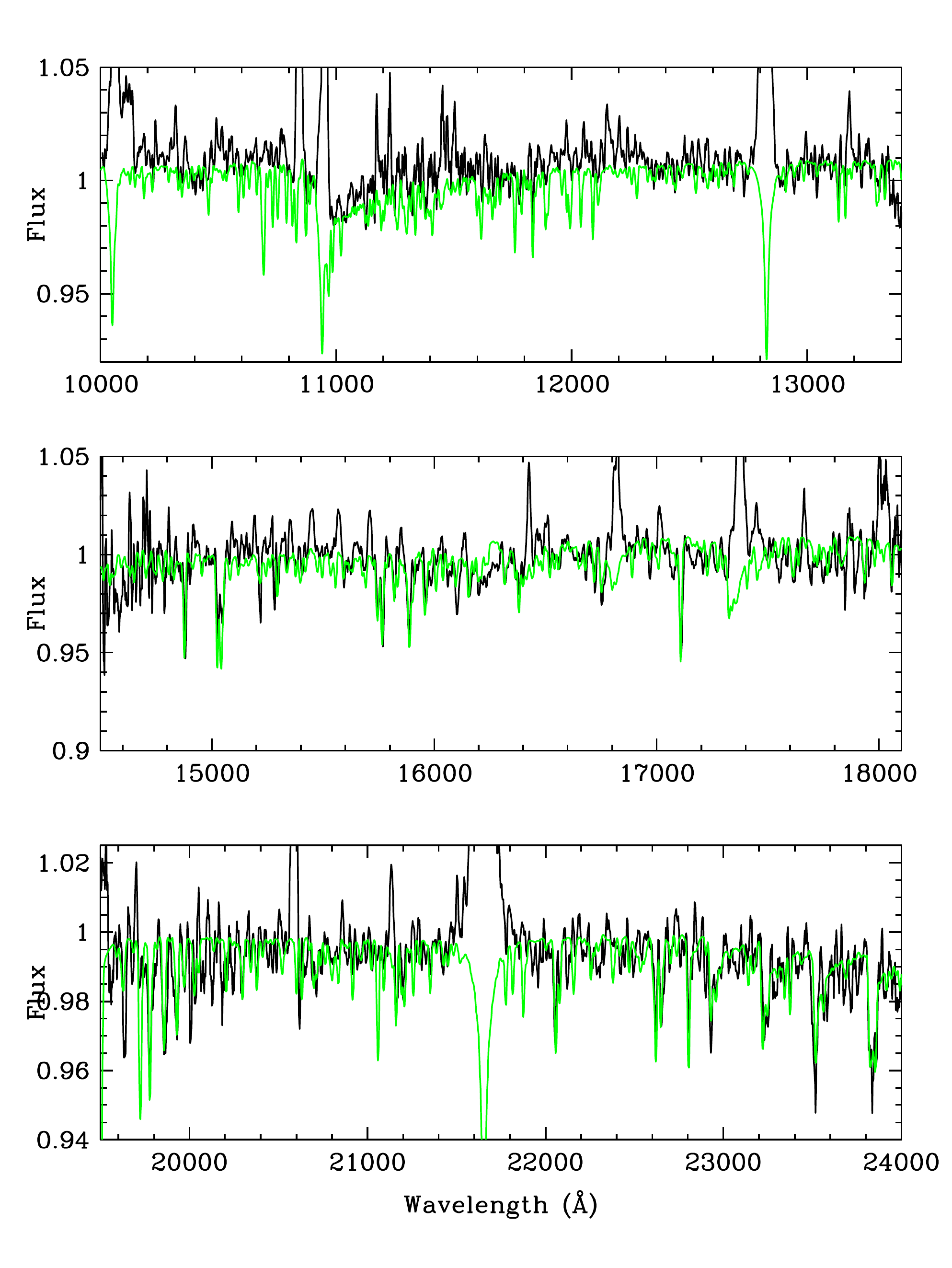}}}
\caption{The $JHK$ spectra of SY Cnc compared to a (highly contaminated)
synthetic spectrum that has T$_{\rm eff}$ = 5250 K, [Fe/H] = 0.0, and [C/Fe] = 
0.0. Both model and source spectra have been boxcar smoothed by 5 pixels.}
\label{sycncjhk}
\end{figure}

\renewcommand{\thefigure}{22}
\begin{figure}[htb]
\centerline{{\includegraphics[width=15cm]{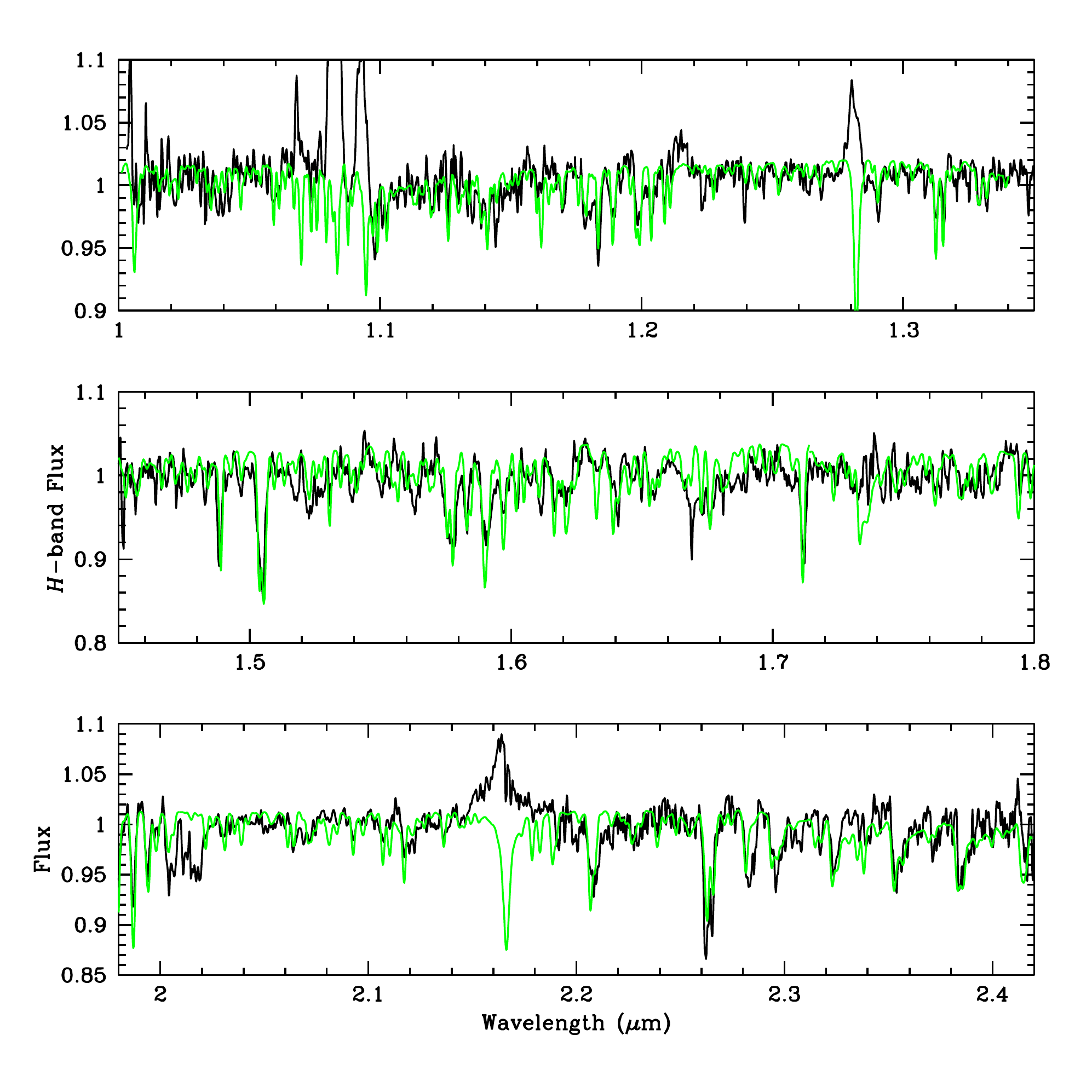}}}
\caption{The Gemini GNIRS spectra of EM Cyg (black). A synthetic
model spectrum is plotted in green, and has T$_{\rm eff}$ = 4500 K,
[C/Fe] = $-$0.5, and [Mg/Fe] = $-$0.5. }
\label{emspec}
\end{figure}

\renewcommand{\thefigure}{23}
\begin{figure}[htb]
\centerline{{\includegraphics[width=15cm]{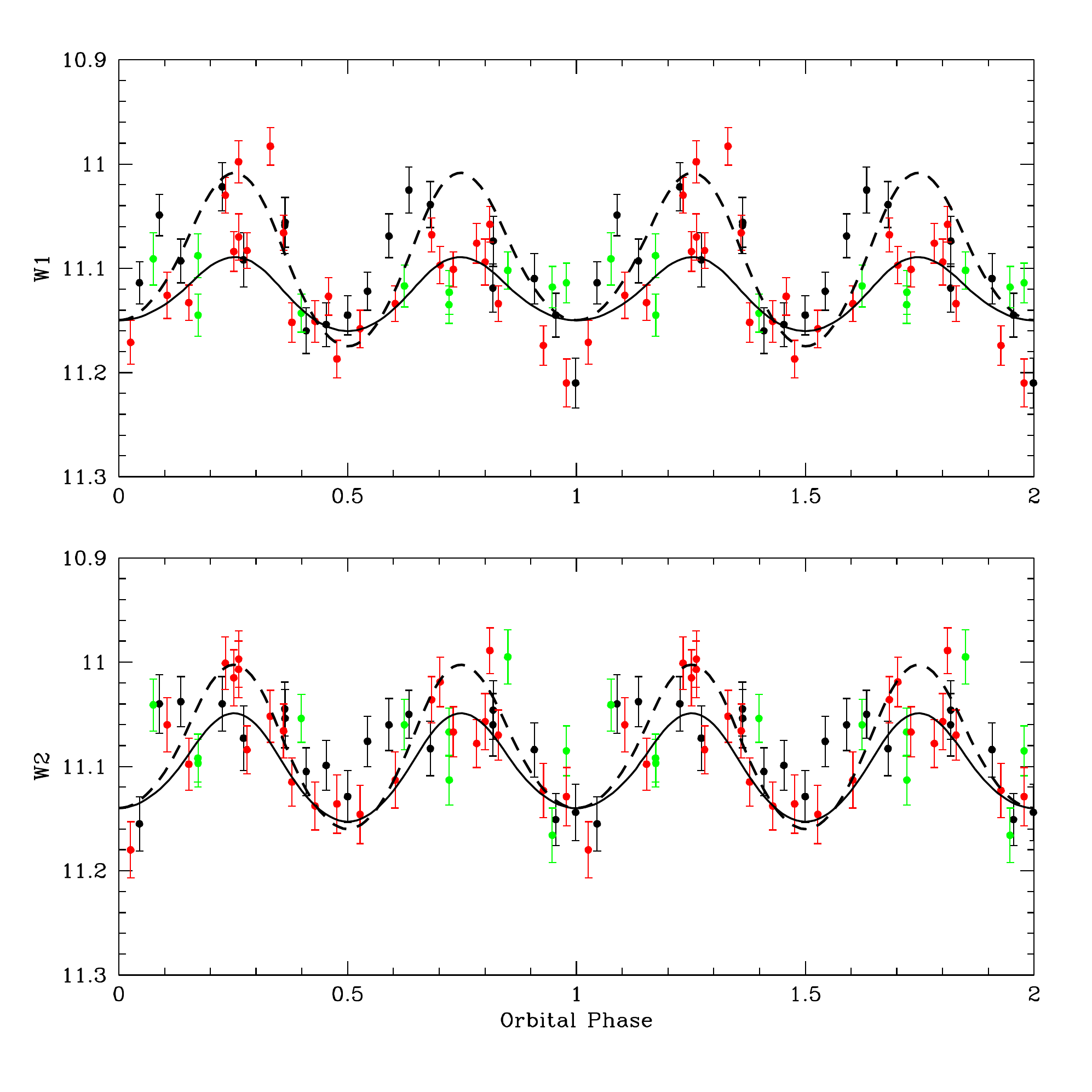}}}
\caption{The $WISE$ (black circles), and $NEOWISE$ light curves for EM Cyg.
There are two epochs of $NEOWISE$ data plotted here. The one in green is
for 2014 Oct 24, and the one in red is for 2015 Oct 18. The solid curve
is the light curve model assuming a contaminating source. The dashed curve
assumes no contamination.}
\label{emcyglc}
\end{figure}

\renewcommand{\thefigure}{24abc}
\begin{figure}[htb]
\centerline{{\includegraphics[width=15cm]{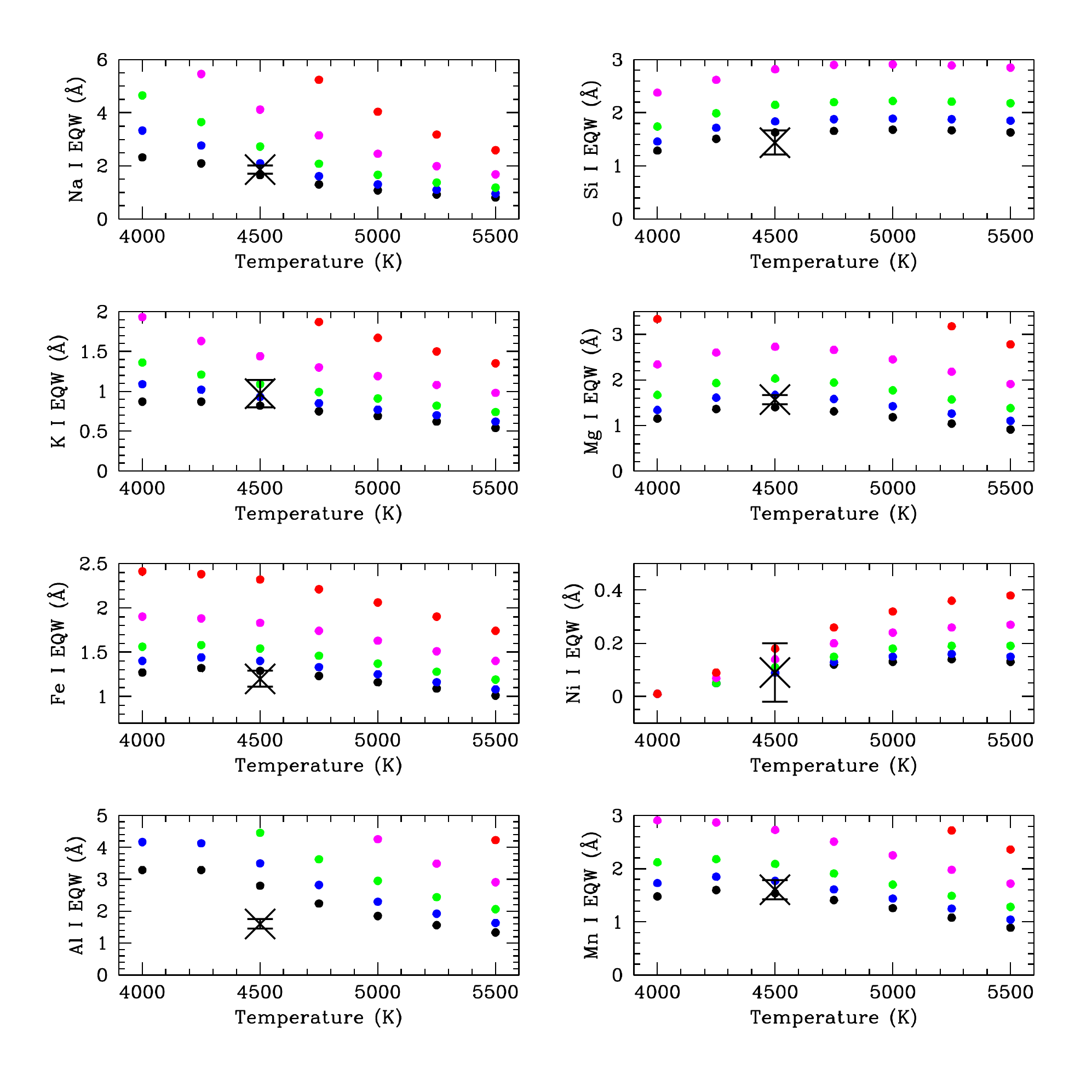}}}
\caption{The EQW measurements for EM Cyg in the $J$-band compared to a
solar abundance pattern grid.}
\label{emeqw}
\end{figure}

\renewcommand{\thefigure}{24b}
\begin{figure}[htb]
\centerline{{\includegraphics[width=15cm]{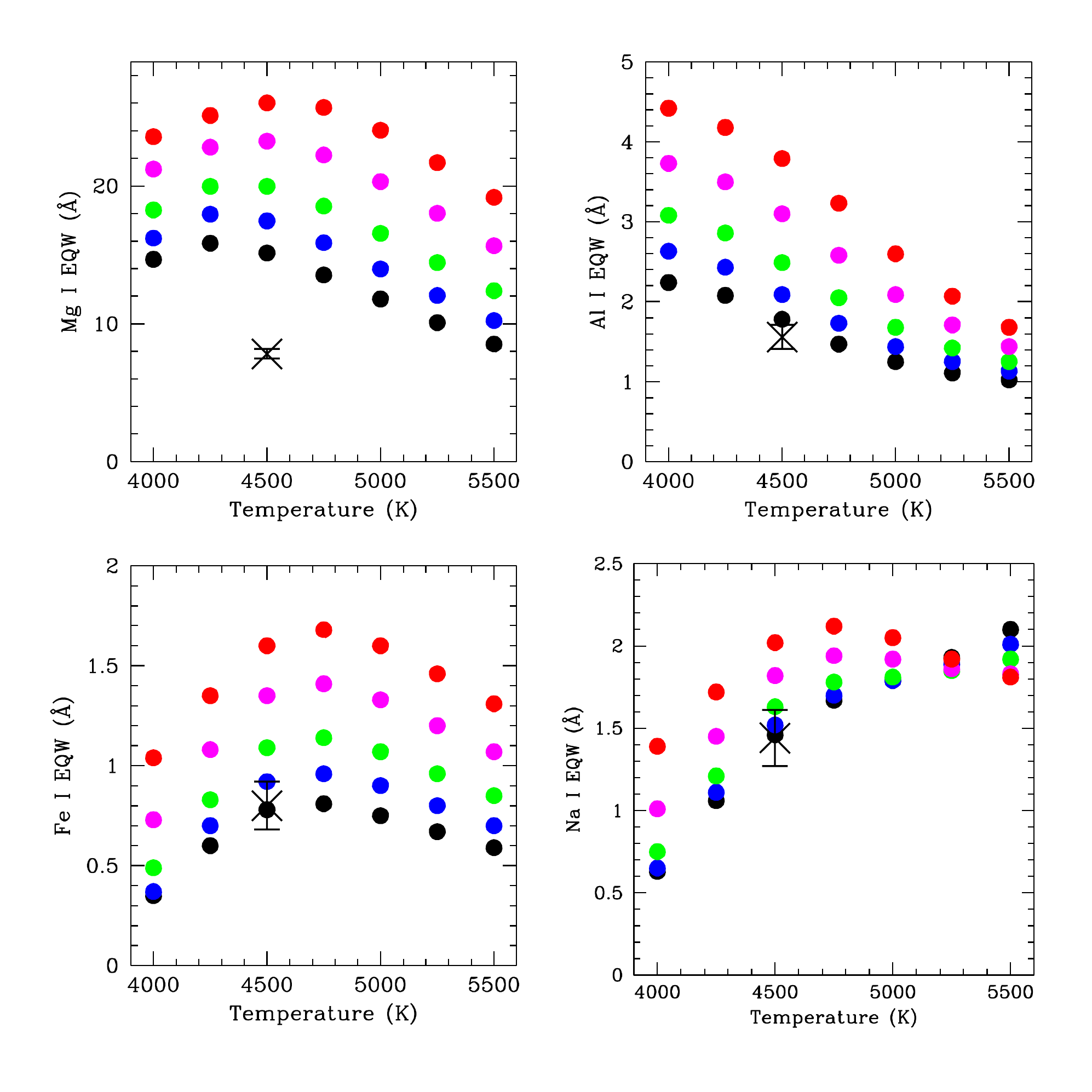}}}
\caption{The same as panel $a$, but for the $H$-band.}
\end{figure}

\renewcommand{\thefigure}{24c}
\begin{figure}[htb]
\centerline{{\includegraphics[width=15cm]{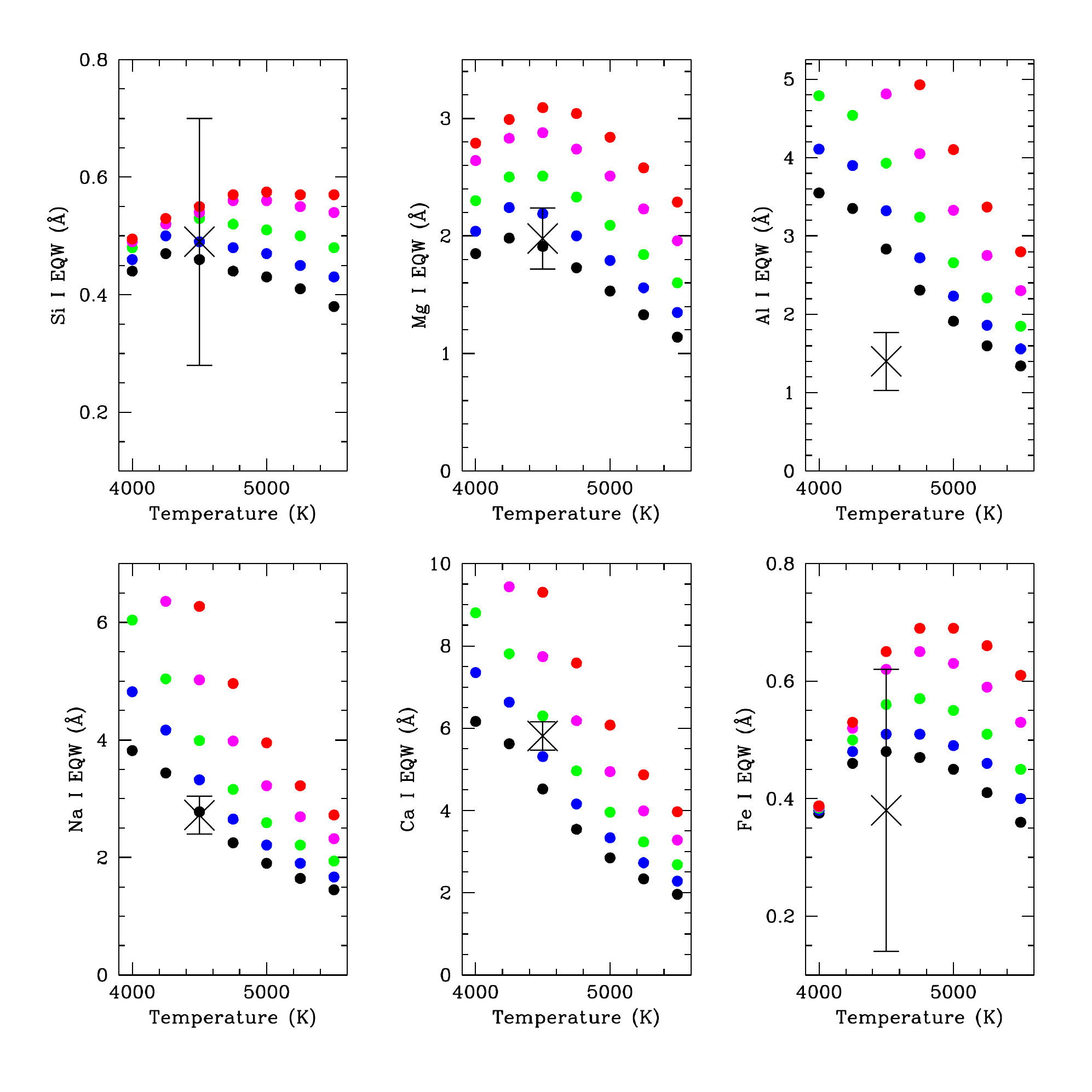}}}
\caption{The same as panel $a$, but for the $K$-band.}
\end{figure}

\renewcommand{\thefigure}{25abc}
\begin{figure}[htb]
\centerline{{\includegraphics[width=15cm]{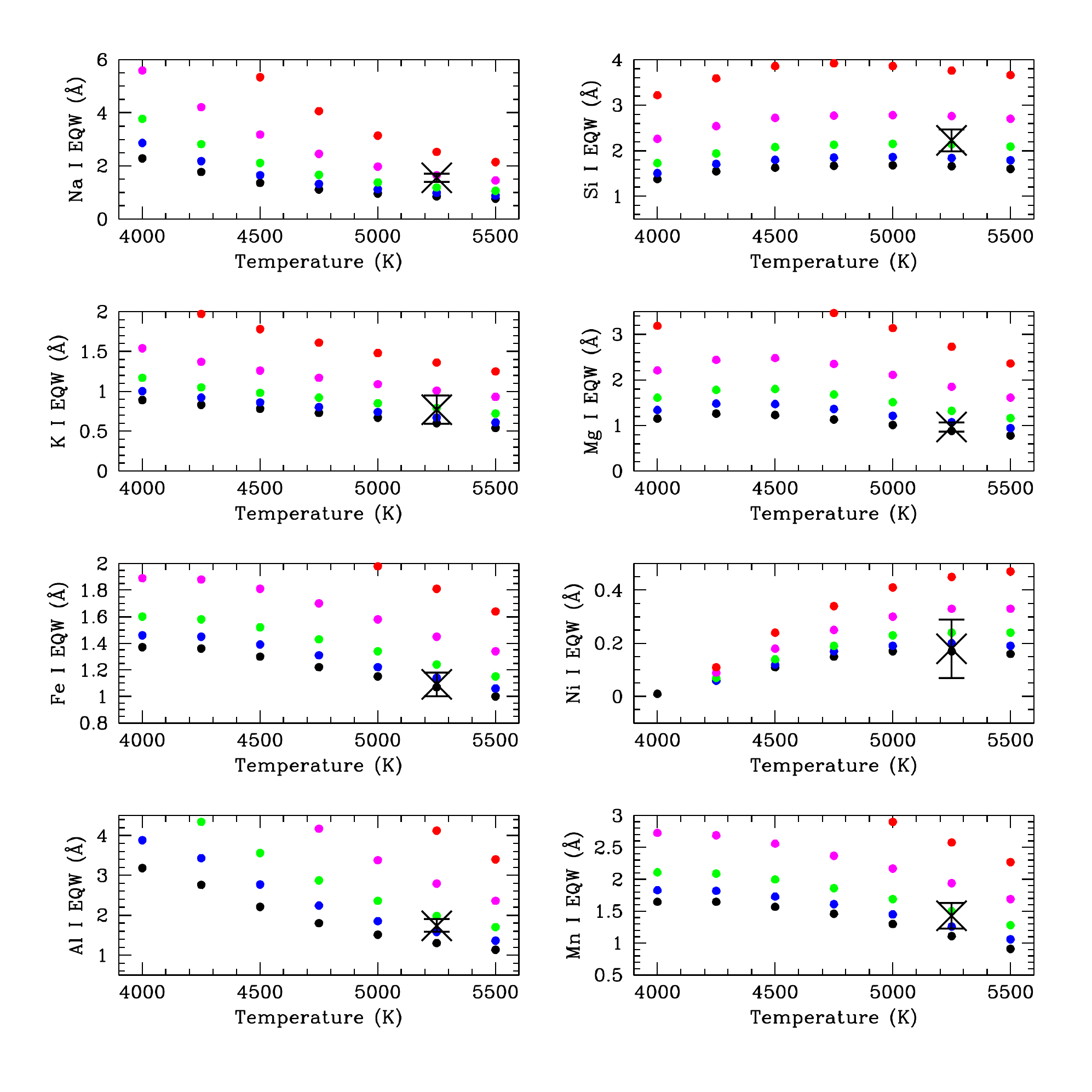}}}
\caption{The fit of the $J$-band EQW measurements for EY Cyg to those for the
synthetic spectra with log$g$ = 4.0, and [Fe/H] = 0.0.}
\label{eycyg}
\end{figure}

\renewcommand{\thefigure}{25b}
\begin{figure}[htb]
\centerline{{\includegraphics[width=15cm]{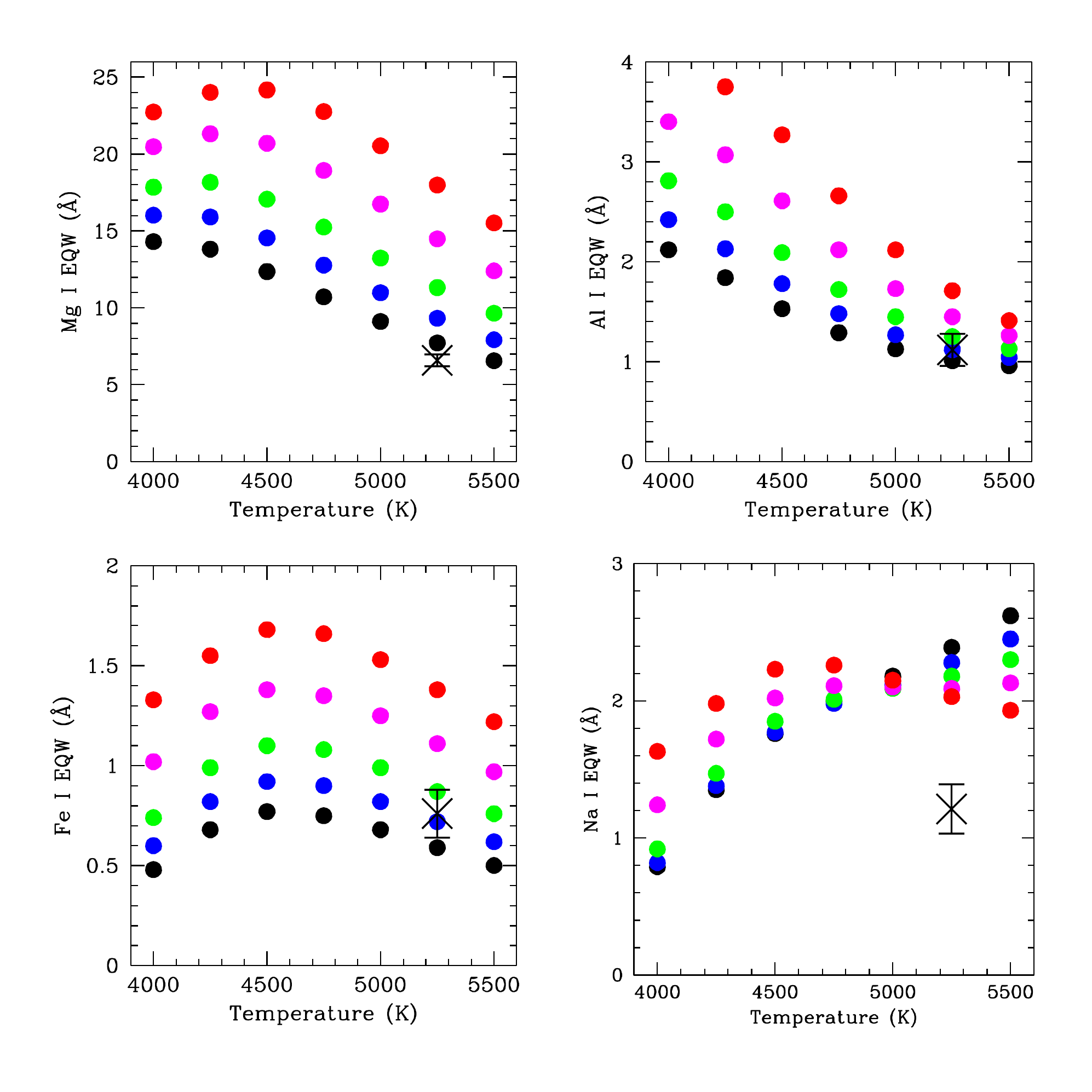}}}
\caption{The same as panel $a$, but for the $H$-band.}
\end{figure}

\renewcommand{\thefigure}{25c}
\begin{figure}[htb]
\centerline{{\includegraphics[width=15cm]{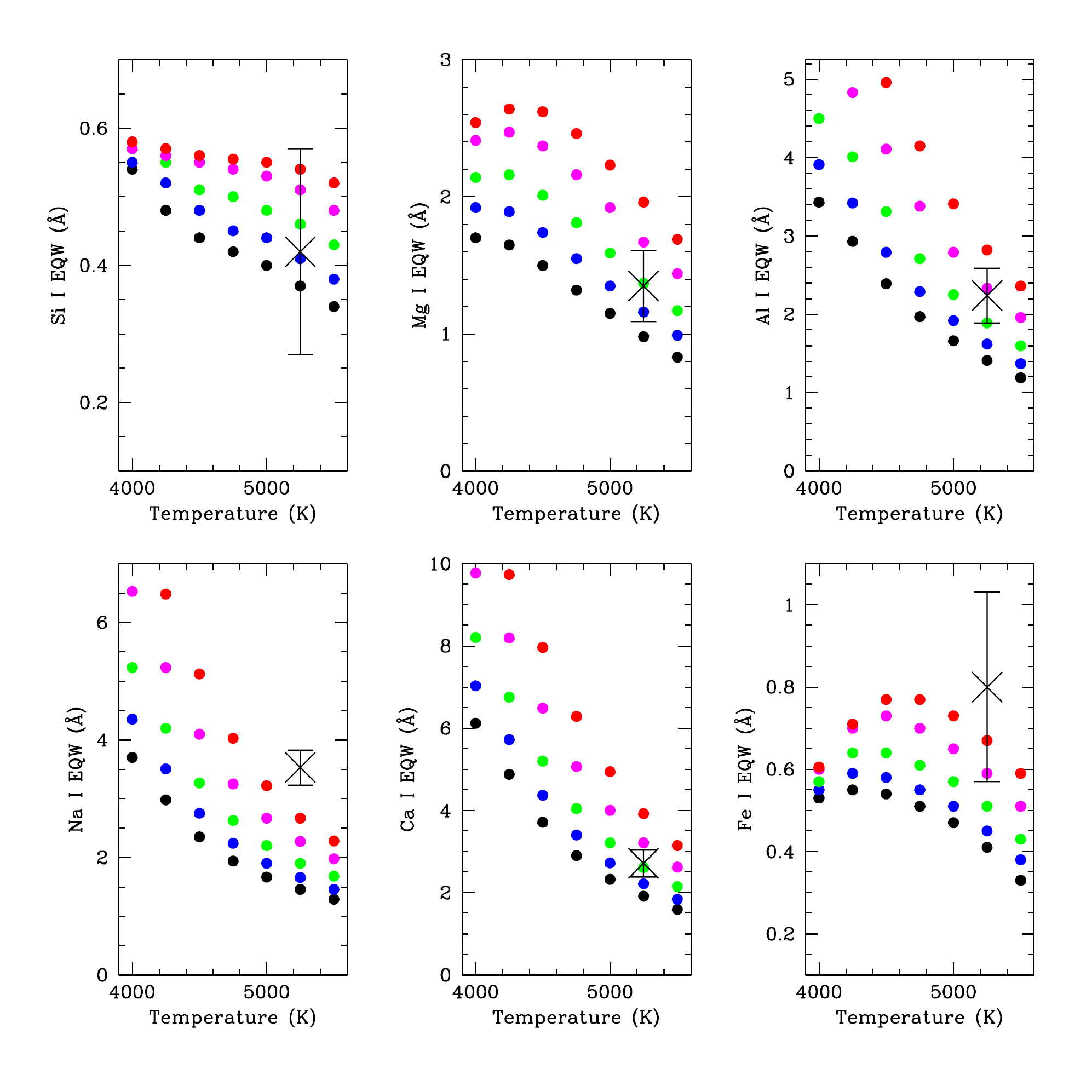}}}
\caption{The same as panel $a$, but for the $K$-band.}
\end{figure}

\renewcommand{\thefigure}{26}
\begin{figure}[htb]
\centerline{{\includegraphics[width=15cm]{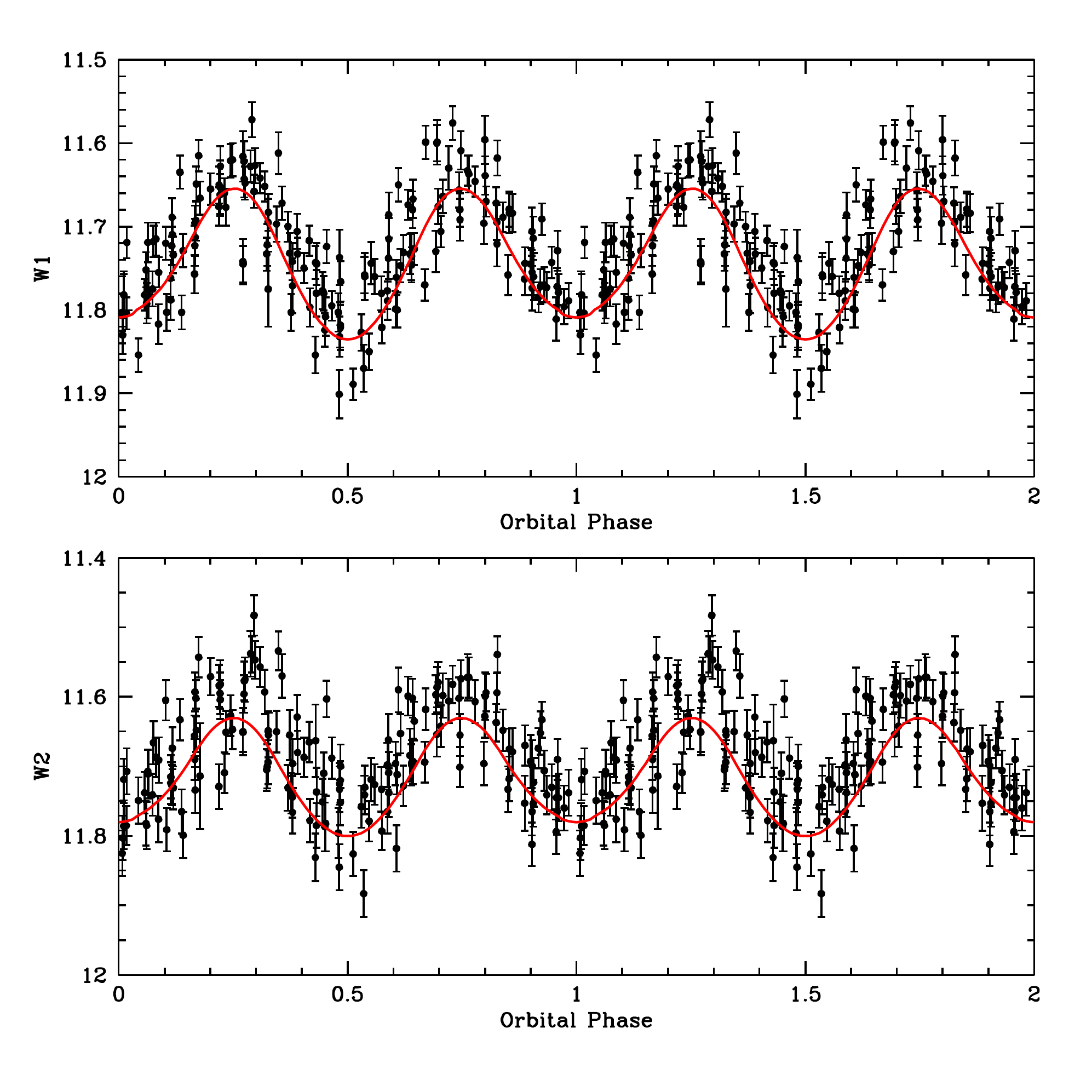}}}
\caption{The $WISE/NEOWISE$ light curves of V508 Dra. The light curve model,
 with an orbital inclination of $i$ = 75$^{\circ}$, is plotted in red.}
\label{v605dralc}
\end{figure}

\renewcommand{\thefigure}{27}
\begin{figure}[htb]
\centerline{{\includegraphics[width=15cm]{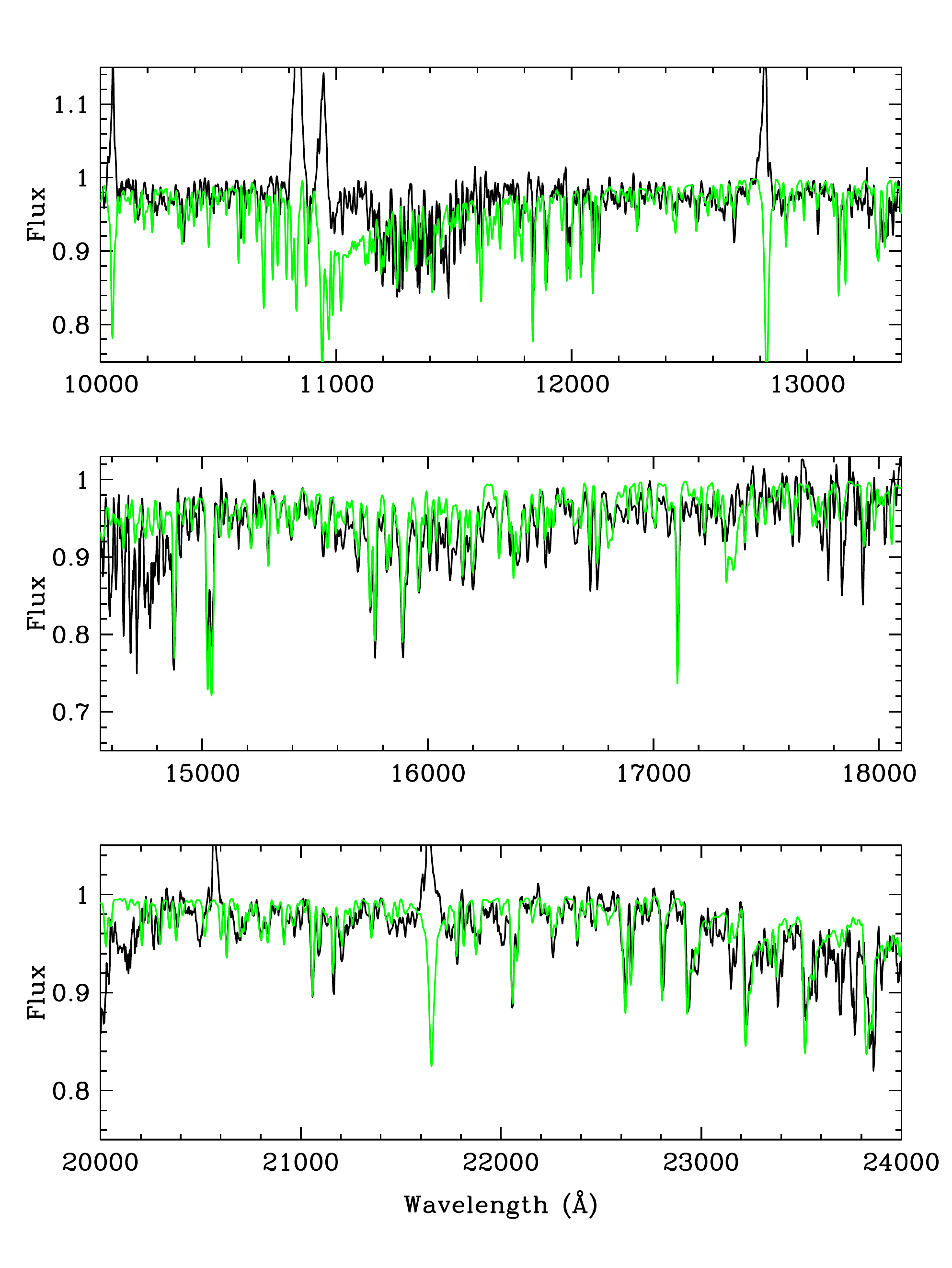}}}
\caption{The $JHK$ spectra of V508 Dra (black), overplotted by a synthetic
spectrum with T$_{\rm eff}$ = 4500 K, log$g$ = 4.0, [Fe/H] = 0.0, and [C/Fe]
= $-$0.3 (green).}
\label{v605drajhk}
\end{figure}

\renewcommand{\thefigure}{28}
\begin{figure}[htb]
\centerline{{\includegraphics[width=15cm]{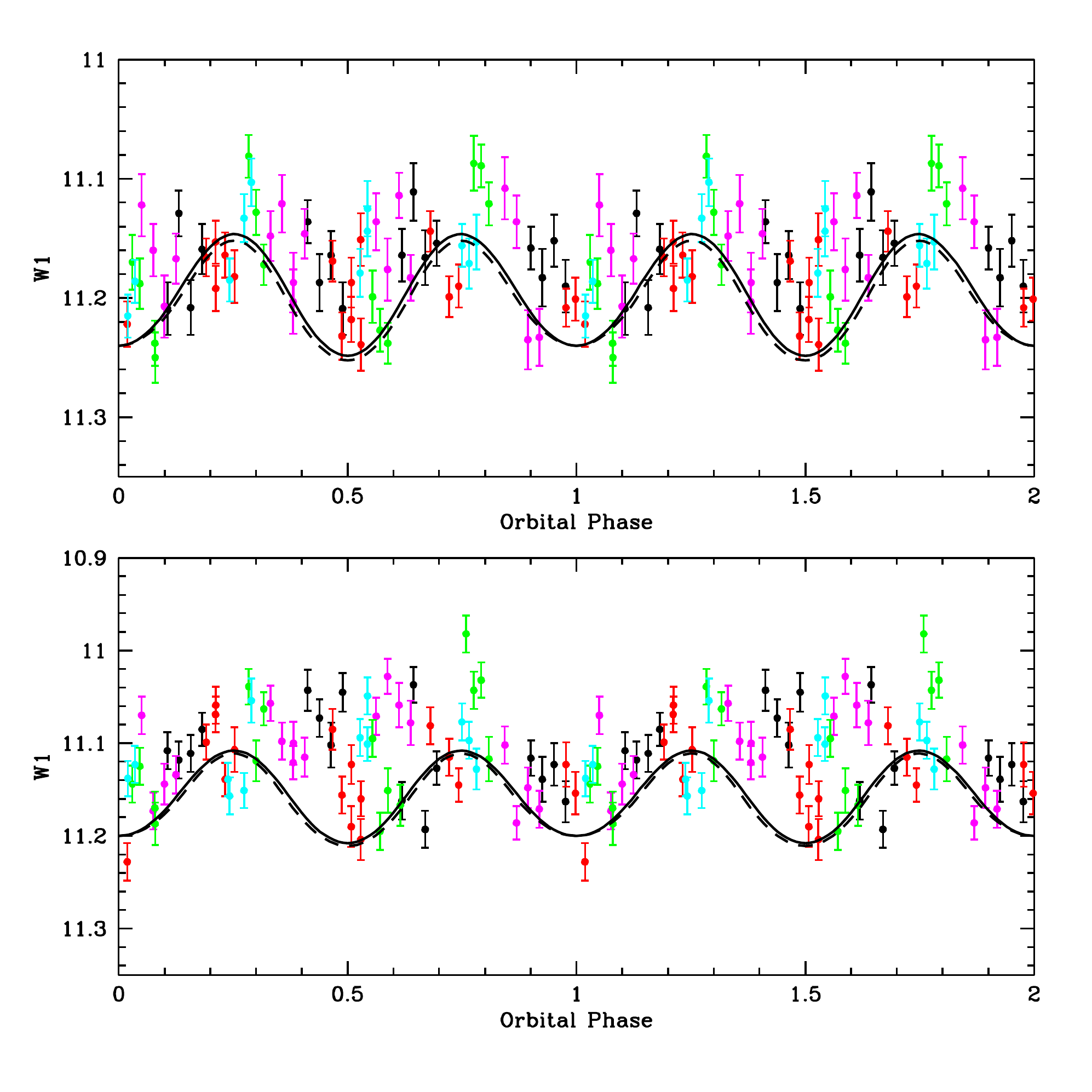}}}
\caption{The $WISE/NEOWISE$ light curves of AH Her. The black points are the
$WISE$ data for the epoch of 2010 February 22. We have calculated the
mean values in both $W1$ and $W2$ for all five epochs of data, and have
offset them to match the values for the initial $WISE$ epoch. The magenta
points are for the second $WISE$ epoch (2010 August 23), red represents
the first $NEOWISE$ epoch (2014 February 22), green is for 2015 August
17, cyan is for 2016 February 22. The solid line is a light curve model 
for $i$ = 46$^{\circ}$. The dashed line is a light curve model with $i$ = 
58$^{\circ}$, with a contaminating source that supplies 30\% of the flux.}
\label{ahherlc}
\end{figure}

\renewcommand{\thefigure}{29}
\begin{figure}[htb]
\centerline{{\includegraphics[width=15cm]{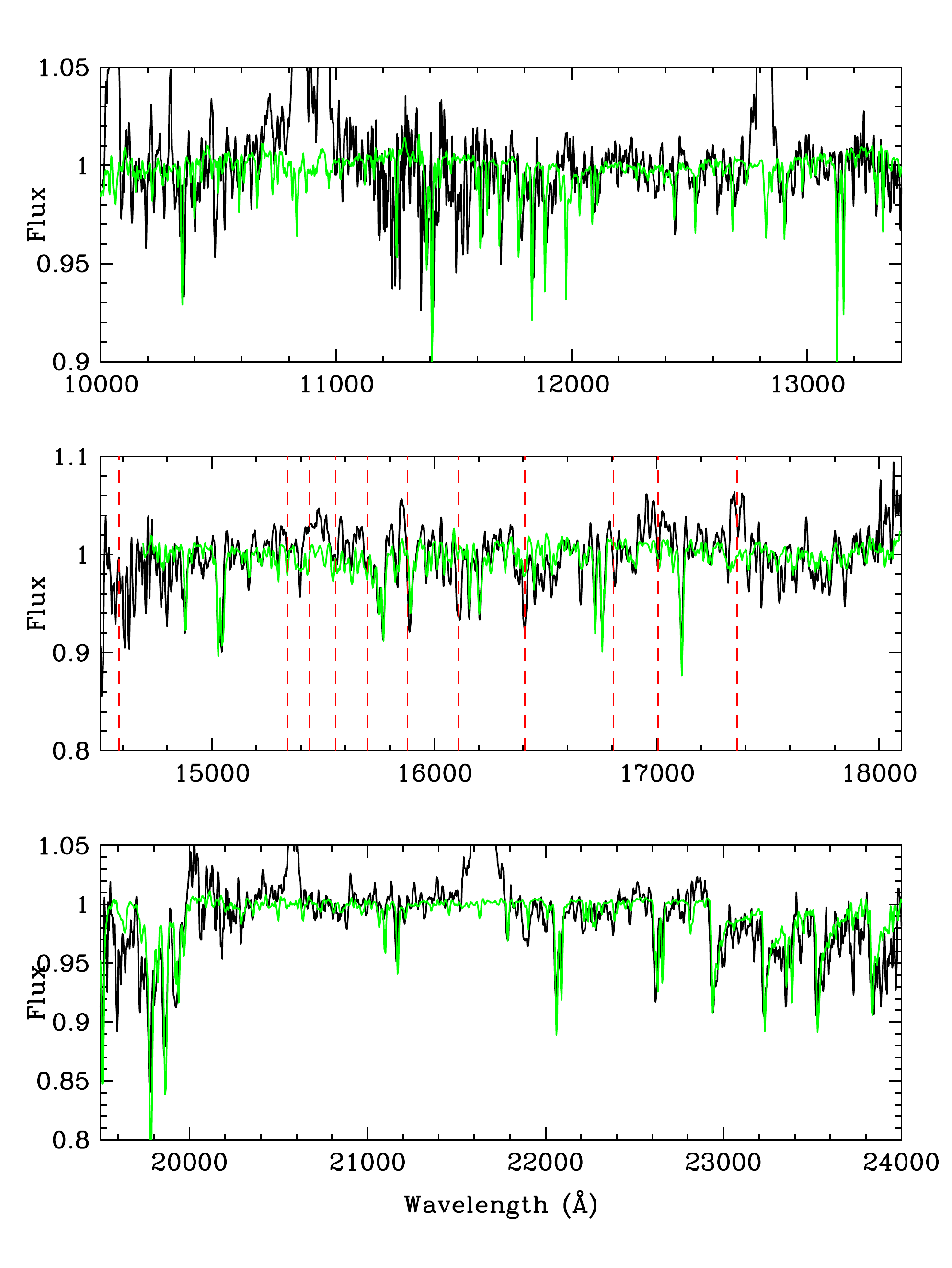}}}
\caption{The TripleSpec data for AH Her (black). We overplot the spectra
of the M1.3 dwarf HD 42581 from the IRTF Spectral Library (green). In the panel
for the $H$-band, we locate the H I Brackett series as vertical red dashed
lines. The line at $\lambda$17005 \AA ~is due to He I. The left most line
represents the Brackett limit.}
\label{ahherjhk}
\end{figure}

\renewcommand{\thefigure}{30abc}
\begin{figure}[htb]
\centerline{{\includegraphics[width=15cm]{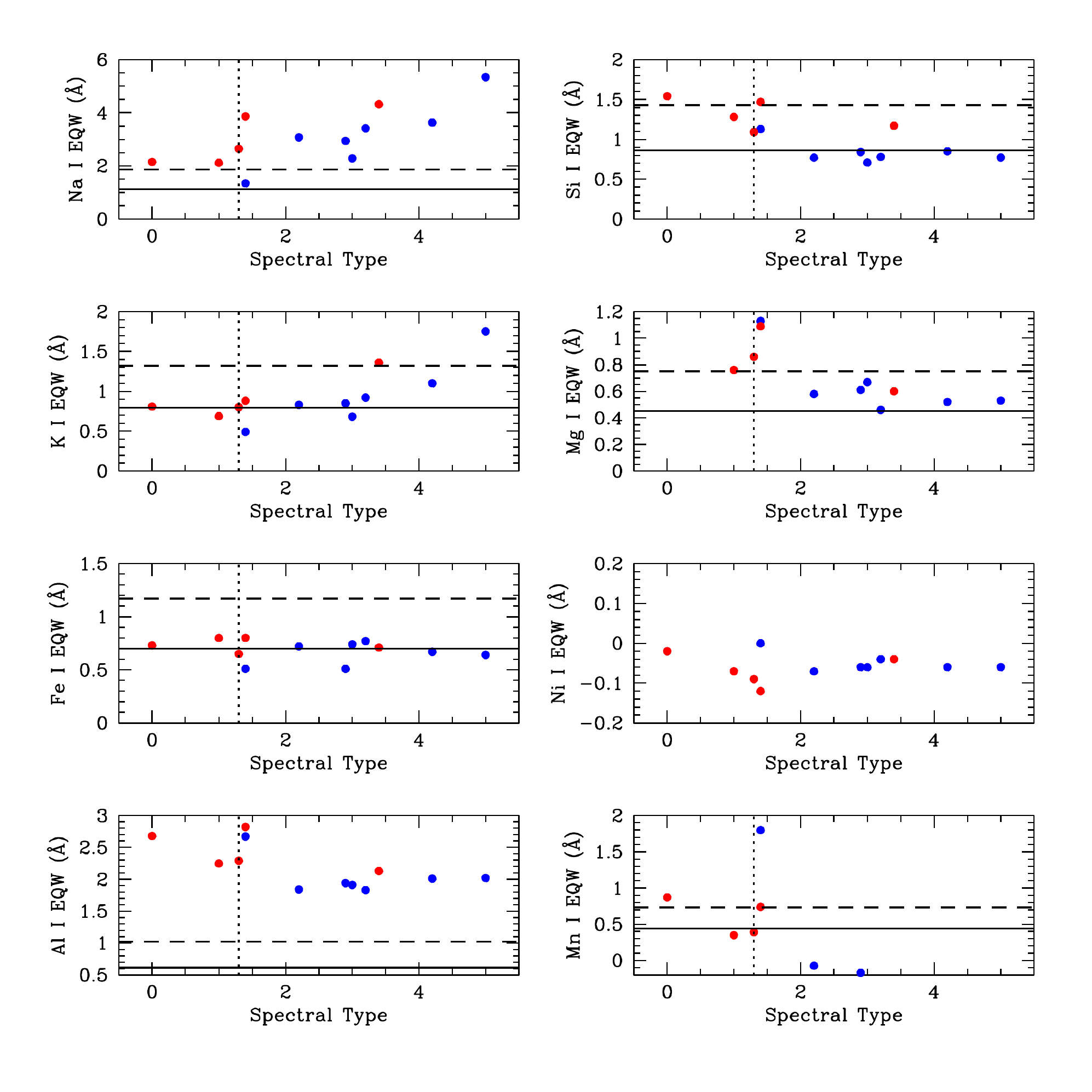}}}
\caption{The EQW measures for AH Her in the $J$-band, plotted as a solid 
horizontal line, compared to those of M dwarfs (see Fig. \ref{rxmdwf}). The
horizontal dashed line would be the true EQW measures for AH Her if they
were not diluted by a 40\% level of contamination. The vertical dotted line
is the derived spectral type.}
\label{ahhermdwarf}
\end{figure}

\renewcommand{\thefigure}{30b}
\begin{figure}[htb]
\centerline{{\includegraphics[width=15cm]{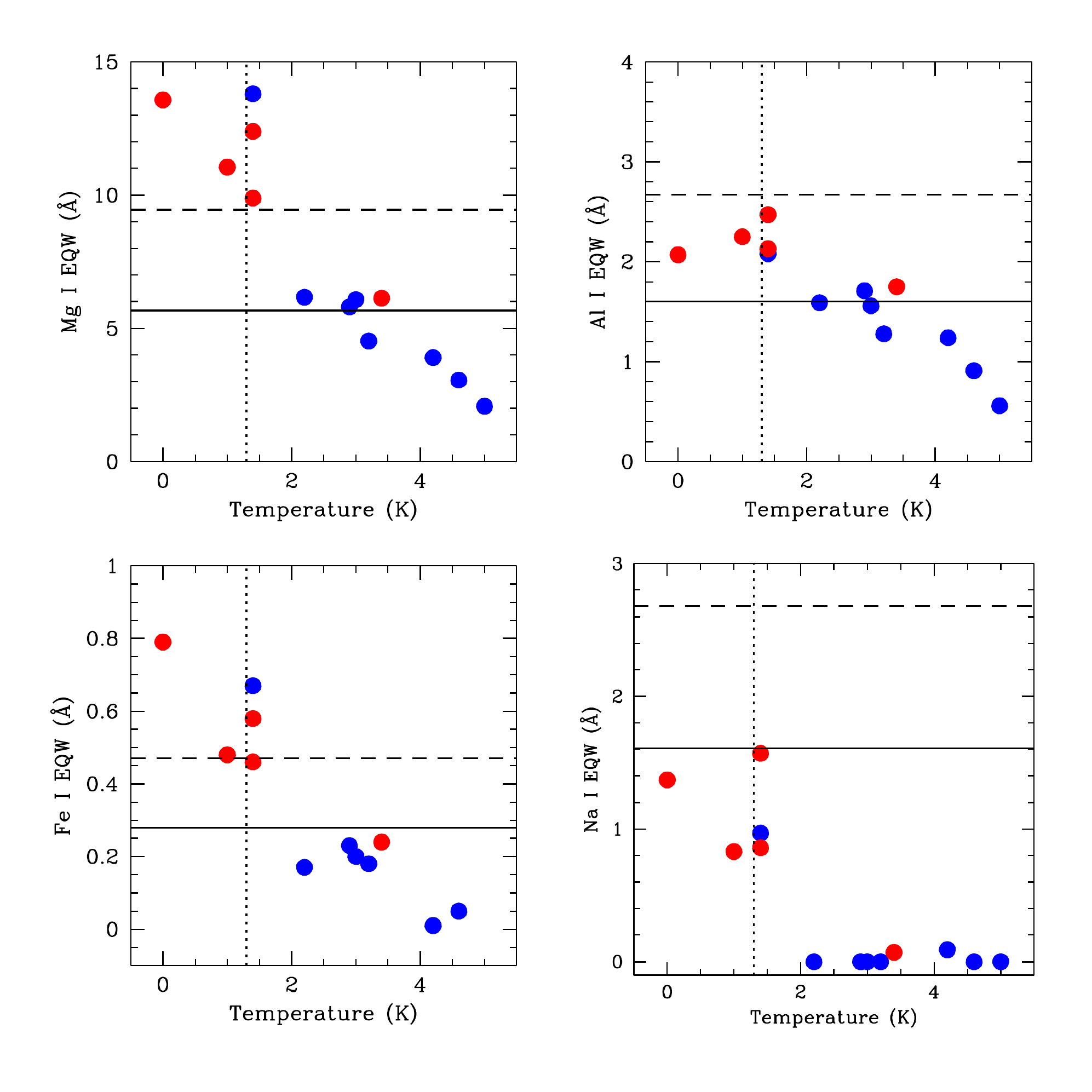}}}
\caption{The same as panel $a$, but for the $H$-band.}
\label{ahhermdwarfH}
\end{figure}

\renewcommand{\thefigure}{30c}
\begin{figure}[htb]
\centerline{{\includegraphics[width=15cm]{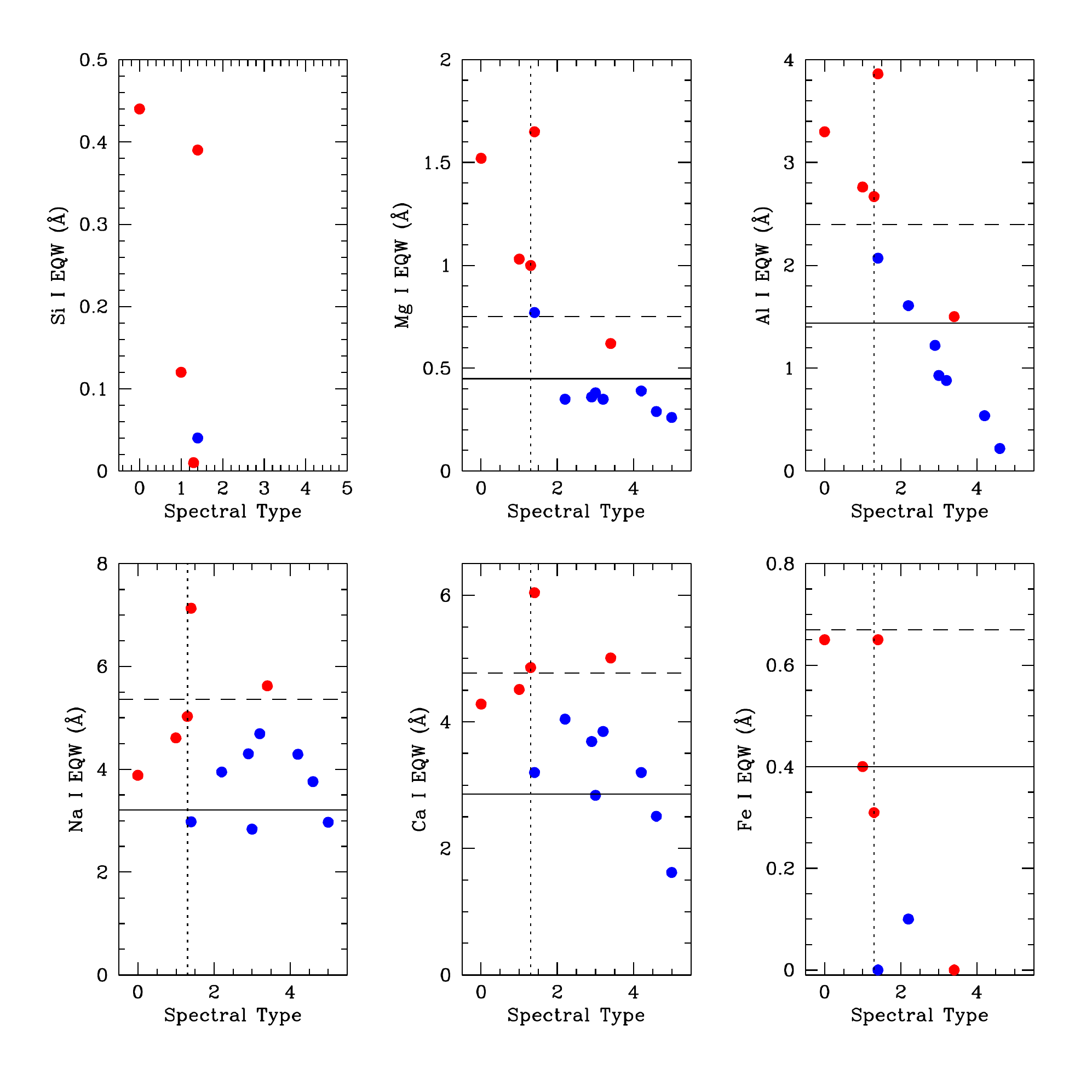}}}
\caption{The same as panel $a$, but for the $K$-band.}
\label{ahhermdwarfK}
\end{figure}

\renewcommand{\thefigure}{31}
\begin{figure}[htb]
\centerline{{\includegraphics[width=15cm]{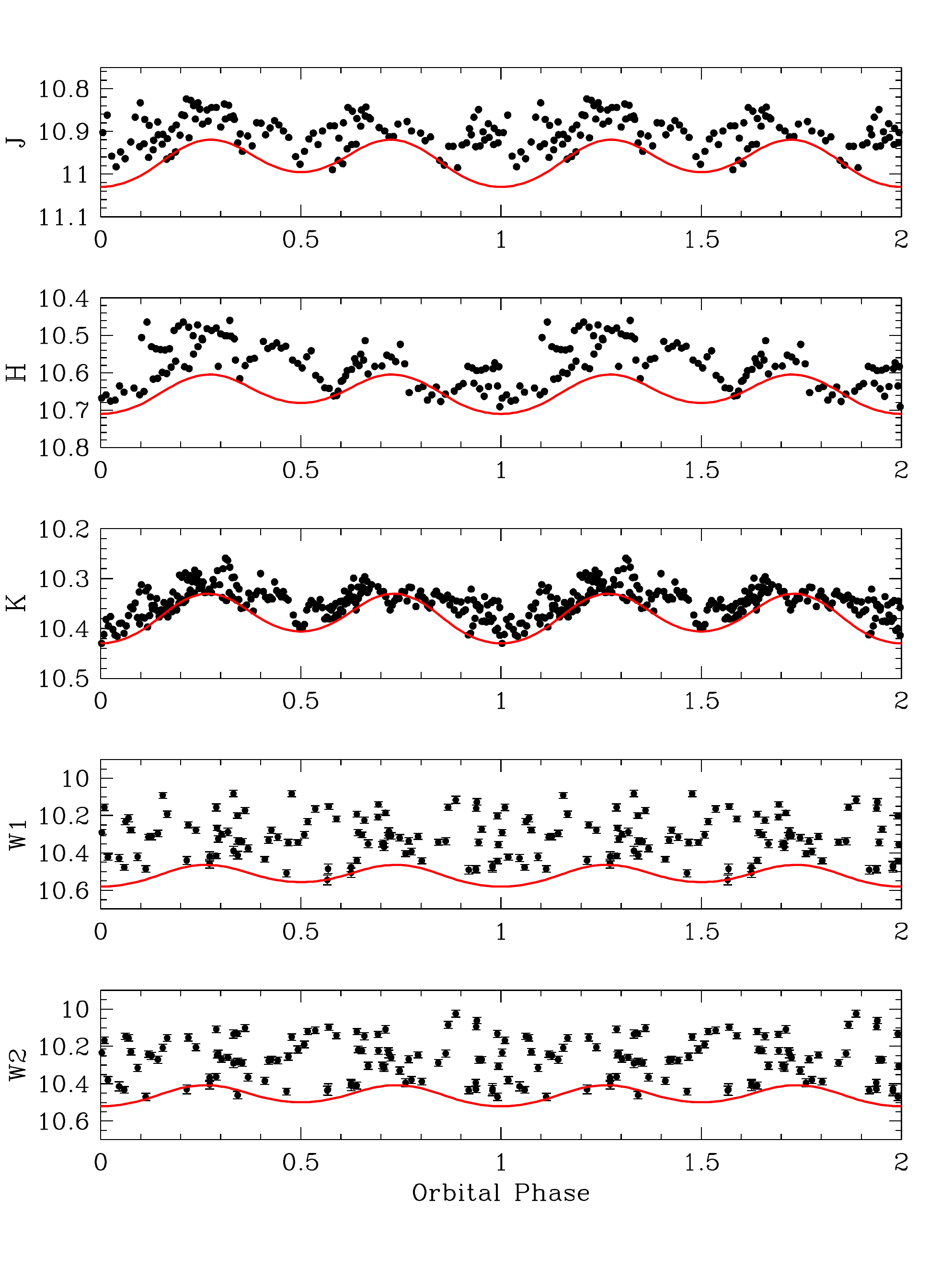}}}
\caption{The $JHK$, $WISE$ and $NEOWISE$ light curves for RU Peg. The
$JHK$ data were obtained with SQIID over three nights, 2003 August 3 through
5. The light curve model (red) has T$_{\rm 1}$ = 45,000 K, T$_{\rm 2}$ = 
5,000 K, $i$ = 50, and a contamination level of 25\%.}
\label{rupeglc}
\end{figure}

\renewcommand{\thefigure}{32abc}
\begin{figure}[htb]
\centerline{{\includegraphics[width=15cm]{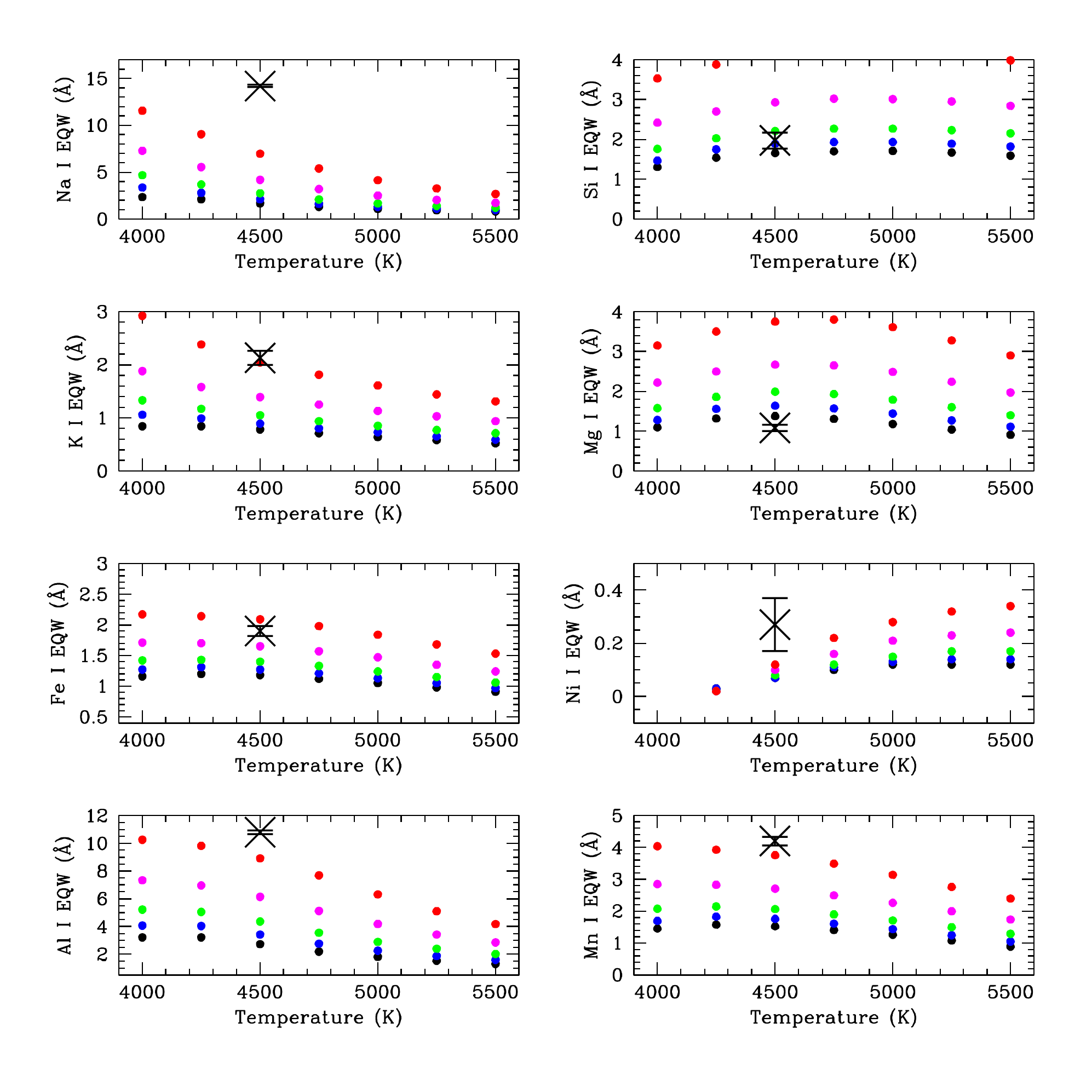}}}
\caption{The equivalent width measurements for QZ Ser in the $J$-band.}
\label{qzser}
\end{figure}
\clearpage

\renewcommand{\thefigure}{32b}
\begin{figure}[htb]
\centerline{{\includegraphics[width=15cm]{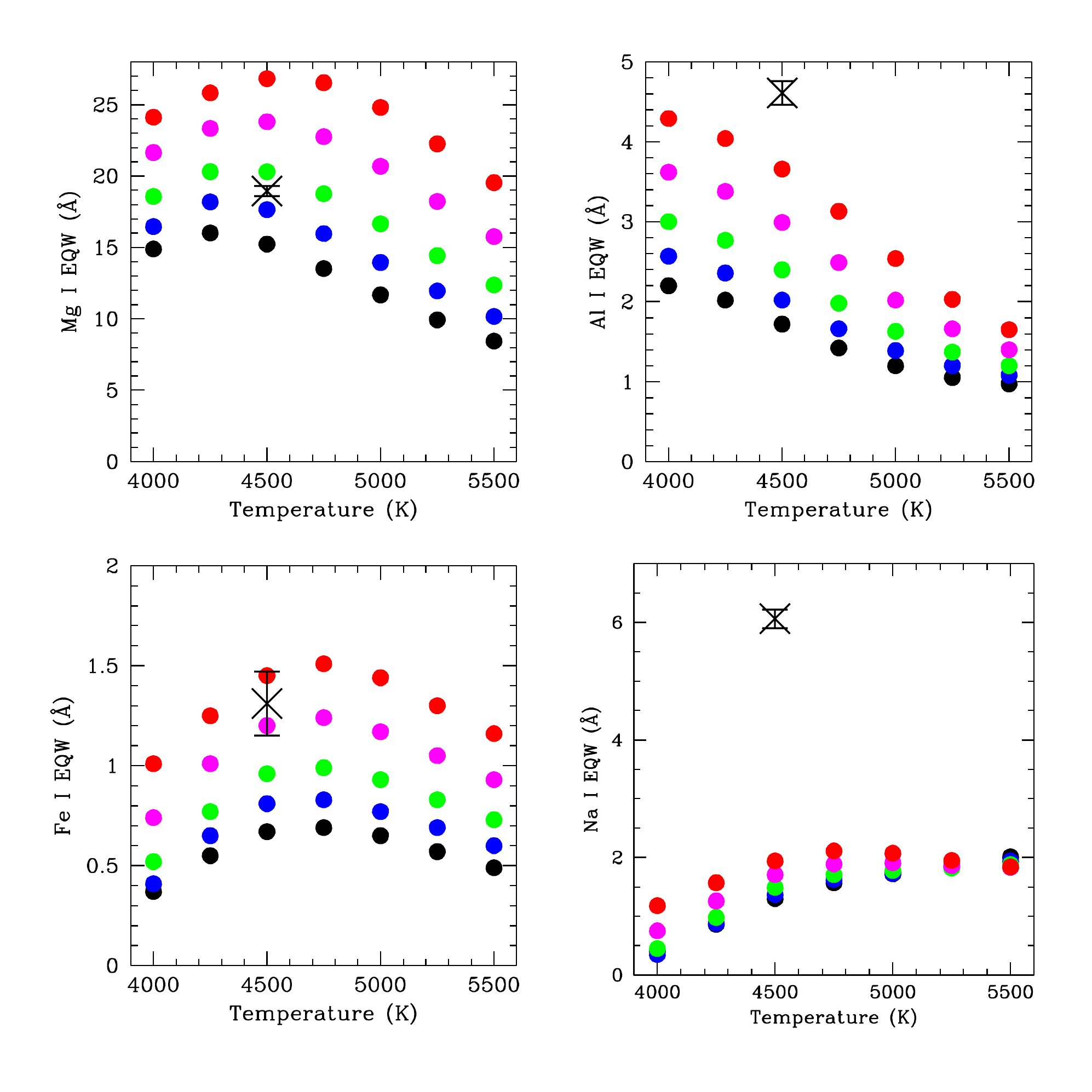}}}
\caption{The same as panel $A$, for the $H$-band.}
\end{figure}

\renewcommand{\thefigure}{32c}
\begin{figure}[htb]
\centerline{{\includegraphics[width=15cm]{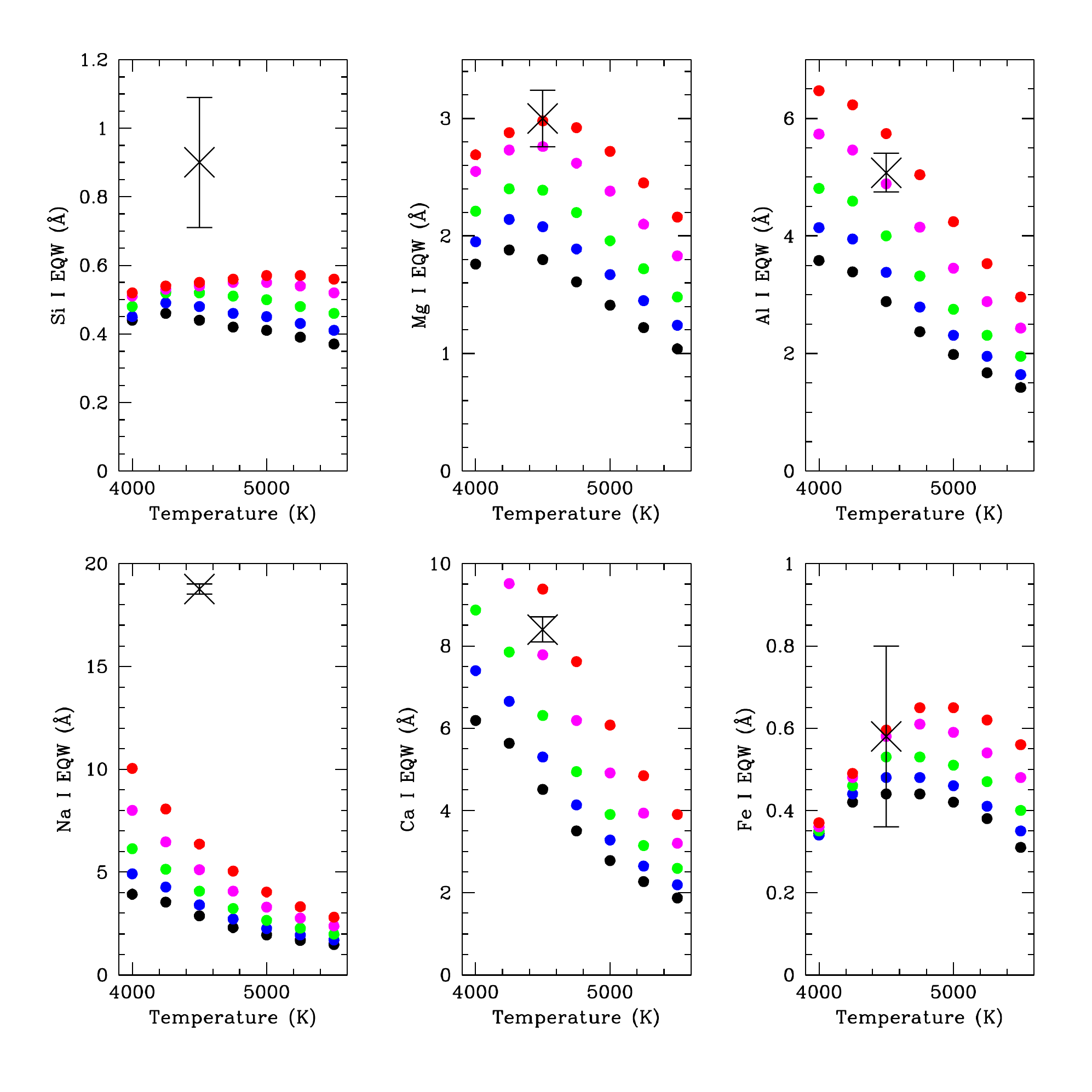}}}
\caption{The same as panel $A$, for the $K$-band.}
\end{figure}
\clearpage

\renewcommand{\thefigure}{33abc}
\begin{figure}[htb]
\centerline{{\includegraphics[width=15cm]{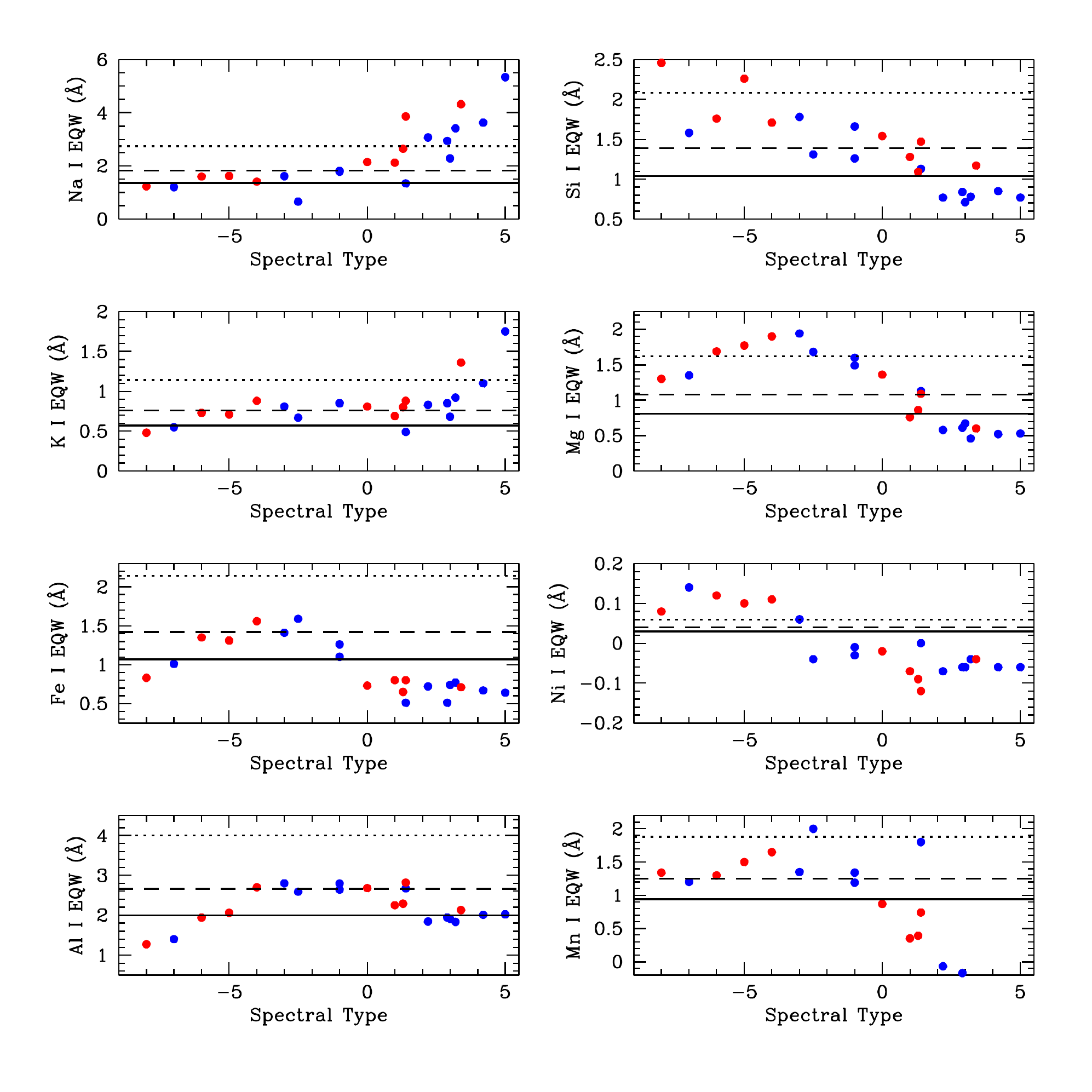}}}
\caption{The equivalent width measurements in the $J$-band for SS Cyg, solid 
lines, compared to those of K and M dwarfs. The latter have been color coded
so that red is [Fe/H] $>$ 0.0, and blue is [Fe/H] $<$ 0.0. The dashed
lines are what the EQWs of SS Cyg would be if a contamination level of 25\%
was removed. The dotted line is for a contamination level of 50\%. 
SS Cyg has a spectral type of $-$4.5.}
\label{sscygmdwarfs}
\end{figure}
\clearpage

\renewcommand{\thefigure}{33b}
\begin{figure}[htb]
\centerline{{\includegraphics[width=15cm]{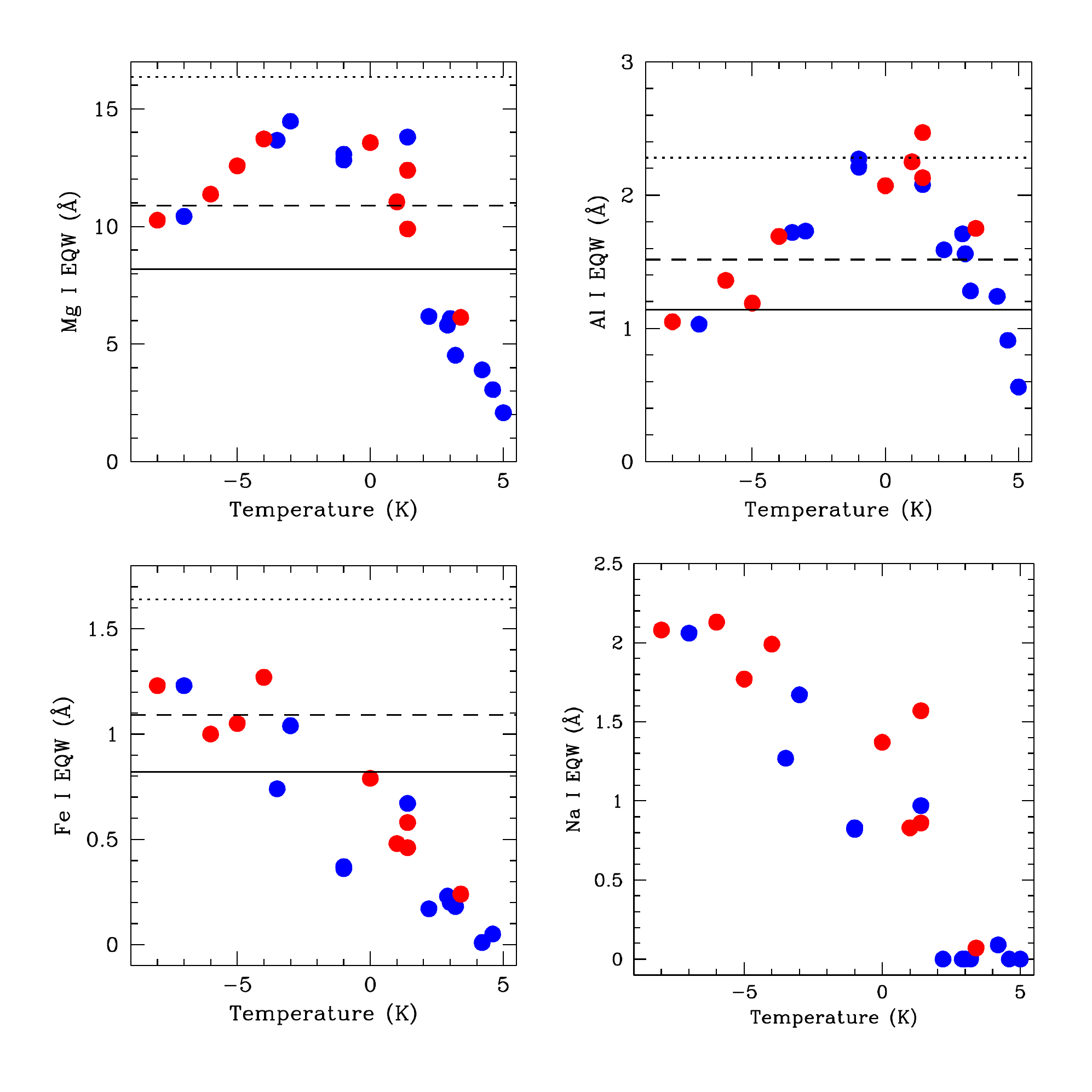}}}
\caption{The same as panel $a$, but for the $H$-band. Due
to an emission feature, the Na I doublet in the $H$-band was unmeasurable.}
\label{sscygmdwarfsH}
\end{figure}
\clearpage

\renewcommand{\thefigure}{33c}
\begin{figure}[htb]
\centerline{{\includegraphics[width=15cm]{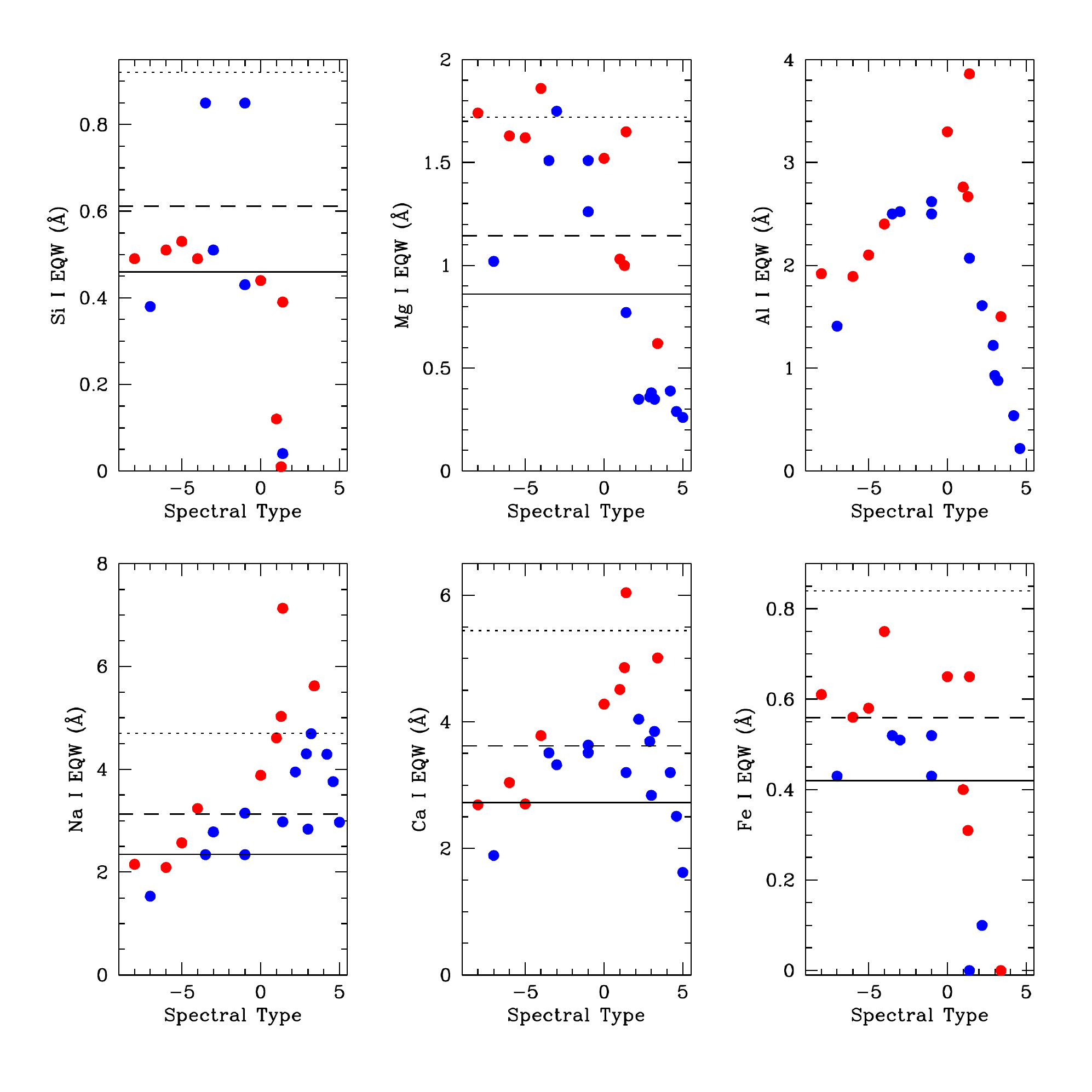}}}
\caption{The same as panel $a$, but for the $K$-band. An He I
emission line corrupts the measurement of the Al I feature at 21100 \AA.}
\label{sscygmdwarfsK}
\end{figure}
\end{document}